\documentclass[bst/sn-mathphys-ay]{sn-jnl}

\usepackage{newtxmath}
\usepackage{amsmath,booktabs}
\usepackage{xcolor}
\usepackage{float}
\usepackage{url}
\usepackage{array}
\usepackage{makecell}
\usepackage{pifont}
\usepackage{subcaption}
\usepackage{threeparttable}
\usepackage[tableposition=below]{caption}
\usepackage{multirow}
\usepackage{graphicx}
\usepackage{adjustbox}
\usepackage{colortbl}
\usepackage{natbib}
\usepackage{longtable}
\newcommand{\cmark}{\textcolor{black}{\ding{51}}}
\newcommand{\xmark}{\textcolor{black}{\ding{55}}}

\begin{document}

\title[Article Title]{Comprehensive Review of Reinforcement Learning for Medical Ultrasound Imaging}


\author[1]{\fnm{Hanae} \sur{Elmekki}}\email{hanae.elmekki@mail.concordia.ca}
\author[1]{\fnm{Saidul} \sur{Islam}}\email{saidul.islam@concordia.ca}
\author[1]{\fnm{Ahmed} \sur{Alagha}}\email{ahmed.alagha@mail.concordia.ca}
\author[2,6]{\fnm{Hani} \sur{Sami}}\email{hani.sami@mail.concordia.ca}
\author[1]{\fnm{Amanda} \sur{Spilkin}}\email{amanda.spilkin@mail.concordia.ca}
\author[1]{\fnm{Ehsan} \sur{Zakeri}}\email{ehsan.zakeri@concordia.ca}
\author[4]{\fnm{Antonela Mariel} \sur{Zanuttini}}\email{antonela-mariel.zanuttini.1@ulaval.ca}

\author*[5,1]{\fnm{Jamal} \sur{Bentahar}}\email{jamal.bentahar@ku.ac.ae}

\author[3]{\fnm{Lyes} \sur{Kadem}}\email{lyes.kadem@concordia.ca}
\author[3]{\fnm{Wen-Fang} \sur{Xie}}\email{wenfang.xie@concordia.ca}
\author[4]{\fnm{Philippe} \sur{Pibarot}}\email{philippe.pibarot@med.ulaval.ca}
\author[5]{\fnm{Rabeb} \sur{Mizouni}}\email{rabeb.mizouni@ku.ac.ae}
\author[5]{\fnm{Hadi} \sur{Otrok}}\email{hadi.otrok@ku.ac.ae}
\author[5]{\fnm{Shakti} \sur{Singh}}\email{shakti.singh@ku.ac.ae}
\author[5,6]{\fnm{Azzam} \sur{Mourad}}\email{azzam.mourad@lau.edu.lb}

\affil*[1]{\orgdiv{Concordia Institute for Information Systems Engineering}, \orgname{Concordia University}, \city{Montreal}, \country{Canada}}

\affil[2]{\orgdiv{Department of Software and IT engineering}, \orgname{Ecole de Technologie Superieure (ETS)}, \city{Montreal}, \country{Canada}}

\affil[3]{\orgdiv{Department of Mechanical, Industrial and Aerospace Engineering},  \orgname{Concordia University}, \city{Montreal}, \country{Canada}}

\affil[4]{\orgdiv{Department of Medicine},  \orgname{Laval University}, \city{Quebec}, \country{Canada}}

\affil[5]{\orgdiv{Department of Computer Science, 6G Research Center},  \orgname{Khalifa University}, \city{Abu Dhabi}, \country{UAE}}

\affil[6]{\orgdiv{Artificial Intelligence \& Cyber Systems Research Center, Department of CSM},  \orgname{Lebanese American University}, \city{Beirut}, \country{Lebanon}}


\abstract{Medical Ultrasound (US) imaging has seen increasing demands over the past years, becoming one of the most preferred imaging modalities in clinical practice due to its affordability, portability, and real-time capabilities. However, it faces several challenges that limit its applicability, such as operator dependency, variability in interpretation, and limited resolution, which are amplified by the low availability of trained experts. This calls for the need of autonomous systems that are capable of reducing the dependency on humans for increased efficiency and throughput. Reinforcement Learning (RL) comes as a rapidly advancing field under Artificial Intelligence (AI) that allows the development of autonomous and intelligent agents that are capable of executing complex tasks through rewarded interactions with their environments. Existing surveys on advancements in the US scanning domain predominantly focus on partially autonomous solutions leveraging AI for scanning guidance, organ identification, plane recognition, and diagnosis. However, none of these surveys explore the intersection between the stages of the US process and the recent advancements in RL solutions. To bridge this gap, this review proposes a comprehensive taxonomy that integrates the stages of the US process with the RL development pipeline. This taxonomy not only highlights recent RL advancements in the US domain but also identifies unresolved challenges crucial for achieving fully autonomous US systems. This work aims to offer a thorough review of current research efforts, highlighting the potential of RL in building autonomous US solutions while identifying limitations and opportunities for further advancements in this field.}




\maketitle

\section{Introduction}
\subsection{Background}
The medical ultrasound (US) scanning, also known as sonography, is the process of using sound-waves to capture non-invasive images from different part of the human body ~\citep{sanches2012ultrasound}. This process typically requires contact between an US transducer (probe) and inside of a body opening or on the skin covering the part of interest, such as heart, lungs, eyes, brain, breast,  kidneys, blood vessels, muscles, fetus, etc ~\citep{chan2011basics}. The US process requires a gel applied on the skin to help in transmitting the sound waves. These waves consists of high-frequency transmission through the probe to the body part, which when reflected would result in a real-time image of the structure through listening to the echoes. The probe is essentially connected to a computing machine that uses an application with a user interface to display the obtained image on the screen ~\citep{chiang2000pc}. The US scanning has been used for more than 20 years, with high records of safety as it is based on non-ionizing radiations as opposed to X-rays or other types of imaging. The scan results are usually required by a medical doctor for a variety of reasons ~\citep{avola2021ultrasound}. For example, diagnosis and decision making for some diseases requires visual representation of the heart or assessing a joint inflammation. In addition, during a surgery, surgeons might require the use of US for guiding through certain procedures. Furthermore, to check the health of the fetus during pregnancy, growing rate, and overall condition, a periodic US scan is necessary. This also applies on checking the well-being of different organs of the human body. The time it takes for sound to travel in addition to the sound amplitude determine the resulting image. There is a variety of parameters that a sonographer can use to configure the quality of the obtained image, including: frequency, wavelength, pulse duration, pulse repetition frequency (PRF), gain, depth, focus, image resolution, etc ~\citep{jiang2023robotic}. More details about each of these parameters, the mode for scanning, and the process of medical US imaging are provided in Section \ref{sec:US_fundamentals}.

There are different ways the US machine is currently controlled, including a trained healthcare provider that performs manual scan, an automated process where the machine is pre-programmed for a certain behavior ~\citep{seitz2020development}, and a remotely controlled probe ~\citep{duan2021tele}. A properly trained health care providers are the only certified entities to perform such scan. In addition to adjusting the probe parameters, there are different aspects of the US that must be handled by the health care providers. For instance, the position of the patient is crucial for obtaining the correct observation of the scanned part. Furthermore, the position and angle of the probe are also crucial in the quality and content of the position obtained. Usually, when the US process starts, the parameters of the probe gets adjusted for a better outcome. After applying the gel, the scanning process starts with a conventional approach of manually controlling the probe and performing actions like sweeping, angling, and rotating. Based on the recommendations provided by the physician or patient's symptoms, the sonographer tailors the scan. The manual scan procedure can be found in hospitals, specialized clinics, or laboratories.

The second approach for US is through an automated process ~\citep{li2021overview}. In such systems, the objective is to reduce or completely remove the need for direct human control during the examination. Typically, the system performs automatic adjustments of the scanning parameters and is responsible for capturing images, or even in some cases perform the entire scan. These machines are mostly reviewed by a human expert who supervises the process with minimal to no intervention. An automated process of US would rely on semi-automated and software algorithms to guide the placement of the probe/transducer or adjust the parameters/settings. An example of automated US machine is the Automated Breast Ultrasound (ABUS) ~\citep{zanotel2018automated}, ideally used for women with dense breast tissue. This automated scanning system uses a robotic arm with pre-programmed movements to move over the breast tissue, capture images, and ensure consistent coverage of the breast.

A third example is the remotely controlled US machines, which typically relies on a health care provider remotely controlling a probe connected to the internet to perform the scan, or what is known as teleportation ~\citep{black2024mixed}. The main objective of such machines is to reduce the travel time for sonographers and making sure that the US service is accessible for patients in rural areas or places with limited availability of health care providers. An example of such a remotely operated system is Butterfly iQ+ ~\citep{greenberg2023butterfly}, which is a portable device that can be connected to a smartphone or tablet and controlled remotely by another user, where data can be shared instantly.

The manual US process is facing multiple challenges following the traditional process. First, the limited number of available and trained health care providers for performing US is causing long delays for getting appointments, performing the scan, and receiving diagnosis. These delays can sometimes affect the health status of the patient and could lead to adverse consequences. Second, given the recommendations by the physician for the scan requirements and the multiple parameters and conditions that the provider should control during US could lead in many cases to making human mistakes, which requires the patient to retake an appointment and perform the scanning again. Similarly, autonomous and remotely operated US machines are also facing many challenges, including the reliance on predefined rules and steps for the software to follow, which is not able to generalize properly to cases that could requires adaptation to unexpected or complex anatomical variations, subtle pathology, or dynamic patient conditions ~\citep{avola2021ultrasound}. In addition, the systems would unexpectedly face technical issues, such as a slow connectivity, which may render the scanning process ineffective and inaccurate ~\citep{moran2020preclinical}.

To this end, there has been recently a major focus between academic researchers and medical companies on the development of Artificial Intelligence (AI)-based solutions to be integrated within the US process to address the aforementioned challenges by enabling efficiency and increased availability and accessibility for such services to a wider range of patients ~\citep{kuang2021articles, sriraam2025comprehensive}. The objective of using AI in this context is to build autonomous US systems capable of controlling the US process. Primarily, Reinforcement Learning (RL) solutions has gained most of the attention, due to its ability in building autonomous and intelligent solutions that could perform effective actions while being able to adapt to environment changes and requirements, which is ideal when dealing with US ~\citep{coronato2020reinforcement}. Automation of the RL process does not only have to be in image acquisition ~\citep{ning2021autonomic}, but also can be responsible for improving the quality of the obtained images ~\citep{jarosik2021pixel}, analyzing the results ~\citep{liu2020ultrasound}, and even helping physicians in giving recommendations or observations about the patient's case through decision making and diagnosis. Building an RL agent requires a clear design of the Markov Decision Process (MDP) environment, necessitating a clear definition of the state and action spaces, the reward function, and the transition matrix if accessible ~\citep{singh2022reinforcement}. A state should cover the necessary information for the agent to define its position in the environment. In addition, it is important to define the goal state of the agent, which is the main target to reach. An action covers the different possibilities that agent can select to take a particular decision in the environment. Based on the current state and decision taken, the agent would result in a new state and receive a reward. This reward is typically used to update the agent policy, which ultimately get updated through the RL algorithm. An optimal decision would be defined as providing the best possible sequence of actions to maximize the average/cumulative future reward. More details about the design of the MDP environment and the different characteristics of various possible RL algorithm to obtain an optimal policy is provided in Section \ref{sec:rl_fundamentals}. Due to its adaptability to dynamic changes in the environment and ability to perform a sequence of decisions, makes it ideal for use in different stages and applications of US for different purposes. Thus, the recent literature has shifted its focus from traditional solution to AI-driven mechanism to resolve challenges in existing US solutions through RL.

Despite their enormous potential, the deployment of RL solutions in real-life settings to tackle the aforementioned US limitations still faces many challenges. First and foremost, RL algorithms require vast amount of data to start building the policy, which makes it particularly challenging to directly test on patients. Furthermore, when building the policy from scratch, RL relies on exploration techniques in an attempt to try different actions in the environment. Afterwards, RL uses the collected experience to form a knowledge about the existing environment through continuously updating its policy. During this period of exploration, multiple mistakes are made by the RL agent to learn, which is not tolerable in the sensitive case of US systems, where a direct interaction is required to form the RL agent, especially in the case of US image acquisition. Instead, there exist multiple alternative solutions in the RL literature to mitigate these challenges, including offline learning, bootstrapping techniques, and the use of a simulated environment. 

\subsection{Motivation}
RL holds significant promise for advancing medical US imaging by enabling automation, precision, and adaptability across key tasks. In \textbf{image acquisition}, RL can optimize probe positioning and navigation for capturing high-quality images. For \textbf{image enhancement}, RL aids in improving image clarity by addressing noise and artifacts inherent to US data. In \textbf{image analysis}, RL facilitates automated interpretation, including segmentation and feature extraction. Moreover, in \textbf{decision-making and diagnosis}, RL supports health care providers by integrating image findings with clinical data for accurate and timely decisions. Despite the existing process in leveraging RL for building or supporting US systems, practical deployment faces challenges such as high-dimensional state spaces and real-time requirements. Current surveys lack a dedicated focus on RL for US imaging, overlooking its unique constraints. To address this gap, we present a comprehensive review exploring the role of RL, challenges, solutions, and future research directions in the field of RL-driven US systems for addressing challenges in traditional US solutions as well as the integration of RL. 

\subsection{Related Surveys}
Following the latest literature surveys, there exist three main categories that covers the use of RL to support US. The first category covers the intersection between AI and US. The second category is related to the literature works that leverage RL in healthcare. Finally, the third existing category, is the use of RL specifically for image processing tasks. In each of these categories, there exists a limited number of surveys that comprehensively study the existing scientific contributions. However, there is no survey that explicitly studies the recent advancements in using RL for supporting US systems. Table \ref{tab:surveys} covers the latest showcases a sample of recent surveys in each of the three categories. In this table, we compare the main focus of the paper, the domain it is applied in, and the taxonomy used.

Deep learning has emerged as a highly promising area with significant applications in the medical field ~\citep{javed2024deep,das2024deep}. Recently, the research community have focused on the use of AI solutions for replacing conventional US systems and addressing its challenges. For instance, authors in ~\citep{tenajas2023recent} have built a taxonomy around the use of AI for guiding the US process and acquiring US planes. Their main contribution revolves around surveying the use of AI solutions in general for solving US challenges, including deep learning. In the same context, several recent surveys exist on the use of robots to perform US, acknowledging that robots would be able to solve challenge related to the shortage of trained sonographers. In ~\citep{bi2024machine}, the different types of ML techniques for supporting robotic US are surveyed, including two main streams: \textbf{(1)} the reliance on an expert to perform actions, or \textbf{(2)} completely autonomous robot manipulation. Similar surveys, such as ~\citep{jiang2023robotic, huang2023review}, also review the latest advancements of using ML for supporting robotic US, but the streams in the taxonomy mainly differs, including considerations for teleportation, semi-autonomously, and fully autonomous. The main literature gap we find in existing surveys covering the intersection between US and AI is the limited focus on a single aspect of the US process, while ignoring the remaining steps. Precisely, the US process is composed of four main steps, including: \textbf{(1)} image acquisition, \textbf{(2)} image enhancement, \textbf{(3)} image analysis, and \textbf{(4)} decision-making and diagnosis. The sample of surveys presented in Table \ref{tab:surveys} focus mainly on one stages of these four US steps, do not have a clear methodology, or do not study the different aspects of the RL agent development.

On the other hand, to showcase the importance of RL for solving challenging problems in healthcare, the RL in Healthcare section in Table \ref{tab:surveys} summarizes a sample of survey papers that study the latest advancements in this regard. As we can see from the presented applications, there are surveys that focus on Medical Imaging ~\citep{zhou2021deep} and intelligent healthcare ~\citep{abdellatif2023reinforcement} in general, while others focus on specific diseases such as cancer ~\citep{khajuria2023review} and diabetes ~\citep{tejedor2020reinforcement}.

Outside the healthcare context, there exist a large number of survey papers that focus on the use of RL for image processing problems in general ~\citep{le2022deep}, which in turn emphasizes the success of RL application in this domain, which can be applied for serving other domains such as autonomous US. On a more focused level, we can also pinpoint other surveys that studies the use of RL for supporting visual navigation under the context of image processing ~\citep{zeng2020survey}, which is vital for the application of US, which is an integral part in the four discussed stages. Furthermore, existing surveys on the use of RL for supporting US lacks a through explanation of the full development pipeline of the RL solution. Particularly, there exist five different stages for effective RL solution development, testing, and improvement, which are: \textbf{(1)} data preparation \& processing, \textbf{(2)} problem formulation, \textbf{(3)} simulation environment, \textbf{(4)} implementation \& training, and \textbf{(5)} validation \& fine-tuning. Identifying existing techniques for each of the RL development stages is of paramount importance, especially for realizing the latest advancements and properly identifying existing challenges that require further investigation. For example, building a simulated environment for establishing an initial version of the RL model is essential to address the high exploration challenge of RL agents. Hence, understanding and identifying the existing simulation technologies and the potential for building additional environment is important for the research community. 

In contrast, our work in this survey attempts to address the research gap in the intersection between US and AI, with a particular focus on RL, which have shown immense attention and advancement in the past decade. Particularly, we study the latest impactful research advancements that have leveraged RL as a solution for the four different stages of the US process. Furthermore, our taxonomy details each of the advancements and challenges of the five RL development stages for each of category in the US process. More details about the contributions of this survey are provided in Section \ref{sec:contributions}.




\begin{table*}[]
    \centering
    \caption{Comparison of Related Surveys}
    \begin{adjustbox}{width=\textwidth}
    \begin{tabular}{|p{1.3cm}|p{0.8cm}|p{0.5cm}|p{2.2cm}|p{4.0cm}|p{3.5cm}|p{4.0cm}|}
        \hline
        \textbf{Ref.} & \textbf{Year} & \textbf{RL} & \textbf{Methodology} & \textbf{Main Focus/Scope} & \textbf{Application-specific Domain} & \textbf{Taxonomy} \\ 
         \hline
         \hline
         \multicolumn{7}{|l|}{\textbf{AI for US Scanning}}\\
         \hline
         \hline
         ~\citep{tenajas2023recent} & 2023 & \xmark & \xmark & AI-assisted US scanning, for improving diagnostic accuracy, standard plane recognition, organ identification, and scanning guidance & US & Acquiring US planes using AI and US guidance using AI\\
         \hline
         ~\citep{bi2024machine} & 2024 & \xmark & \xmark & The different types of ML for robotic US & Robotic US & Comparing modular and direct approaches for action reasoning, which either relies on an expert or completely on machine learning\\
         \hline
         ~\citep{jiang2023robotic} & 2023 & \xmark & \cmark & The different AI techniques involved in teleport and autonomous US &
         Robotic US & Teleportation v.s. autonomous\\
         \hline
         ~\citep{huang2023review} & 2023 & \xmark & \xmark & Three mainstreams of robotic US, their techniques, and applications & Robotic US & Teleportation, semi-autonomous, and guided intervention systems\\
         \hline
         \hline
         \multicolumn{7}{|l|}{\textbf{Reinforcement Learning in Healthcare}}\\
         \hline
         \hline
         ~\citep{zhou2021deep} & 2021 & \cmark & \xmark & A background on deep reinforcement learning and its applications in medical imaging & General Medical Imaging & Parametric medical image analysis and solving optimization problems.\\
         \hline
         ~\citep{abdellatif2023reinforcement} & 2023 & \cmark & \cmark & Using reinforcement learning for supporting i-health through IoT and smart sensors & Intelligent healthcare (IoT) & Application domains, system components, and evaluation metrics\\
         \hline
         ~\citep{coronato2020reinforcement} & 2020 & \cmark & \cmark & Reinforcement learning to personalize treatment and advance precision medicine & Intelligent healthcare & Healthcare applications: medical imaging, diagnosis, emergency care, at-home treatment\\
         \hline
         ~\citep{khajuria2023review} & 2023 & \cmark & \xmark & Reinforcement learning techniques for cancer diagnosis and treatment & Cancer & Image classification, segmentation, dynamic treatment regimes, problem modeling and evaluation metrics\\
         \hline
         ~\citep{xu2022deep} & 2022 & \cmark & \cmark & Deep reinforcement learning for medical imaging and radiology therapy treatment & Medical imaging \& Radiation Therapy & Medical image registration, reconstruction, segmentation, organ detection, and radiology therapy planning\\
         \hline
         ~\citep{tejedor2020reinforcement} & 2020 & \cmark & \xmark & Reinforcement learning for blood glucose control algorithms & diabetes blood glucose control & RL methods for blood glucose control: environment formulation \& algorithms \\
         \hline
         \hline
         \multicolumn{7}{|l|}{\textbf{AI for Image Processing}}\\
         \hline
         \hline
         ~\citep{le2022deep} & 2022 & \cmark & \cmark & Deep reinforcement learning and its applications in computer vision & Computer vision & Landmark \& object detection, object tracking, image registration \& segmentation\\
         \hline
         ~\citep{zeng2020survey} & 2020 & \cmark & \cmark & Visual navigation using deep reinforcement learning from virtual to real robots & Visual Navigation & Direct, hierarchical, and multi-task deep reinforcement learning\\
         \hline
         \hline
         \textbf{Our work} & \textbf{2025} & \cmark & \cmark & \textbf{A comprehensive survey of different stages of the RL development pipeline serving each of the US processing stages} & \textbf{Autonomous US} & \textbf{The five stages of the RL development pipeline serving each of the four US processing stages, including image acquisition, enhancement, analysis, and decision-making \& diagnosis}.\\
         \hline
    \end{tabular}
    \end{adjustbox}
    \label{tab:surveys}
\end{table*}

\subsection{Contributions}
\label{sec:contributions}
This survey provides a comprehensive study of the various literature efforts centered on leveraging RL and its capabilities as a solution for US imaging challenges. Specifically, it presents an objective analysis of the entire RL development pipeline, consisting of five stages: \textbf{(1)} data preparation and processing, \textbf{(2)} MDP problem formulation, \textbf{(3)} simulation environment, \textbf{(4)} implementation and training, and \textbf{(5)} validation and fine-tuning. While these steps are interdependent, they can also be treated as separate problems when necessary. Each stage of the RL pipeline is carefully examined in the context of four key categories of problems that constitute the US medical pipeline: \textbf{(1)} Image Acquisition, \textbf{(2)} Image Enhancement, \textbf{(3)} Image Analysis, and \textbf{(4)} Decision-Making and Diagnosis. These steps in the US medical pipeline are sequential, cohesive, and integral to achieving a complete patient diagnosis through US imaging.

To address the gaps in existing surveys, this paper begins by establishing the foundation of RL, including its core principles, key components, and primary algorithms. Following this, the survey introduces a novel taxonomy that highlights recent advancements in the US medical field enabled by RL technology. By aligning the RL pipeline with the US medical pipeline, twenty distinct categories are identified, within which each paper is systematically reviewed. For each study, we detail its contributions, experimental approaches, limitations, and future directions. Additionally, we discuss the broad limitations and challenges across the categorized studies and propose a comprehensive list of research opportunities for the community to explore and build upon.

To the best of our knowledge, this is the first survey to provide a structured and detailed cross-pipeline taxonomy for addressing US medical challenges using RL. The contributions of this survey are summarized as follows:

\begin{itemize}
    \item Offers an in-depth review of US medical challenges and the intersection of recent RL advancements as potential solutions.
    \item Provides a foundational understanding of RL, covering the entire development pipeline from data preparation to validation and fine-tuning. Additionally, it delivers the necessary background to comprehend the stages of the US medical pipeline, from image acquisition to diagnosis, emphasizing the characteristics of these challenges.
    \item Develops an innovative taxonomy that integrates the stages of the US medical pipeline with the phases of the RL development pipeline.
    \item Conducts a multi-class classification and detailed analysis of literature efforts that address US medical challenges through one or more stages of the RL pipeline.
    \item Identifies and discusses key challenges and open research opportunities for advancing the intersection of medical imaging and RL, targeting both medical and engineering domains.
\end{itemize}

We believe that this survey will serve as a cornerstone for promoting a unified understanding of the RL pipeline and its application to US imaging challenges. By mapping the relationships between RL advancements and US medical needs, this work provides a platform for the research community to build upon and promotes a holistic view of the interdisciplinary efforts required to address these complex problems.

\subsection{Survey Methodology and Outline}
In this section, we describe each of the steps followed to gather the literature work and categorize them according to the proposed taxonomy. The Preferred Reporting Items for Systematic Reviews and Meta-Analyses (PRISMA) framework ~\citep{Pagen71} has been used to perform a systematic and effective literature search across the web. The steps followed in this survey according to PRISMA include: (1) Initial Literature Search, (2) Removal of Duplicates, (3) Literature Screening, (4) Full Review, (5) Categorization \& Taxonomy. Each of these steps is detailed in the following passage:

\begin{enumerate}
    \item \textit{Initial Literature Search}: The process of obtaining high quality and recent journals and conferences from medical sources while focusing on the integration between US and RL involves the following steps to identify: the digital libraries, the search keywords and combinations, a set of search queries, and a reliable source of articles (surveys, journals, magazines, and conferences).
The digital libraries selected include the following academic databases: IEEE Xplore, ScienceDirect (Elsevier), and ACM Digital Library and Google Scholar. These databases are selected as they are the main publishers for medical-related journals. In addition, the search keywords used to collect relevant articles from these databases are medical keywords combined with the ``Reinforcement Learning" keyword. These medical keywords include ``Health", ``Medical", ``Medicine", ``Healthcare", ``Disease", ``Pathology", ``Pathologic", ``Ultrasound", ``Organ", ``Pandemic", ``Diagnosis", ``Tumor", and ``Surgery". Afterward, the search queries are built. In each search query, the keyword for search within the title is ``Reinforcement Learning", while the rest of keyword combinations are used for abstract search. Additionally, the search only considers recent articles for the range of years between 2020 and 2024. The following is the main query used for collecting the initial set of articles:
    
    \begin{itemize}
    \item ``Reinforcement Learning" AND (Health OR Medical OR Medicine OR Healthcare OR Disease OR Pathology OR Pathologic OR Ultrasound OR Organ OR Pandemic OR Diagnosis OR Tumor OR Surgery)
    \end{itemize}
    This search resulted in around 592 initial articles.

    \item \textit{Removal of Duplicates}: 
    In this step, we categorize the articles based on the specific problem they address. To do this, we first read the abstract of each paper to filter out those that are not directly related to US. The remaining papers are then classified into one of four categories: `RL for US Image Acquisition', `RL for US Image Enhancement', `RL for US Image Analysis', and `RL for US-image-based Decision-making and Diagnosis'. During this process, duplicated papers addressing the same issue with similar solutions are removed. This filtration step reduces the number of articles to approximately 204, which are then moved on to the next stage of Literature Screening.
    \item \textit{Literature Screening}: At this stage, screening efforts have taken place to identify the articles that are directly related to US, while using RL as the main solution to overcome the considered problem. This process considers the title, abstract, conclusion, and publication year. As mentioned earlier, only articles that were published in the past four years are considered. The number of articles after the screening stage are reduced to around 93 articles. 
    \item \textit{Full Review}: The resulting articles from the literature screening step are subject to a full review at this stage. During the full review process, a thorough review has been given to: problem statement, contributions, proposed system/architecture/solution, experimental setup, dataset, experiment results, and analysis. Following the detailed review, pre-prints were prioritized for removal, since they did not go through a peer review process. The resulting number of articles considered for categorization are around 70 articles. As illustrated in Figure \ref{fig:statistic}, the distribution of the papers selected for this survey is presented. The figure clearly demonstrates that the interest in RL within the field of US has significantly increased in recent years, emerging as a prominent area of research.

        \begin{figure}
    \centering
    \includegraphics[width=0.9\linewidth]{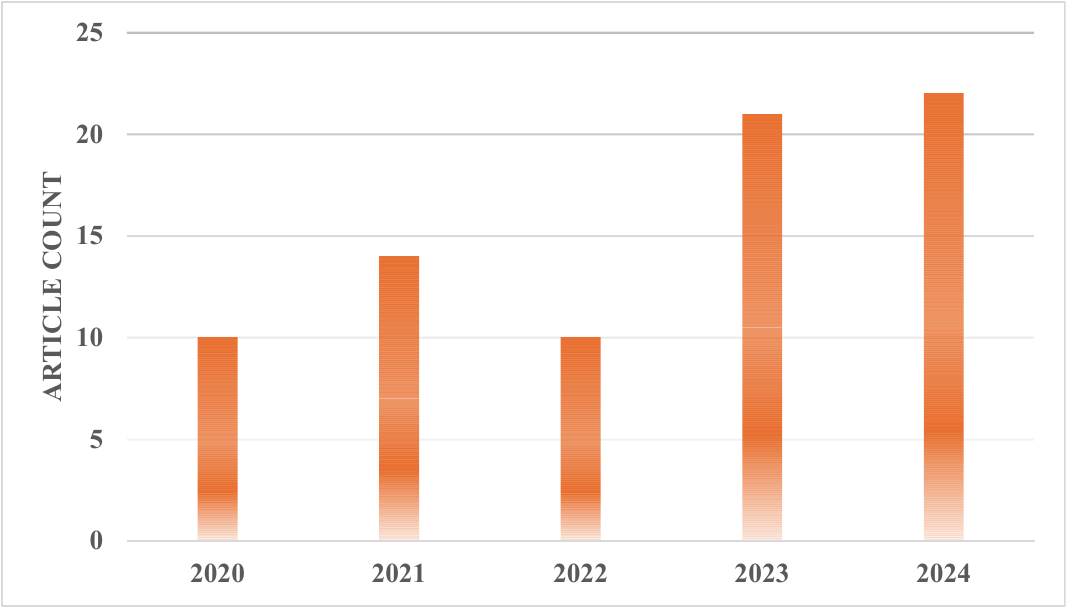}
    \caption{Statistics of the Articles Investigated in this Survey Paper.}
    \label{fig:statistic}
\end{figure}

    \item \textit{Categorization \& Taxonomy}:
    In this phase, the filtered articles are assigned to their corresponding categories form the taxonomy. In the US categorization, there exist four categories, which include: Image Acquisition, Image Enhancement, Image Analysis, and Decision-Making \& Diagnosis. Each of the four queries used for the Initial Literature Search is related to a US category in our proposed taxonomy. To further distribute the papers, a RL categorization is considered, covering the five main steps for RL development. This split has resulted in 20 different classification of the articles. Some of the articles have been repeated in different categorizations in case the proposed solution includes several steps of RL. 
\end{enumerate}

\begin{figure}
    \begin{center}
    \centering
\includegraphics[width=0.7\linewidth]{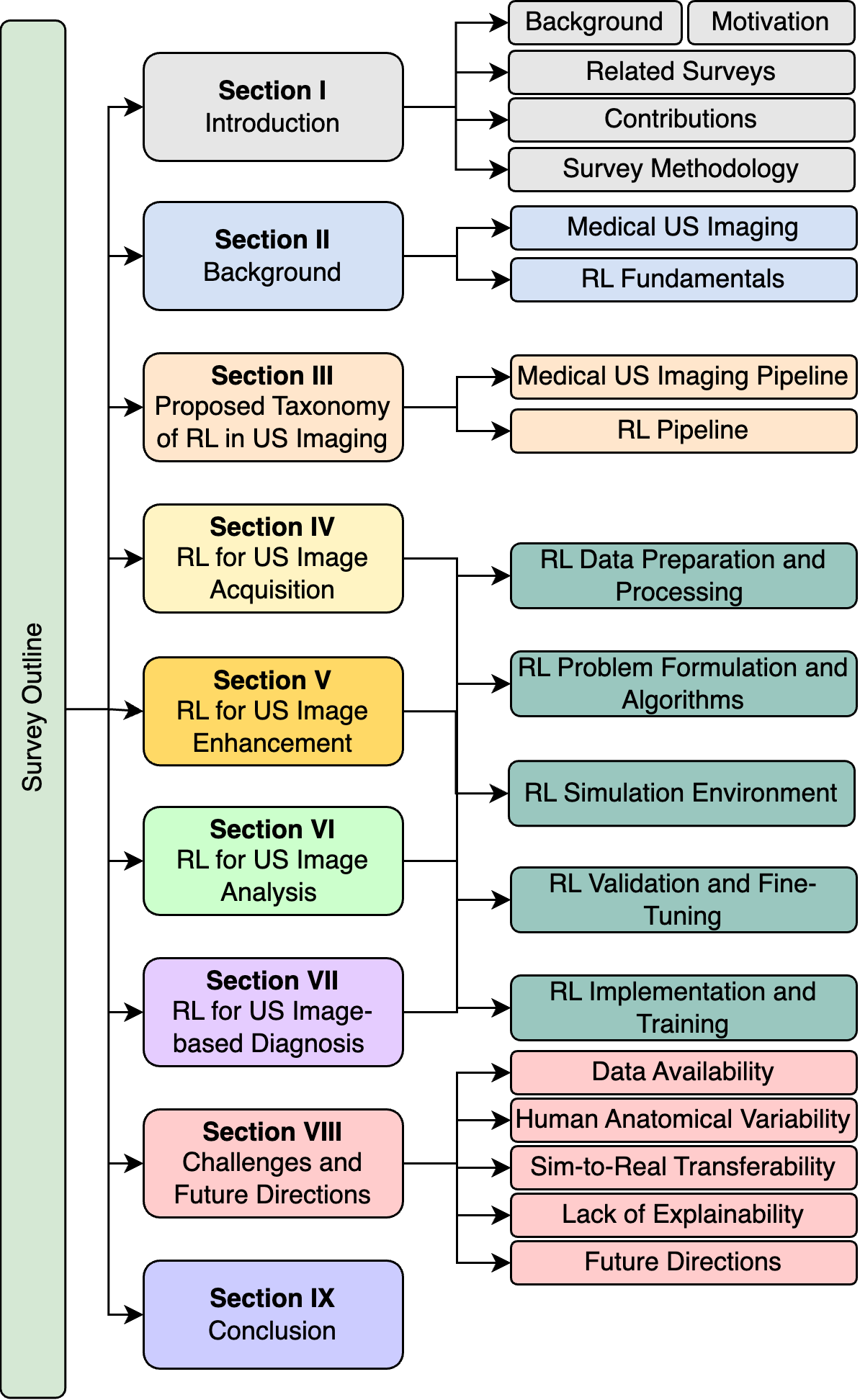}
   \caption{Survey Outline (Ultrasound (US) Imaging \& Reinforcement Learning (RL)).}
   \vspace{-30pt}
    \label{fig: survey outline}
    \end{center}
\end{figure}

The remainder of the survey is organized as follows. Section \ref{Sec:Background} gives background information about medical US imaging and RL. The proposed taxonomy is presented and discussed in Section \ref{Sec: Taxonomy}. Sections \ref{Sec: RLImageAcq}-\ref{Sec: RL Diagnosis} survey the existing literature with regards to the RL pipeline in each of the US imaging stages. Section \ref{Sec: Challenges} discusses the challenges and future directions, while Section \ref{Sec: Conc} concludes the survey. Figure  \ref{fig: survey outline} gives a detailed presentation of the survey outline.

\section{Background}
\label{Sec:Background}
This section provides background information on US imaging and RL. It first introduces the fundamentals of US imaging along with its different modalities and its typical pipeline. The section then transitions to RL and its key principles and formulations, as well as key RL algorithms. Together, covering these topics aims to create an understanding of the key concepts underlying the survey.

\subsection{Medical US Imaging}
\label{sec:US_fundamentals}

US imaging is an important technique for medical diagnosis. It is known for its real-time imaging capabilities, non-invasive nature, and usability in various medical applications. It is commonly used for tasks such as the visualization of soft tissues (i.e. fat, muscle, etc), blood flow monitoring, and the evaluation of heart functions. This section presents an overview of the fundamentals of US imaging, its different modalities, and the common pipeline that guides its use in medical practices.

\begin{figure}
    \centering
    \includegraphics[width=0.6\linewidth]{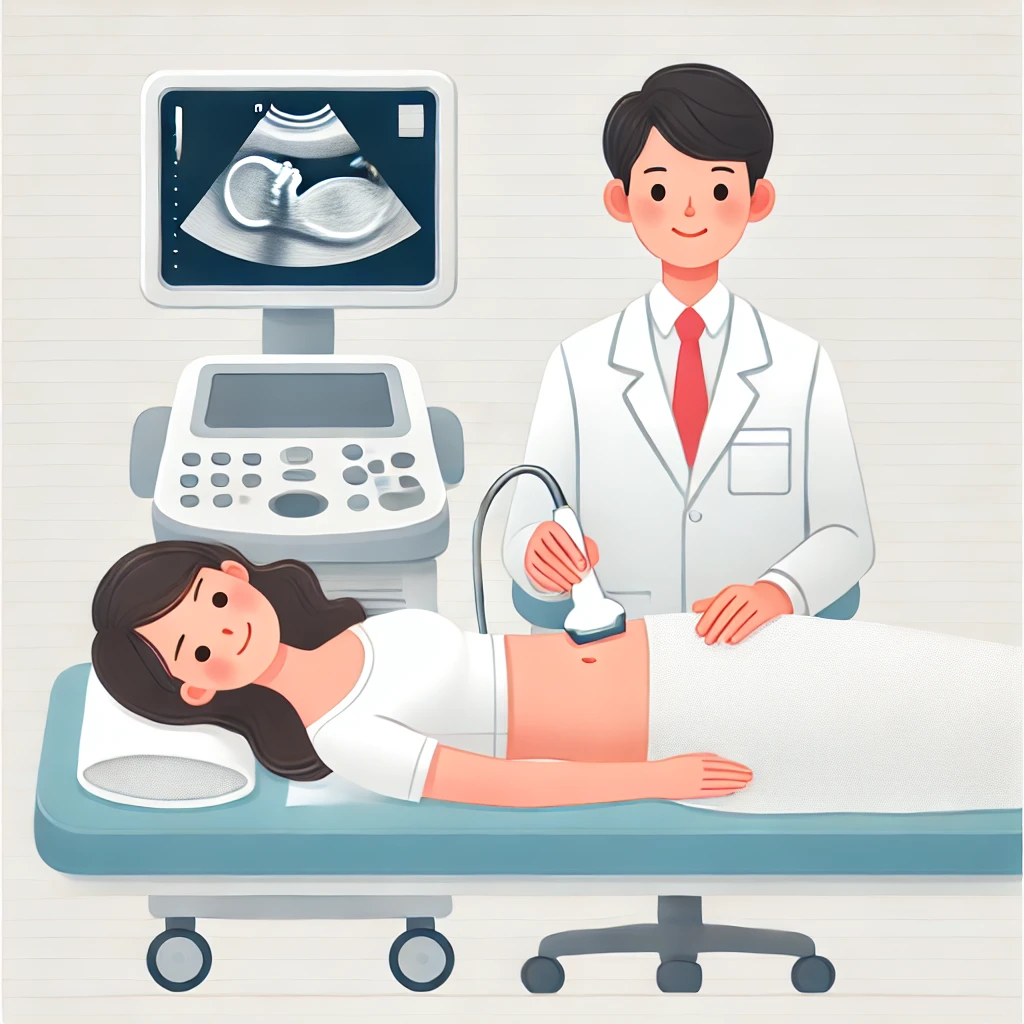}
    \caption{Ultrasound Imaging Illustration}
    \label{fig:US Imaging Illustration}
\end{figure}

\hfill
\subsubsection{Basics of US Imaging}
\hfill

In US imaging, a transducer is used to produce US (high frequency) waves to visualize the internal anatomical structures of the body in a non-invasive manner. A transducer is used to generate and emit these sound waves into the body, which interact with tissues and organs based on their acoustic properties. Fractions of the sound waves are typically reflected by boundaries between tissues of different acoustic impedances, and such reflected waves (echos) then received by the transducer and processed to form the image. The magnitude of reflection depends on the difference of impedance between such tissues, which helps in distinguishing the different tissue types (soft tissues, bones, fluids, etc).

Despite its numerous advantages, US imaging faces several challenges that need to be addressed to improve its effectiveness. US imaging typical exhibits low spatial resolution and many artificats due to ultrasonic diffraction ~\citep{komatsu2021towards}. This is mainly because sound waves get attenuated as they travel through tissues, which leads to reducing the signal strength and results in lower image quality. The interpretation of US images is complicated by the existence of artifacts, such as speckle noise and refraction, which requires efficient image processing and enhancement to enhance the quality of the image. Furthermore, accurate acquisition and diagnosis of US images relies heavily on the operator's expertise and skills, which could vary between different operators, introducing more variability to the obtained images. Addressing these limitations is key to achieving more reliable and higher-quality US scanning.

\hfill
\subsubsection{US Imaging Modalities}
\hfill

\begin{figure*}[ht]
    \centering
    \begin{subfigure}[b]{0.3\textwidth}
        \centering
        \includegraphics[width=\textwidth]{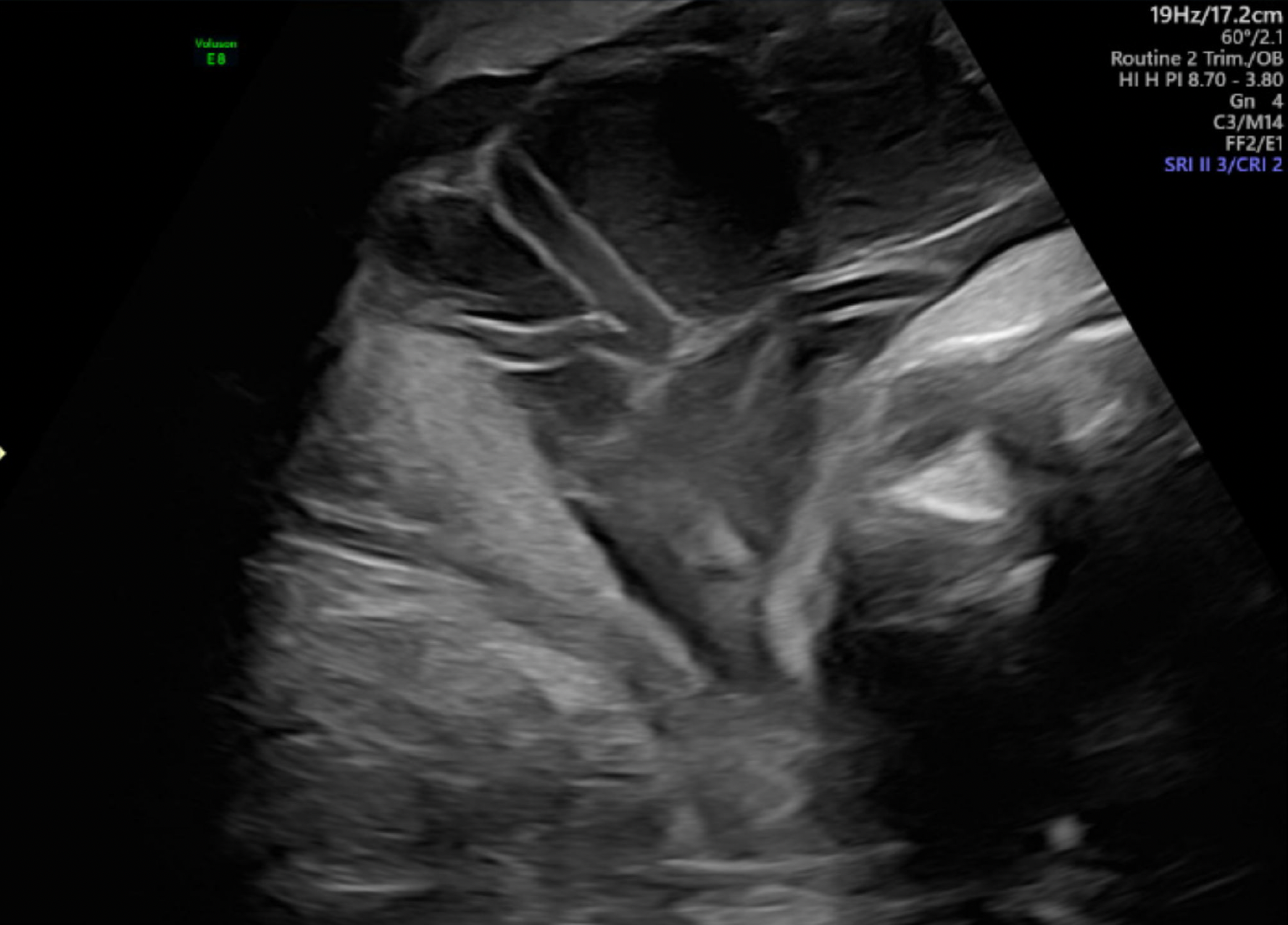}
        \caption{2D US}
        \label{fig:2D Ultrasound}
    \end{subfigure}
    \hfill
    \begin{subfigure}[b]{0.3\textwidth}
        \centering
        \includegraphics[width=\textwidth]{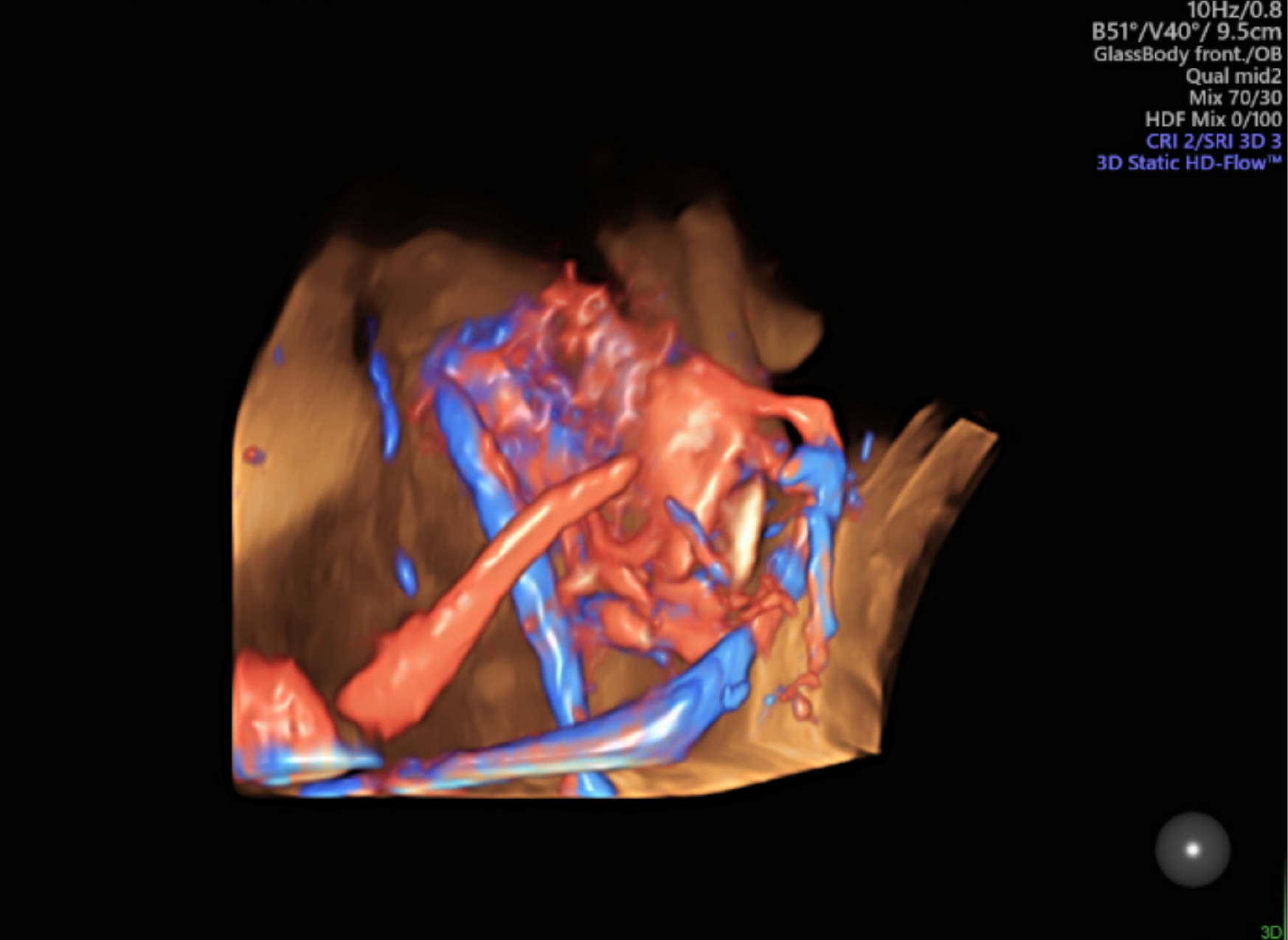}
        \caption{3D US}
        \label{fig:3D Ultrasound}
    \end{subfigure}
    \hfill
    \begin{subfigure}[b]{0.3\textwidth}
        \centering
        \includegraphics[width=\textwidth]{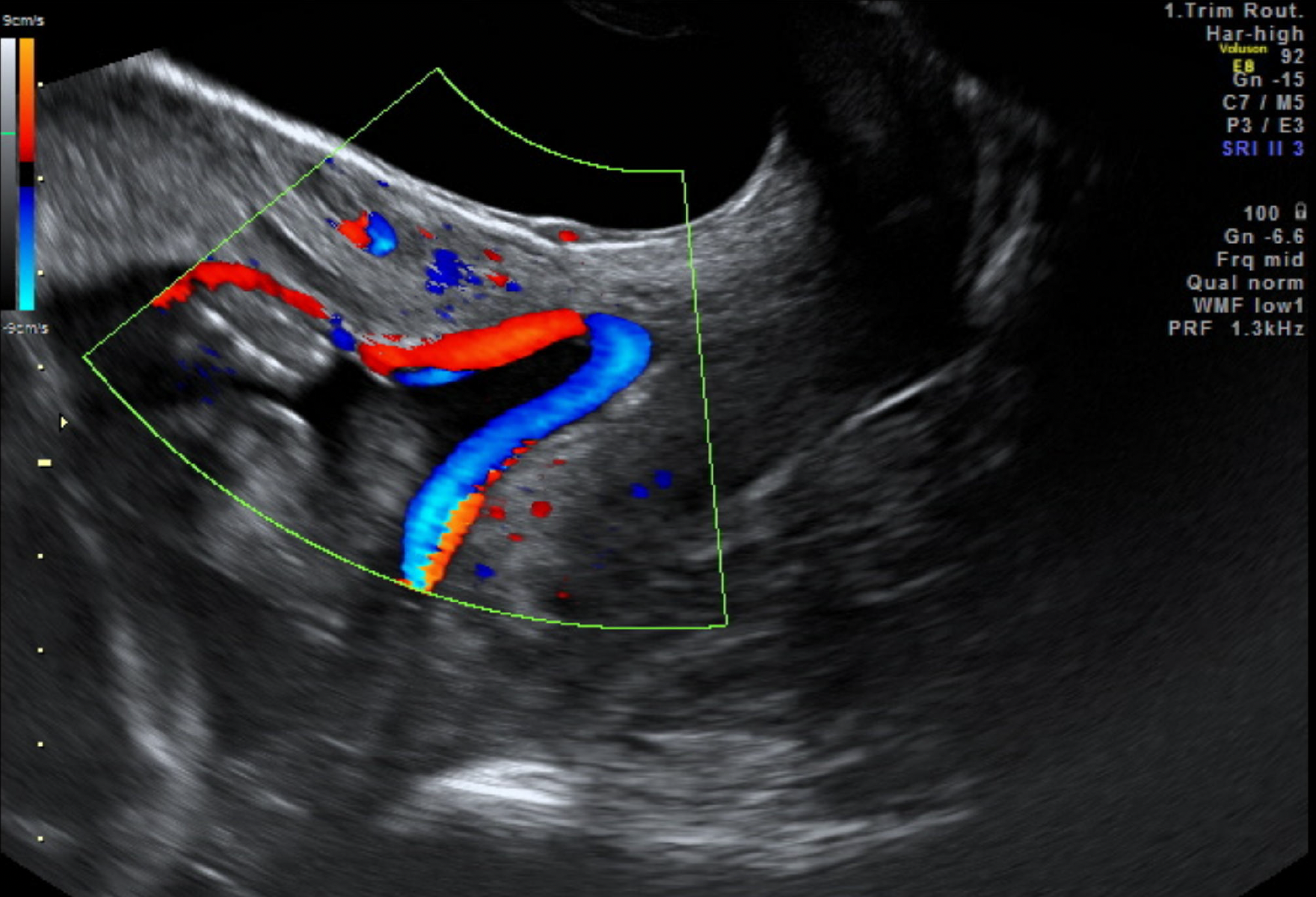}
        \caption{Color Doppler US}
        \label{fig:Color Doppler US}
    \end{subfigure}
    \caption{The Different Modalities of Ultrasound (US) Imaging ~\citep{bohilțea2022clinically}.}
    \label{fig:US Modalities}
    \vspace{-20pt}
\end{figure*}

US imaging can come in different modalities, each providing certain features beneficial to specific diagnostic needs and medical applications. The most common modalities are Two-dimensional (2D) US, Three-dimensional (3D) US, and Doppler US, as illustrated in Figure \ref{fig:US Modalities}.

2D US is the commonly used modality in practice. It captures a single plane of real-time grayscale images of internal anatomical structures of the body. Due to its simplicity and speed, it is popular for dynamic imaging, where practitioners can observe moving structures like the heart. 2D US, however, has some limitations when it comes to capturing complex anatomica structure, which is mainly due to its inability to provide comprehensive volumetric information.

3D US extends the capabilities of 2D US imaging by capturing volumetric data. This enables more detailed representations of the internal anatomical structures in the body. This modality can be acquired as static 3D volumes or as real-time 4D imaging, where time acts as a 4th dimension. Generally, 3D US requires advanced equipment and high computational resources, in addition to advanced expertise for acquiring and analyzing 3D US images. Additionally, when compared to 2D US, 3D US imaging has lower spatial and temporal resolution, which could present challenges when analyzing the images for diagnosis.

While 2D/3D US imaging aim to create static real-time images of the internal body structures, Doppler US is another component of US technology that focuses on measuring and visualizing the movement of fluids (e.g. blood) within the body. It uses the Doppler effect, which detects the frequency shift of sound waves as they get reflected off moving objects like red blood cells. This allows gathering information about the blood flow such as its patterns, speed, and direction in vessels. On common type of Doppler US is color Doppler, which displays blood flow direction and velocity using color coding. Doppler US is essential in diagnosing many problems related to blood flow, such as cardiovascular conditions including heart valve diseases and blocked arteries.

\subsection{Reinforcement Learning Fundamentals}
\label{sec:rl_fundamentals}
Reinforcement Learning (RL) is a machine learning paradigm for sequential decision-making, which aims to optimize an agent's interactions with an environment ~\citep{sutton2018reinforcement}. RL is based on rewarded interactions with the environment, where an agent takes an action based on the current state of the environment, and receives a reward (feedback) that guides the learning process. This makes RL suitable for problems involving dynamic systems with sequential decision-making, such as robot swarms ~\citep{alagha2024adaptive, alagha2023blockchain}, video games ~\citep{ye2020towards}, and computer vision ~\citep{alagha2023multiagent, zheng2023learning}. This section discusses the fundamentals of RL and its key algorithms, and presents a general pipeline followed when developing and implementing RL solutions.

Throughout the RL process, the aim is to learn an optimal policy that guides the agent's interactions with an environment such that the cumulative rewards are maximized ~\citep{sutton2018reinforcement}. This iterative process involves four main components: policy, reward, value function, and a model of the environment. These components formulate a MDP, which is a mathematical model for sequential decision-making that governs the dynamics of the environment.

The agent's \textbf{policy} ($\pi$) is responsible for mapping a state ($s$) of the environment into an action ($a$). The policy can be deterministic (i.e. $\pi(s) = a$) giving a specific action per state, or stochastic (i.e. $\pi(a|s)$) giving a probability distribution over the possible actions for a given state. The policy governs the decision-making of an agent, as it determines what action the agent should take based on the state of the environment. The main aim of an RL algorithm is to find the optimal policy ($\pi^*$) that maximizes the expected cumulative reward.

The \textbf{reward} function ($R$) gives feedback to the agent assessing the quality of its actions. The agent's behavior is modified such as the cumulative reward (return) is maximized, which is defined as the discounted sum of future rewards given as:

\begin{equation}
    G_t = \sum_{k=0}^\infty \gamma^k R_{t+k}
\end{equation}
where $t$ is the current timestep and $\gamma$ is the discount factor that balances the importance of immediate vs future rewards. A reward function could be sparse or shaped. Sparse rewards are simple functions that give minimal feedback to the agent in certain time steps (typically when the task is finished), which makes the learning challenging. Shaped reward functions give more frequent feedback to the agent throughout the learning process, which speeds up the learning process, even though they require intricate design and, sometimes, computational capabilities.

The \textbf{value function} estimates the future rewards given a certain starting state (or state-action pair). It helps the agent assess the potential of states/actions, which can guide the decision-making process that aims to maximize the expected cumulative reward. The state-value function ($V(s)$) for a given policy $\pi$ estimates the expected cumulative reward if the agent starts from state $s$ and follows $\pi$ in its decision-making. It is generally expressed as:

\begin{equation}
    V(s) = \mathbb{E}_\pi \left[ \sum_{t=0}^\infty \gamma^t R_{t+1} \mid s_0 = s \right]
\end{equation}

On the other hand, the action value function ($Q(s, a)$) estimates the expected return if an agent takes action $a$ in state $s$ then follows policy $\pi$ onward. It is expressed as:

\begin{equation}
    Q(s, a) = \mathbb{E}_\pi \left[ \sum_{t=0}^\infty \gamma^t R_{t+1} \,\middle|\, s_0 = s, a_0 = a \right]
\end{equation}

Both the state and action value functions can be expressed recursively using Bellman equations, given as:

\begin{equation}
    V^\pi(s) = \mathbb{E}_\pi \left[ R(s, a) + \gamma V^\pi(s') \right]
\end{equation}

\begin{equation}
    Q^\pi(s, a) = \mathbb{E}_\pi \left[ R(s, a) + \gamma Q^\pi(s', a') \right]
\end{equation}

The two value functions are central to many different RL algorithms, and are used to iteratively improve the policy throughout the training process.

Finally, the \textbf{model} encapsulates the dynamics of the environment, including the state transition function ($P(s'|s,a)$) and the reward function ($R(s, a)$). The state transition function defines the probability of transitioning into a new state $s'$ from the current state $s$ if action $a$ is taken. It basically models the dynamics of the environment, which vary from one problem to another. Some state transition functions are deterministic (i.e. an action in a certain state always leads to the same new state), while others are stochastic (i.e. there is a probability distribution over possible new states given a specific action in the current state). In model-based RL, the agent uses this model for decision-making and simulating future states and outcomes. In contrast, the agent in model-free RL relies instead on direct interaction with the environment to learn optimal policies.

The four components of RL are typically modeled within an MDP that formulates the interactions between the agent and the environment. An MDP is defined by a tuple $(S, A, P, R, \gamma)$, where:

\begin{itemize}
    \item $S$: The set of all possible states in the environment.
    \item $A$: The set of actions available for the agent.
    \item $P(s'|s, a)$: The transition probability function, representing the likelihood of moving to state $s'$ after taking action $a$ in state $s$.
    \item $R(s, a)$: The reward function, which assigns a scalar feedback signal to the agent after taking action $a$ is state $s$.
    \item $\gamma$: The discount factor, where $0 \leq \gamma \leq 1$, which balances the importance of future rewards compared to immediate rewards.
\end{itemize}
The agent’s objective in an MDP is to learn an optimal policy $\pi^*(a|s)$ that maximizes the expected cumulative reward ($G_t$).

The standard MDP formulation discussed above could be further extrapolated to fit the nature of the problem and its requirements. For example, Partially Observable Markov Decision Processes (POMDPs) extend MDPs for environments where agents have no full knowledge of the environment, and can partially observe the state. Another extrapolation is Multi-Agent MDPs (MMDPS), which consider multiple agents co-existing in the same environment and collaborate or compete to achieve tasks. In Hierarchical MDPs (H-MDP), the decision-making is structured into sub-tasks, where outcomes from some tasks trigger and gear the decision-making in other tasks. Another form is Constrained MDPs (CMDP), where additional constraints such as resource limitations and resource requirements are introduced to limit the decision-making within certain thresholds. The different variations of MDPs are essential to be considered according to the problem in hand and its requirements.


\section{Proposed Taxonomy of RL in US Imaging}
\label{Sec: Taxonomy}

\begin{figure*}
    \begin{center}
    \centering
\includegraphics[width=1.0\linewidth]{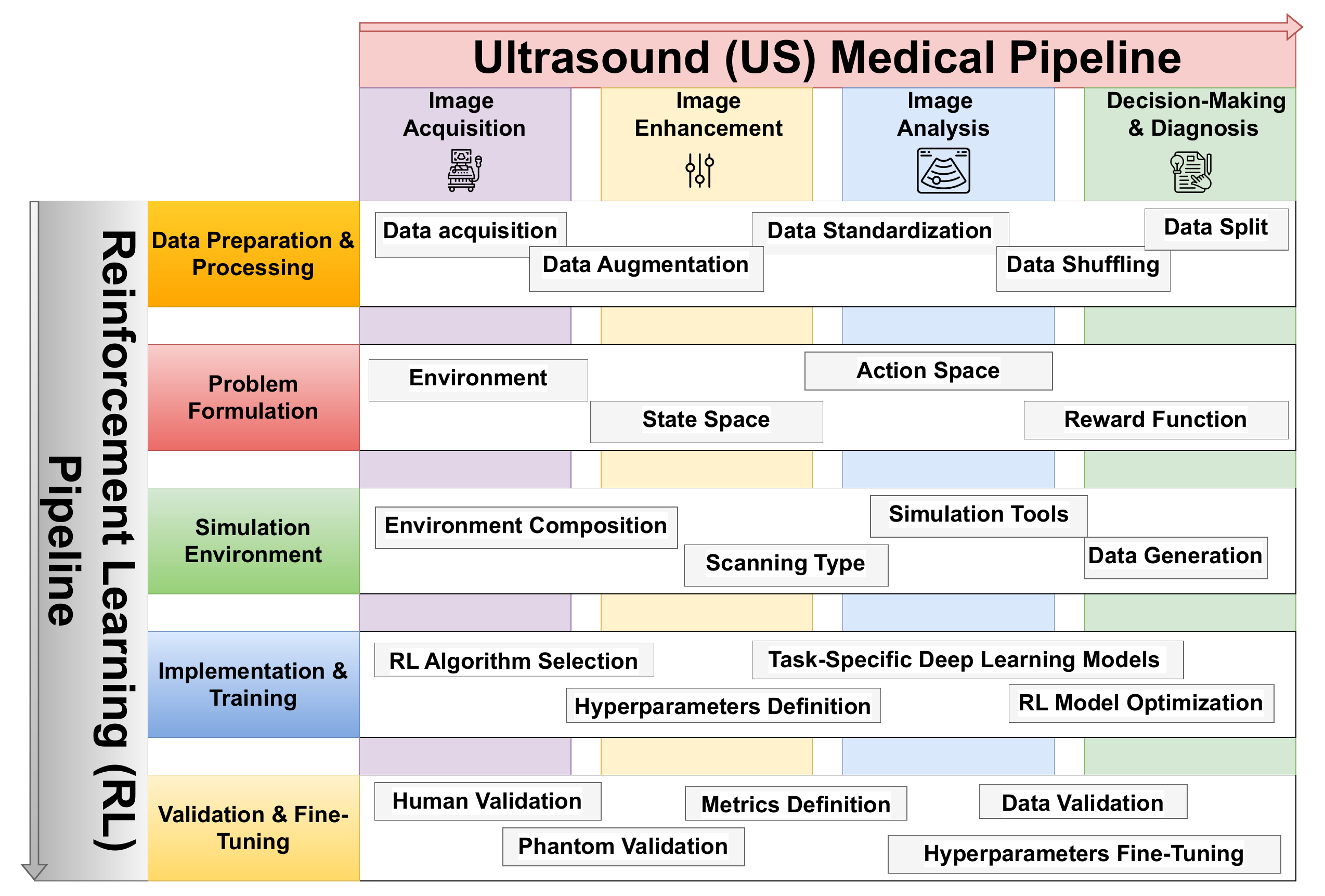}
    \caption{Taxonomy of Reinforcement Learning (RL) in Ultrasound (US) Imaging.}
    \vspace{-30pt}
    \label{taxonomy}
    \end{center}
\end{figure*}

The proposed taxonomy aligns the stages of US imaging with a structured RL pipeline, which is applied to each of the US imaging stages. The taxonomy is summarized in Figure \ref{taxonomy}, which shows the stages of US images, where in each stage the entire RL pipeline is applied. The two pipelines are further detailed in the sections below.

\subsection{Medical US Imaging Pipeline}

A typical medical US imaging process follows a pipeline that can be summarized in 4 stages, being 1) US image acquisition, 2) image enhancement, 3) image analysis, and 4) decision-making and diagnosis. 

In the \textbf{US image acquisition} stage, real-time US scans are captured using the US probe (transducer), after preparing and positioning the patient. During this process, several transducer settings, such as frequency, gain, and depth, are varied aiming to get the optimum resolution and image clarity. The stage is generally carried out by highly trained sonographers, as it requires expertise in correctly placing the probe, adjusting its settings, and understanding the target anatomical structure. Typically, this stage often faces challenges such as operator dependency and variable image quality. 

Following the acquisition of US images, they typically undergo \textbf{image enhancement} steps to improve their quality and reduce artifacts. As discussed previously, artifacts like speckles can cover and obscure important anatomical details, which could affect the quality of the information extracted and ultimately lead to inaccurate diagnosis. Image enhancement techniques, such as adaptive filtering, super-resolution, and denoising, can be specifically designed and adapted to US images to increase their quality and remove noise, while preserving important details. 

After image enhancement, the \textbf{image analysis} stage aims to extract information from the US images that are relevant to the medical application. This includes tasks such as measuring parameters such as blood flow, image segmentation to extract anatomical structures, and detecting abnormalities such as lesions and blockages. Extracting such information is critical to the diagnosis of medical conditions and diseases. Generally, the variability of anatomical structures and the quality of US images, as well as the variance in sonographers' skills, pose as challenges in the quality of information extracted from US scans. 

The final stage in the US medical imaging pipeline is \textbf{decision-making and diagnosis}, where the gathered information are analyzed and interpreted into a medical diagnosis and recommendations. Some tasks in this stage occur in real-time, as in interventional procedures, where US imaging is used to guide clinicians in precisely positioning instruments (like needle and catheters) to the target location. As this stage heavily relies on human expertise in analyzing the collected information, it has high potential for human errors and variance in diagnosis due to variabilities in expertise.

\subsection{Reinforcement Learning Pipeline}

While different problems require unique RL solutions, a typical pipeline can be followed when designing any RL solution. This pipeline consists of 5 different stages, being \textbf{1)} data preparation and preprocessing, \textbf{2)} problem formulation, \textbf{3)} environment simulation, \textbf{4)} implementation and training, and \textbf{5)} validation and fine-tuning. This section explores these 5 stages and highlights the key aspects of each stage. \textbf{Data preparation and preprocessing} is the first stage of the pipeline, which involves collecting and processing the data needed to formulate and define the problem, as well as implement the RL solution. In certain problems, like in video games, data gathering is entirely done in simulations. For other real-world applications, like in medical US imaging, obtaining realistic datasets becomes crucial, as the agent's learning depends significantly on the quality of its observations. To this, it becomes essential to collect and gather data that can then be used to formulate the problem environment, within which the agent can make decisions. For most applications, the data needed to train RL algorithms come in the form of trajectories, where each trajectory represents a state, the action taken, and the resultant next state. 

The second stage of the pipeline is \textbf{problem formulation}, which is the process of defining the components of the RL problem as per the problem of interest. This includes the intricate design and definition of the states of the environment, the valid and invalid actions for the agent, the reward function, and the dynamics of the environment governed by the transition probability function. These components have a significant impact on the learning process, and in many cases require continuous re-design and modification to enhance the performance of the learning.

Generally, RL algorithms require thousands to millions of interactions with the environment to develop optimal policies. Generating such interactions in the real physical world is impractical due to time and cost constraints. Alternatively, a common approach is to generate a \textbf{simulation environment} that mimics the real-world environment and allows quick generation of thousands of interactions between the agent and the environment in a timely manner. For example, in sensitive applications like medical imaging, training a RL agent in the real-world (using rewarded interaction) is often impractical due to time and safety concerns. Instead, data can be collected to create a simulated environment (e.g. the agent traverses a US image to find a certain structure) within which the agent can act freely and develop an efficient policy. With recent technological advancements, simulators can capture physics-based models, realistic visualizations, and stochastic elements that mimic real-world uncertainties.

Given the problem formulation and a simulated environment, \textbf{RL implementation and training} can make use of the many different existing RL algorithms to optimize the decision-making of the agent. Regardless of the RL algorithm, the learning process is generally similar, where the agent interacts with the environment, collects experiences (actions and rewards), and updates its policy based on the observed reward. Common RL algorithms can be broadly categorized into value-based methods, which focus on learning value functions to derive the policy, and policy-based methods, which directly optimize the policy. Early value-based methods, such as Q-Learning and SARSA, and policy-based methods, such as REINFORCE, laid the foundations of RL algorithms and proved effective in handling simple RL problems. Following the advancements of deep learning, RL was extrapolated into Deep RL (DRL), where several RL algorithms started incorporating deep neural networks. Deep Q-Networks (DQN) extended Q-Learning by using neural networks to approximate the value functions, which enabled tackling more complex and high dimensional problems. Other popular algorithms include Proximal Policy Optimization (PPO) and Trust Region Policy Optimization (TRPO), which are policy-based methods, proved efficient in handling complex environments with discrete and continuous action spaces. The choice of the right RL algorithm depends on several factors, such as the problem's state and action spaces and computational constraints.

Following the training stage, and during the \textbf{validation and fine-tuning} stage, the performance of the trained RL model is extensively evaluated on new unseen scenarios, either in simulation or in the real physical world. While real-world training is not viable due to time constraints and patient safety, fine-tuning in the real world is  possible, where few experiences and interactions can be used to slightly adjust and update the model parameters to ensure improved performance. Fine-tuning could sometimes involve re-designing some parts of the RL problem, such as the reward function, if the model performance does not meet expectations in the real world. Techniques like hyperparameter tuning and transfer learning are commonly used to enhance the learning process and the performance of the model.

\section{RL for US Image Acquisition}
\label{Sec: RLImageAcq}

In this section, we will explore the use of RL in US image acquisition, focusing on both manual and autonomous scanning techniques.
Manual US scanning remains the standard practice and requires significant human expertise and precision, particularly for complex procedures such as cardiac imaging, where multiple views from different angles are necessary. However, manual scanning is prone to several challenges, including operator variability, fatigue, and the repetitive nature of the tasks, which can result in inconsistent outcomes. Moreover, the shortage of trained US professionals, especially in rural areas, has created a growing demand for more efficient and scalable solutions. Despite the advantages of manual scanning, these limitations underscore the urgent need for innovation in the field.
To address these challenges, autonomous US scanning has emerged as a promising solution. By automating the scanning process, RL-based methods can reduce variability, improve efficiency, and standardize procedures across different practitioners. Autonomous systems can perform repetitive tasks with high precision, offering significant improvements in the quality and consistency of imaging. RL has been particularly useful in enabling these systems to learn optimal scanning strategies through interactions with the environment, overcoming many of the limitations associated with manual scanning. 

AI and robotics have further accelerated the development of autonomous US scanning. The integration of AI tools, powered by RL, has shown promise in automating scanning tasks, improving decision-making, and adapting to patient-specific variations. Robotics, too, has seen significant progress, with robots capable of mimicking human behavior and automating scanning tasks ~\citep{zakeri2024ai, zakeri2025robust}. For instance, RL has been applied to the autonomous scanning of organs such as the liver, where challenges like the presence of ribs and bones complicate image acquisition ~\citep{bi2024autonomous}, or vessels that require precise targeting ~\citep{bi2022vesnet}. In surgical settings, RL-based robots can improve the effectiveness of pre-planning by autonomously adjusting the robot’s position and orientation to capture the desired US view, while maintaining constant contact force and velocity ~\citep{lin2023deep, li2024adaptive}.

In the following sections, we will delve deeper into studies investigating how RL has been used to develop autonomous scanning solutions. While recent advancements in manual US scanning are more limited, one notable example includes the use of RL for needle tracking using US scanning ~\citep{macuradynamic}. However, the primary focus of current research is on autonomous scanning, where RL plays a key role in improving the efficiency, accuracy, and robustness of the US scanning. We will explore these developments in detail, as well as examine the stages involved in implementing RL models, from data acquisition and processing to model validation and fine-tuning.


\vspace{1em}
\subsection{RL Data Preparation and Processing}

Data preparation and processing for RL in US image acquisition involves collecting the necessary data needed to train the RL agent in performing the acquisition task. In the case of autonomous US scanning, this data helps the robot navigate autonomously from its starting position to the target scanning area. On the other hand, for manual US scanning, the data assists the RL model with tasks such as object tracking, as demonstrated in the work by ~\citep{macuradynamic}, where RL is applied to track needles using US images.

For RL applications, the data that guides the agent's learning is structured in the form of states and actions. The state typically includes information such as the US image, the probe's position, contact force, and torque, which represent the current configuration of the system. The action refers to the possible moves or adjustments the agent can make, such as altering the probe's position or pressure. The next state is the resulting state after the action is taken, which may include changes in the US image and the agent’s interaction with the environment. Table \ref{tab:autonomous_scanning_datasets} provides a detailed overview of the datasets used in existing literature. From this table, in autonomous US scanning, the data collected often includes US images, which are used to guide the RL agent's decision-making process. However, to enable more effective and informed decisions, this data alone is typically not sufficient. Additional information such as the probe's position, contact force, and torque are often included as part of the state space, helping the RL agent understand its current state and environment. The detailed formulation of the RL problem in terms of state, action, and reward is described in the following section.

Additionally, the table reveals that most existing studies rely on private datasets due to ethical restrictions on sharing US images from human volunteers. In certain cases, research utilizes animal data ~\citep{macuradynamic,stevens2022accelerated} or medical phantoms that simulate human organs, depending on the organ under examination, such as the breast ~\citep{yao2024decision}, thyroid ~\citep{luo2024multi}, abdomen ~\citep{deng2024portable}, or kidney ~\citep{hu2024probe}, among others. Due to the limited availability of public US image datasets, some studies make use of private or public datasets from other imaging modalities, such as computed tomography (CT) scans, to create 3D volumes from which US images are subsequently derived ~\citep{bi2024autonomous, lin2023deep, hu2024probe, amadou2024goal, li2023rl}. This approach enables the generation of simulated US images that closely resemble real US images. Notably, only few studies were identified that implemented online RL without relying on dataset acquisition for offline training ~\citep{duan2024safe, raina2024coaching}.

To introduce variability into the training dataset, several studies ~\citep{yao2024decision, amadou2024goal,hase2020ultrasound} have employed data augmentation techniques, such as rotating, flipping, and transformation of US images. For example, in ~\citep{macuradynamic}, these techniques were used to enable tracking of the needle as it enters the tissue from different angles. Additionally, some authors apply data resizing to standardize the input dimensions ~\citep{bi2024autonomous,deng2024portable}. The authors in ~\citep{bi2024autonomous}, resized rib cage to a generic cylindrical coordinate system based on the average body size in the dataset to account for size differences between patients. During RL training, the dataset is often also shuffled to enhance the model's robustness and efficiency. In general, the dataset is typically divided into 80\% for training and 20\% for testing, or alternatively, 70\% for training, 10\% for validation, and 20\% for testing.

The availability of datasets is crucial for developing RL-based solutions for US image acquisition. From the reviewed works, it is evident that the datasets used are generally private, highlighting the need for the development of public datasets. Additionally, encouraging research on generating simulated US images presents a promising avenue for expanding available datasets, although these simulated images may not capture all the nuances present in real US images. The next section discusses the RL problem formulation, detailing how this data is used to define the RL agent’s states.

\begin{table*}[!t]
  \centering
    \caption{Overview of Data Sources for RL Model Development -  US Image Acquisition}
\resizebox{\textwidth}{!}{
  \begin{tabular}{|p{2.0cm}|p{2.0cm}|p{2.0cm}| p{1.5cm}| p{12cm}|}
\hline
    Reference & Scanned Organ &  Dataset Type & Data Source &  Description \\ \hline

 ~\citep{su2024fully} & Thyroid & Private dataset & Human & Over 180 volunteer adults participants (18+ years old) for both sexes
\& including individuals with thyroid problems. 3D thyroid model: use 500 head and neck CT images from patients (250 male and 250 female aged between 7 and 82)\\
\hline

 ~\citep{li2021autonomous} & Spine & Private dataset & Human &  41 3D US volumes of L1-L5 lumbar vertebrae from 17 healthy male volunteers (ages 20-26). Dataset volume is 350 × 397 × 274 with voxel size of 0.5 × 0.5 × 0.5 mm³\\
\hline

~\citep{bi2024autonomous} & Liver &  Private dataset & Simulation & Simulated US created from public CT dataset (3D-IRCADb-01) which contains 20 liver tumors and 8 rib cages manually selected (2 rib cages used for model testing)\\
\hline

 ~\citep{yao2024decision} & Breast &  Private dataset & Phantom & 2 breast phantoms, each with 25 lesion locations. A grid created for each phantom. Each bin contains 5 US images\\
\hline

~\citep{bi2022vesnet} & Blood vessels & Private dataset & Human, Phantom & 4,421 US images acquired from a vessel phantom (images \& background).
 1,041 US human images.
1,266 images from 3D volume of the carotid artery reconstructed using human data\\
\hline

~\citep{amadou2024goal} & Heart & Private dataset & Simulation & Use LIDCIDRI a public dataset, which contains 929 patient CT scans, to generate simulated US images \\
\hline

~\citep{li2023rl} & Heart & Private dataset & Simulation & Simulate US images using publicly thoracic CT dataset SegTHOR which contains 40 real human subjects.
\\
\hline

~\citep{ning2023autonomous} & Hand Vessels & Private dataset & Human & 15k Clinical radial artery US and corresponding Doppler images from 20 subjects scanned by 2 sonographers \\\hline

~\citep{ning2021force} & Arm, Thigh \& Abdomen & Private dataset & Human, Phantom &  US images and contact force data collected during experiments with both phantoms and volunteers\\ \hline

~\citep{ning2021autonomic} & Abdomen \& Lumbar spine &  Private dataset & - & US images, RGB images, probe positions \\ \hline


~\citep{lin2023deep} & Heart &  Private dataset & Simulation & 1k simulated US images using PLUS toolkit and Slicer3D based on positions and orientations of the probe, and 3D anatomical volume \\ \hline

~\citep{shen2023towards} & Heart & Public dataset & Human & Left ventricular 3D US data (apical 5-chamber, short-axis apical level view, short-axis papillary muscle level view) from the Cardiac Dataset Atlas Project. Volume size is 214×204×214 voxels \\ \hline

~\citep{chen2021learning} & Kidney & Private dataset & Human & 5195 sets of data collected from 5 individuals. Each set contains 7-dimensional motion vector (4 quaternion posture dimensions and 3 force dimensions), corresponding US image, and a label indicating whether the image clearly shows the kidney area. Dataset including both good and bad trajectories \\ \hline

~\citep{hu2024probe} & Kidney & Public \& private datasets & Human, Siumlation, Phantom & 520 annotated US images from the Open Kidney Dataset. Synthetic US images generated from 54 CT volumes. Using 8 CT volume and US image from the phantom for model testing\\ \hline

~\citep{ning2023inverse} & Hand, Kidney, Spine & Private dataset & - & 500 episodes of expert demonstrations with data on the probe's state and the actions \\ \hline

~\citep{luo2024multi} & Thyroid & Private dataset & Phantom & Data collected from physical thyroid phantom\\ \hline

~\citep{deng2024portable} & Abdomen & Private dataset & Phantom & 3k US image acquired from the phantom (2k similar images taken from the same position, 1K dissimilar images) \\ \hline

~\citep{li2024adaptive} & Abdomen & Private dataset & Simulation & Data created in the simulation environment\\ \hline

~\citep{hase2020ultrasound} & Spine & Private dataset & Human & US images acquired from 34 volunteers with 11 sweeps per scan\\ \hline

~\citep{duan2024safe} & Spine & - & - & Data acquisition is not conducted. Online RL is applied\\ \hline

~\citep{stevens2022accelerated} & Blood vessels & Private dataset  & Simulation, Phantom, Animal & Intravascular ultrasound (IVUS) image obtained from simulated wire targets, wire phantoms, and in-vivo data from a porcine model.\\ \hline

~\citep{raina2024coaching} & Urinary bladder & - & - & Online Training following coach feedback \\ \hline

~\citep{shida2024robotic} & Heart & Private dataset & Human & 7422 US images annotated with the mitral valve. Data acquired from 24 males in their teens to twenties (height: 168.5±8.5 cm; weight: 65±17 kg)\\ \hline

~\citep{li2024action}  & Blood vessels & Private dataset &Simulation, Phantom & Simulation data and vascular Phantom data. Data includes the positions of the probe and corresponding US images\\ \hline

~\citep{macuradynamic}  & Human tissue & Private \& public dataset & Phantom, Simulation, Animal & Private datasets: data from blue phantom and 40 frames of tofu (simulating tissue). Public dataset: US4US dataset includes data from blue phantom and chicken breast\\ \hline
\end{tabular}
}
\label{tab:autonomous_scanning_datasets}
\vspace{-10pt}
\end{table*}

\subsection{RL Problem Formulation and Algorithms}


As mentioned earlier, US image acquisition can be categorized into two main types: manual US scanning and autonomous US scanning. In this section, we will discuss each category and how the RL problem formulation is defined for each.

In the context of autonomous US scanning, the formulation of the RL problem involves defining four key elements: \textbf{(1)} the environment, which represents the scanning area; \textbf{(2)} the state space, which includes the state of the robot within the environment; \textbf{(3)} the action space, consisting of the robot's movements; and \textbf{(4)} the reward function, which is used to either penalize or reward the robot during the US scanning task. A detailed overview of the RL problem formulation in related works on US image acquisition is provided in Table \ref{tab:data_acquisition}. This table summarizes the scanned organs in each study, the category of reward function used, the RL algorithms used, and the evaluation metrics. The following discusses each element of the RL problem formulation in the studies reviewed.

\begin{table*}[!t]
  \centering
    \caption{Overview of Related Work about US Image Acquisition}
    \resizebox{\textwidth}{!}{
  \begin{tabular}{|p{2.0cm}|p{2.0cm}| p{2.0cm}| p{4.0cm}| p{8cm}|}
\hline
    Reference & Organ & Reward Function & RL Algorithm & Metrics \\ \hline

 ~\citep{su2024fully} & Thyroid  & Dense  & MDP, DQN & Average reward \& episode length, Expert feedback, Image quality, Robot stability \& velocity \\
\hline

 ~\citep{li2021autonomous} & Spine & Dense  & POMDP, DQN & Position error, Orientation error, SSIM, Success rate, Average number of steps\\
\hline

~\citep{bi2024autonomous} &  Liver  &  Dense & MDP, Double Dueling DQN & Succes rate, Average steps, Shadow avoidance, Attenuation minimization  \\
\hline

 ~\citep{yao2024decision} & Breast  & Dense & MDP, Broad RL \& Q-learning & Accuracy, F1-Score, Dice Coefficient, Average Return and trials, Average Steps, Learned state-value function, Computation time per image \\
\hline

~\citep{bi2022vesnet} & Vessels & Dense  & POMDP, A2C  & Success Rate, Average number of steps, Position error, Orientation error  \\
\hline

~\citep{amadou2024goal} & Heart  &  Dense  & Actor-Critic  & Position error, Angle error \\
\hline

~\citep{ning2023autonomous} & Hand Vessels  & Sparse  & MDP, PPO & Segmentation accuracy, probe displacement, SSIM, Confidence map  \\
\hline

~\citep{ning2021force} & Arm, Thigh \& Abdomen  & Dense  & MDP, PPO & Stability \& completeness of US image,  Contact force \& torque, Scanning trajectory  \\
\hline

~\citep{lin2023deep} & Heart  & Dense & MDP, DQN &  MSE, US image quality, Error distribution of the reference points\\
\hline

~\citep{shen2023towards} & Heart  & Dense & MDP, Dueling double DQN (D3QN) & LPIPS loss, Dis, Ang, SSIM, and MS-SSIM \\
\hline

~\citep{chen2021learning} & Kidney   & Sparse  & TD3 & Training \& validation loss \\
\hline

~\citep{hu2024probe} & Kidney  & Sparse  & POMDP, DQN & Reachability, Image quality \\
\hline

~\citep{ning2023inverse} & Hand, Kidney, Spine   & Inverse RL  & PPO, AIRL &  Posture change, Force change \& Contact position change curves, Structural similarity  \\
\hline

~\citep{luo2024multi} & Thyroid  & Dense & MDP, PPO  & Success rate
of completing the task, the average deviation from the
target pose, SSIM, Episode rewards \\
\hline

~\citep{deng2024portable} & Abdomen & Sparse  & MDP, PPO, Actor-critic & Success rate, Average steps, SSIM, NCC \\
\hline

~\citep{li2024adaptive} & Abdomen  & Dense & MDP, PPO & Velocity, Contact force, Probe spatial position variation, Loss, Variance \\
\hline

~\citep{hase2020ultrasound} & Spine  & Sparse & POMDP, DDQN, Dueling DQN, Prioritized Replay Memory  & Correctness, Reachability, State value estimate maps \\
\hline

~\citep{duan2024safe} & Spine  & Dense & CMDP,
Safe RL & Effectiveness, Clarity of the US images, Confidence probability\\
\hline

~\citep{raina2024coaching} & Urinary bladder  & Sparse & Off-policy Soft Actor-Critic (SAC), POMDP  & Number of High-Quality Images (HQI) sampled, First instance of HQI sampling, Errors in probe motion \\
\hline

~\citep{shida2024robotic} & Heart & Dense & MDP, DQN & Loss of confidence score Rate, Number of movement, Confidence score, Accuracy, Efficiency\\
\hline

~\citep{li2024action} & Blood vessels & Dense  & MDP, A2C & Dice coefficient, SSIM, Mean Intersection over Union (mIoU), Success rate.  \\
\hline

~\citep{stevens2022accelerated} & Blood vessels & Dense & POMDP, Actor-Critic, & MSE, MAE, PSNR, SSIM \\
\hline

~\citep{macuradynamic} & Human tissue & Dense  & MDP, DQN  & Image quality, Needle angle, Reward function \\
\hline 

~\citep{ning2021autonomic} & Abdomen \& Lumbar spine & Dense  & MDP, PPO & Success Rate, Force Control, Execution Time \\
\hline

~\citep{li2023rl} & Heart  & Dense  & MDP, DQN & Pose Error, Success Rate, Total Reward, SSIM,
Average Number of Steps \\
\hline

\end{tabular}
}
\label{tab:data_acquisition}
\end{table*}

In the literature, the formulation of the RL problem for autonomous US scanning can vary based on the specific case being studied and its complexity. Common approaches include modeling the problem using RL based on MDP, POMDP, or Constrained Markov decision process (CMDP) models. The most commonly used formulation for RL in US image acquisition is the MDP, where the RL agent fully observes its state in the scanned environment. In this context, the state can include factors such as the scanning area, the probe's position, image quality, and other variables, depending on the specific scanning objectives and the complexity of the task. For instance, in ~\citep{su2024fully}, the authors developed an RL-based solution for autonomous thyroid scanning, formulating the RL problem as an MDP, even though the robot is guided solely by the US images observed in the scanned environment. In contrast, the work in ~\citep{ning2021autonomic} defines a more detailed RL problem, where the robot observes not only the US image but also other information, such as the entire scene and the force applied by the robot. This is important because the human body is considered an unknown environment, and both the scene image and the applied force are crucial for handling body deformations and movement. Similarly, in ~\citep{yao2024decision}, which focuses on robotic breast scanning, the authors incorporate additional information into the robot's observations to ensure the acquisition of high-quality US images, including data about lesion presence and confidence scores. A detailed discussion of the state definitions across these studies will be presented later in this section. 

In contrast, with a POMDP, the agent is unable to fully observe the environment and can only make partial observations. In the work by ~\citep{raina2024coaching}, where the objective is to autonomously scan the urinary bladder, the state is defined by the US image, and the RL problem is modeled as a POMDP. This is due to the agent's inability to fully observe the exact position of the target state, which is defined as achieving high-quality US images of the urinary bladder. Instead, a human coach provides corrective feedback, helping the robot learn how to reach the desired goal state. Similarly, in ~\citep{hu2024probe}, where the authors developed an RL-based model for autonomous kidney scanning, the RL problem is also modeled as a POMDP because the agent cannot fully observe its state, including the exact scan plane and the anatomical target.
Another example of applying a POMDP approach is found in ~\citep{bi2022vesnet}, where the robot cannot directly observe the position of the US probe relative to the target (e.g., a blood vessel). In these scenarios, the robot has access only to partial information, such as the US image, action history, and segmented areas. Similar approaches are employed in other works ~\citep{stevens2022accelerated, hase2020ultrasound, li2021autonomous}, where the state remains partially observable, necessitating the use of a POMDP.
The final approach used in RL problem formulation is CMDP, where the objective is not only to optimize the expected reward but also to ensure that certain constraints, such as safety-related ones, are respected. For instance, in ~\citep{duan2024safe}, CMDP is applied to ensure an optimal schedule for the contact force during US imaging of the spine, while also maximizing the expected reward.

Another important aspect in the formulation of the RL problem is the definition and use of states. In RL for autonomous US scanning, single-state representations are commonly used, where the state is defined by a single element that characterizes the environment at a given timestep. In many studies, this is typically a single US image, which serves as the sole input for the RL agent. For instance, in works such as ~\citep{duan2024safe}, ~\citep{raina2024coaching} ~\citep{su2024fully}, and ~\citep{shen2023towards}, the state consists of a single US image, and the agent's task is to interpret the image and determine the next action, such as adjusting the probe’s position or applying scanning techniques to improve the quality of the image. This approach simplifies the learning process but may limit the agent's ability to understand the broader context, such as temporal changes in the scan. Additionally, probe position can also serve as a state, as seen in ~\citep{amadou2024goal} and ~\citep{hase2020ultrasound}, where the agent uses information about the probe's location to navigate and scan effectively. The agent’s actions are then determined based on the current US image or probe position, which typically provides sufficient context for relatively simple scanning tasks. On the other hand, multi-state representations aim to provide a more comprehensive view of the environment by combining multiple components into the state. This approach is useful for more complex scanning tasks, where additional context is necessary to improve decision-making. For example, in ~\citep{li2021autonomous}, the state includes multiple stacked US images, allowing the agent to observe the temporal progression of the scan and adjust its actions accordingly. Other studies expand the state further by incorporating 3D volumes ~\citep{bi2024autonomous}, force and torque data ~\citep{ning2023autonomous, ning2023inverse}, or even the confidence score of the scan ~\citep{yao2024decision}. For instance, in ~\citep{yao2024decision}, the state is a 9D vector that includes the expert’s score, lesion presence, and confidence map values. Similarly, in works like ~\citep{deng2024portable} and ~\citep{chen2021learning}, the state combines US images, action history, and probe position or force/torque values to give the RL agent a more detailed and holistic understanding of the scanning environment. This allows the agent to make better-informed decisions, accounting for various aspects of the scan such as anatomical structure, probe movement, and physical interactions with the body. The choice of state representation directly influences how the RL agent interacts with its environment and learns from it. In each of these studies, the RL agent’s actions are guided by the specific type of state representation, which helps the agent understand its position relative to the target state (such as obtaining a high-quality image or reaching the target anatomical region). The detailed design of each state representation and its role in the RL process is crucial for improving the efficiency and accuracy of the autonomous US scanning task.

Another key component to consider in the formulation of the RL problem is the definition of the action space. In the context of autonomous US scanning, the action space can either be discrete or continuous, and the precise formulation of the action space depends on the specific objectives of the study and the type of system being developed. Several studies have modeled the action space with distinct approaches based on the type of US task at hand. In discrete action spaces, actions are limited to specific, predefined movements. For example, in ~\citep{su2024fully}, the actions include simple directional movements such as moving the probe left, right, or to the center for thyroid scanning, while in ~\citep{li2021autonomous}, actions consist of translating the probe along the x- and y-axes by a certain distance and rotating it around the x, y, and z axes in fixed increments, used for spine scanning. Similarly, ~\citep{bi2024autonomous} defines four discrete actions including translations perpendicular to the probe center-line and rotations, focused on liver scanning. Another example is ~\citep{yao2024decision}, where actions involve pressing the probe down, lifting it up, or rotating it left or right for breast scanning. For more specialized tasks like scanning blood vessels, ~\citep{li2024action} defines 9 discrete actions that modify the elliptical properties of the blood vessel, such as its position and shape.

In contrast, continuous action spaces allow for more flexibility in probe movement. In ~\citep{luo2024multi}, for thyroid scanning, the agent makes continuous position and orientation adjustments to the US probe. Similarly, in ~\citep{ning2021autonomic}, continuous actions are used to adjust probe position and orientation, applicable for scanning the abdomen and lumbar spine. Other studies, like ~\citep{chen2021learning}, define continuous actions to control probe rotation along multiple axes and apply continuous force in the x, y, and z directions for kidney scanning. Furthermore, ~\citep{ning2021force} employs torques applied to the probe in the x and y directions for tasks like scanning the arm, thigh, or abdomen, while ~\citep{raina2024coaching} combines position, orientation, and forces for scanning the urinary bladder. Other works like ~\citep{bi2022vesnet}, created a continuous  action space includes rotational movements around the x and y axes as well as translations along the z-axis for scanning blood vessels, while in ~\citep{ning2023autonomous}, the actions involve increasing or decreasing the values of four joints of a robotic arm to adjust the probe position for heart scanning. These variations in action space design reflect the complexity and specificity of the US scanning tasks, with discrete actions typically used for simpler movements and continuous actions employed to achieve finer control over probe positioning and force application.

In the context of these studies, the reward function is defined in various ways, reflecting the specific scanning objectives and system constraints of each approach. Dense rewards are commonly used in most studies, where the RL agent receives continuous feedback based on real-time performance. For instance, ~\citep{su2024fully} defines a reward structure that instructs the RL agent to stop movement if the thyroid gland is not visible, or to reposition the arm leftward if the probe is incorrectly positioned until the optimal state is achieved. In ~\citep{li2021autonomous}, the agent receives +10 when it reaches the goal pose, but faces penalties of -0.5 for tilting the probe beyond 30° or -1 for moving outside the patient’s body. A dynamic reward structure is applied in ~\citep{bi2024autonomous}, which encourages rapid coverage of the target volume, minimizes shadowed areas, and ensures the probe remains close to the target to reduce attenuation. Similarly, ~\citep{yao2024decision} defines rewards based on US image quality, factoring in expert scores, lesion area, and the confidence map’s gravity center.

Some studies focus on optimizing specific performance metrics. For example, ~\citep{bi2022vesnet} rewards the agent for maximizing segmented vessel area and minimizing the distance to the goal, with termination occurring when the vessel’s bounding rectangle is achieved. In ~\citep{amadou2024goal}, rewards are given based on how close the transducer is to the goal state, particularly in heart scanning. The reward function in ~\citep{ning2023autonomous} penalizes large torques to ensure a stable posture, while ~\citep{ning2021force} penalizes torque misalignment and force imbalances, promoting continuous scanning during the procedure. In ~\citep{lin2023deep}, the RL agent receives rewards for minimizing the loss function, and the process halts once the error reaches a threshold (6 mm). Similarly, ~\citep{shen2023towards} quantifies differences between the target and current scan planes to determine the reward.

Other studies employ sparse rewards, offering feedback only upon achieving a particular objective. In ~\citep{raina2024coaching}, the reward is based on the image quality rating generated by a CNN network, with additional coach and trajectory rewards influencing the agent’s actions. In ~\citep{deng2024portable}, rewards are assigned for successive instances of image similarity, with a penalty of -2 when the agent reaches the edge of the target area. The task terminates when the agent remains within the target image for 10 consecutive steps. Similarly, ~\citep{hase2020ultrasound} emphasizes penalizing incorrect stopping behaviors to encourage exploration and avoid stagnation, discouraging repetitive back-and-forth movements.

\begin{figure}[htbp]
    \begin{center}
    \centering
\includegraphics[width=0.7\linewidth]{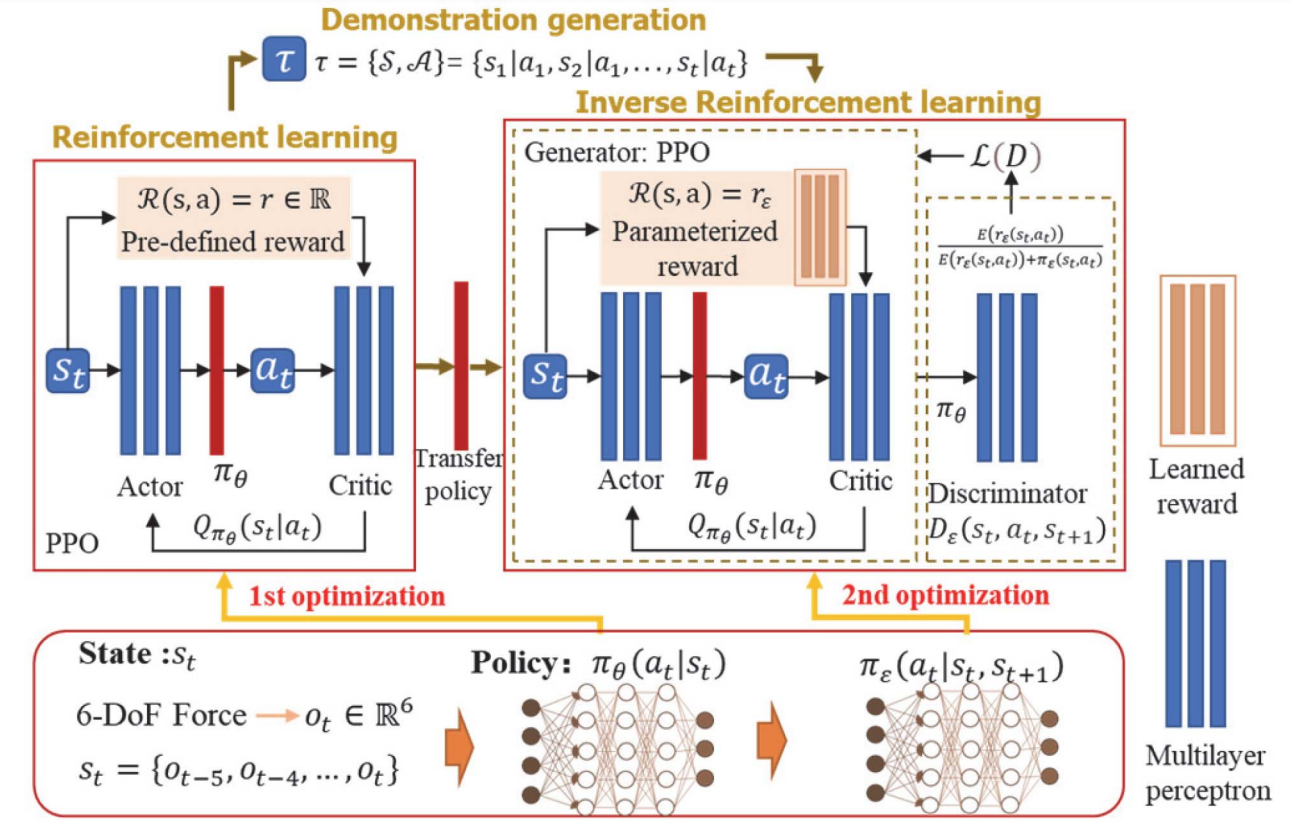}
   \caption{Optimizing Reinforcement Learning (RL) Policy with Predefined Rewards and Expert Demonstrations ~\citep{ning2023inverse}.}
   \vspace{-30pt}
    \label{airl}
    \end{center}
\end{figure}

Other approaches combine multiple performance metrics. For example, ~\citep{luo2024multi} integrates the distance to the target, US image quality, and force control into its reward function, whereas ~\citep{li2024adaptive} focuses on maintaining proper position and orientation, minimizing force fluctuations during scanning. In ~\citep{shida2024robotic}, the reward function is designed to minimize search time while avoiding local solutions, offering positive rewards for optimal solutions and penalties for suboptimal or redundant actions. In ~\citep{li2024action}, rewards are based on the agent’s ability to move the probe closer to the desired ellipse, with large errors penalized (-2) and small errors rewarded (+10). Similarly, ~\citep{stevens2022accelerated} uses metrics like Mean Square Error (MSE) and the Structural Similarity Index (SSIM) to reward the agent based on the reconstruction quality of IVUS images compared to the ground truth.
Lastly, some studies focus on specific anatomical targets or more complex system dynamics. In ~\citep{ning2021autonomic}, the reward encourages proper probe positioning and force control, with penalties for excessive force or boundary violations, particularly in abdomen and spine regions. Similarly, ~\citep{li2023rl} applies a reward structure that rewards goal completion (+10) and penalizes boundary violations (-1), while also factoring in position, orientation, and compliance for continuous adjustment. Another approach highlighted in the literature is Adversarial Inverse Reinforcement Learning (AIRL), as demonstrated in ~\citep{ning2023inverse}. AIRL seeks to bridge the gap between simulated and real environments by refining a pre-trained RL policy with predefined rewards using expert demonstrations. Figure \ref{airl} provides a clear illustration of the proposed solution.

\begin{figure}[htbp]
    \begin{center}
    \centering
\includegraphics[width=0.8\linewidth]{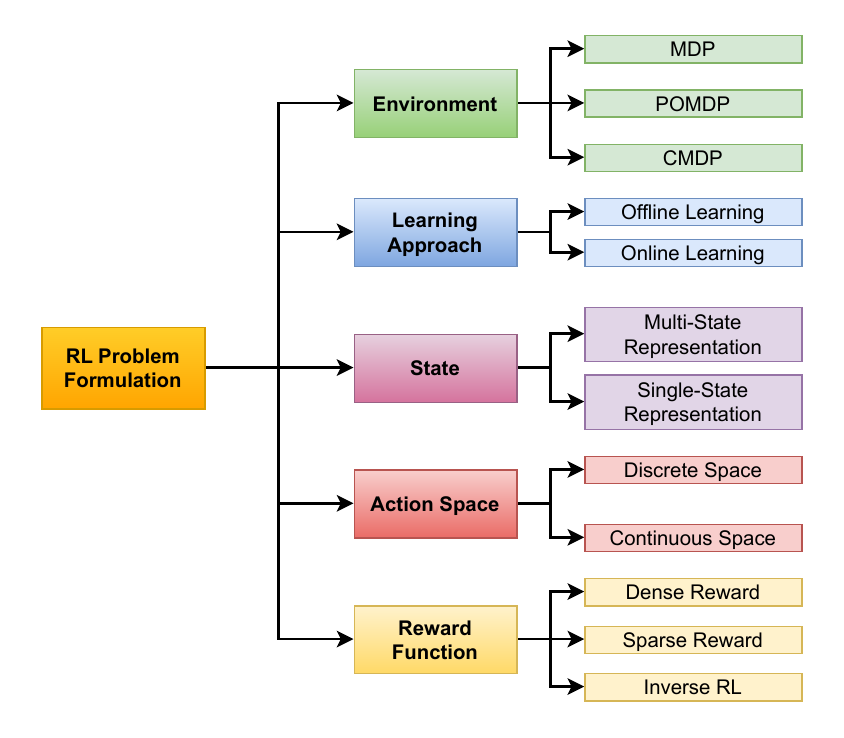}
   \caption{Reinforcement Learning (RL) Problem Formulation - Ultrasound (US) Image Acquisition. The Abbreviation MDP Refers to Markov Decision Process; POMDP refers to Partially Observable MDP; CMDP Refers to Constrained MDP.}
      \vspace{-30pt}
    \label{us_acquisition_drl}
    \end{center}
\end{figure}

Figure \ref{us_acquisition_drl} provides an overview of the various approaches used for autonomous US scanning across the surveyed studies. It categorizes the approaches based on the type of RL environment, the learning methods employed, as well as the design of state and action spaces, and the formulation of the reward function. 
In discussing the state, action, and reward spaces, it becomes evident that the design choices vary considerably depending on the specific objectives of the US scanning task and the anatomical target. For states, many studies incorporate multimodal inputs, combining US images with sensor data (e.g., probe position, force, and torque) to provide a richer representation of the scanning environment ~\citep{bi2024autonomous,yao2024decision,li2021autonomous}. Action spaces also reflect this diversity, with some studies using discrete movements (such as translations or rotations) ~\citep{bi2024autonomous,su2024fully,li2021autonomous}, while others adopt continuous adjustments to probe position or orientation ~\citep{amadou2024goal,ning2023autonomous,ning2021force}. Regarding the reward function, most studies use dense rewards, which allow for more efficient learning by providing continuous feedback based on the agent’s real-time performance ~\citep{bi2024autonomous,su2024fully,li2021autonomous}. However, some studies use sparse rewards, giving feedback only when certain goals are achieved, like obtaining a clear image ~\citep{raina2024coaching,ning2023autonomous,chen2021learning}.  Overall, these varied approaches highlight the complexity and adaptability of autonomous US scanning systems in handling different imaging tasks and scanning conditions.

In the specific case of manual US scanning for tracking a needle during medical procedures such as anesthesia or biopsy, as discussed in ~\citep{macuradynamic}, the state is represented by the US image corresponding to the Plane Wave (PW) angles selected by the agent. These angles refer to the direction in which the US waves move through the tissue. The action space is discrete, allowing the agent to make specific movements or adjustments. The reward function is designed to balance the accuracy of line detection, which is critical for precise needle positioning, with the fidelity of the compounded US image, ensuring clear visualization of the needle’s location. This is particularly important during procedures like biopsies, where accurate needle guidance is crucial, aiding in real-time visualization and safe needle insertion.

\subsection{RL Simulation Environment}

Training and validating RL models require extensive interactions with the environment, as the model learns from the trajectories and feedback generated during these interactions. This makes it challenging to implement in real-world scenarios, particularly in autonomous US scanning, where the robot performs the scan and could potentially harm individuals. Therefore, it is essential to develop simulation environments to train, validate, and fine-tune the RL model without concerns for safety and ethics. In the literature, there are several examples of creating such simulation environments.

To explore the approaches used for simulating environments in US acquisition, it is crucial to understand the composition of the environment being simulated. In the case of manual US scanning, the environment consists of four primary components: the patient (or scanned area), the US probe, the US machine, and the human in control of the scan. To accurately capture the position of the probe, various sensors such as inertia sensors and force or torque sensors can be attached to the probe. For example, these sensors can help track the probe's position and orientation, as demonstrated in ~\citep{chen2021learning}. Although manual US scanning is performed by an operator, simulations can replicate the real-world conditions in which the operator works. In the study by ~\citep{macuradynamic}, the simulation environment mimics these conditions by providing feedback based on the agent’s actions. Specifically, the simulation environment generates US images based on the PW angles chosen by the agent. It extracts the relevant PWs from recorded data and gives the resulting image to the agent. These simulations help analyze operator behavior by simulating the feedback and interactions during actual scanning and provide a controlled environment for training and studying US acquisition.

\begin{figure}[htbp]
    \begin{center}
    \centering
\includegraphics[width=0.6\linewidth]{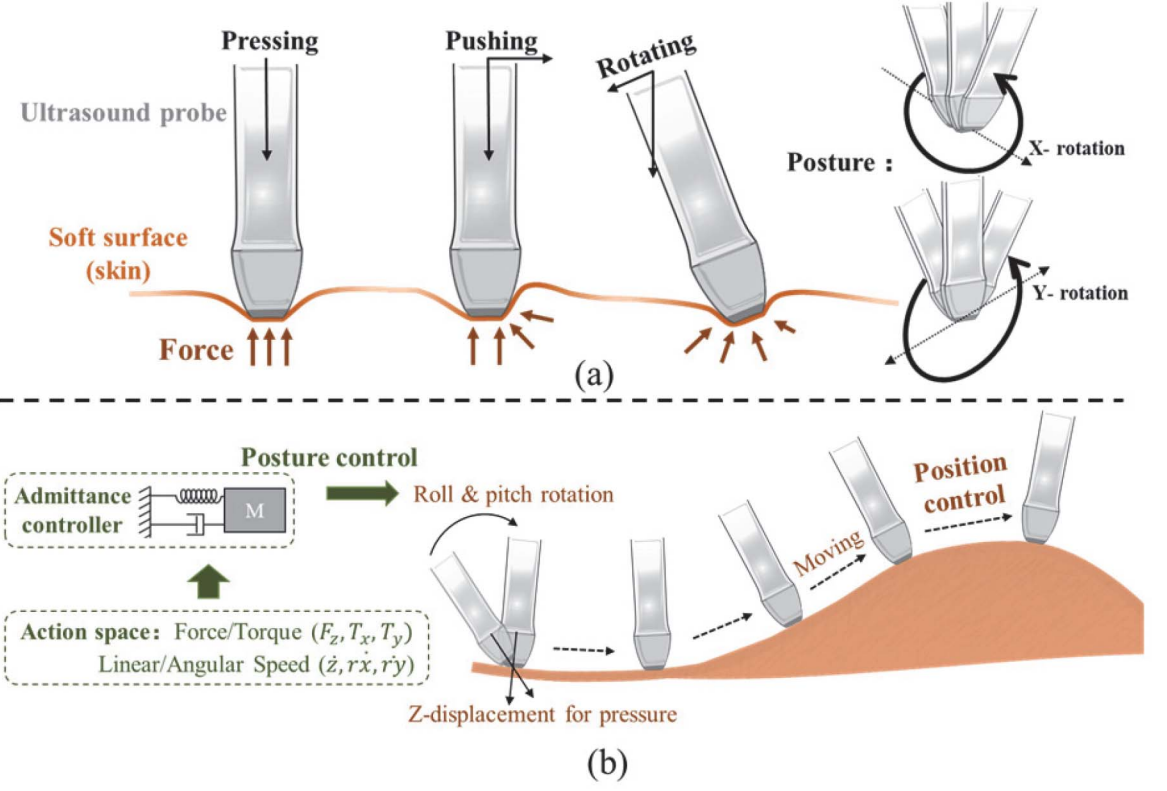}
   \caption{Probe Movement During Ultrasound Scanning ~\citep{ning2023inverse}.}
      \vspace{-30pt}
    \label{DOF}
    \end{center}
\end{figure}

In autonomous US scanning, there are four primary components: the robot, the US machine, the US probe, and the patient or the target scanned area. Other accessories are sometimes added, such as cameras for detecting human body joints ~\citep{su2024fully} or RGB cameras to visualize the environment ~\citep{ning2021autonomic}, allowing the RL agent to gain a more precise understanding of its state. There are additional components installed on the robot to ensure its control, but these are not covered in this survey. Regarding robots, the concept of degree of freedom (DoF) is important to understand, as it represents the number of movable joints in the robot. Each DoF corresponds to a specific movement the robot can perform in a particular direction or a rotation around an axis. For example, a robot with 6 DoF has the ability to move and rotate in three-dimensional space.
Another important factor to consider in robot simulation is the concept of torque and force. The torque refers to the rotational force that causes an object to rotate around an axis, while the force is a push or pull that can change the motion of an object or cause deformation. These physical values are essential for understanding how the robot interacts with its environment, particularly when it needs to perform complex actions such as human scanning. Figure \ref{DOF} shows examples of the potential movements of the US probe on the skin.

Additionally, US scanning can be categorized into two types based on the probe placement: extra-corporeal scanning, where the US probe examines the external surface of the human body, and intra-corporeal scanning, where a specialized probe is inserted into the body to perform more complex scans. Examples include transesophageal echocardiography (TEE) for heart imaging ~\citep{amadou2024goal,li2023rl} and interventional endoscopic US (EUS) ~\citep{adler2018interventional} for diagnosing post-surgery complications, such as fluid collections in the pancreas. In the literature, most research on autonomous US scanning primarily focuses on extra-corporeal scanning. Table \ref{tab:robots} provides details on the robots, US probes, US machines, and the DoF used in various works related to autonomous US scanning. From this table, we observe that most studies focus on extra-corporeal scanning, indicating a gap in the literature regarding intra-corporeal scanning. This gap can be attributed to the delicacy and invasiveness of the task, which requires more sophisticated technology and techniques, as well as the difficulty in creating realistic simulation environments for such procedures. Furthermore, the table highlights the diversity of robots, US machines, and probes used in the literature, demonstrating the complexity involved in creating a generalized simulation environment. The varying DoF across studies also show how the complexity of the task is addressed differently in each case.

\begin{table*}[!t]
  \centering
    \caption{Autonomous US Scanning: Overview of Robots, US Machines, Probes, DoF, and Scanning Types}
    \resizebox{\textwidth}{!}{
  \begin{tabular}{|p{2.0cm}|p{3.5cm}|p{5.5cm}| p{4.0cm}| p{0.5cm}| p{3.0cm}|}
\hline
    Reference & Robot Name &  US Machine & US Probe & DoF & US Scan Type  \\ \hline

 ~\citep{su2024fully} & UR3 Robot & US control board & 2D linear US probe & 6 & Extra-corporeal scanning  \\
\hline

 ~\citep{li2021autonomous} & KUKA LBR iiwa 7
R800 Robot &  Wisonic Clover diagnostic US machine & C5-1B convex US transducer & 6 & Extra-corporeal scanning  \\
\hline

~\citep{bi2024autonomous} & - &  - & Virtual probe  & 4 &  Extra-corporeal scanning \\
\hline

 ~\citep{yao2024decision} & UR5 Robot & Affiniti 30, Philips Inc., Holland  & L12-4, Philips Inc., Holland & 4 & Extra-corporeal scanning \\
\hline

~\citep{bi2022vesnet} & KUKA LBR iiwa 7
R800 Robot & Cephasonics, California, USA. \newline ACUSON Juniper US System, Siemens AG, Erlangen, Germany  & CPLA12875, Cephasonics, California,
USA. \newline 12L3, Siemens AG, Erlangen, Germany & 3 & Extra-corporeal scanning   \\
\hline

~\citep{amadou2024goal} & - & - & - & &  Intra-corporeal scanning \\
\hline
  
~\citep{li2023rl} & - & - & - & 3 &  Intra-corporeal scanning\\
\hline

~\citep{ning2023autonomous} & UR3 Robot &   Mindray DC 6E II (Mindray, China) & 2D linear probe (L12-4s, Mindray, China) & 6 & Extra-corporeal scanning \\
\hline 

~\citep{ning2021force} & UR3 Robot &   Wireless UProbe ultrasonic device, HengTeng, China & Wireless US probe & 6 & Extra-corporeal scanning \\
\hline

~\citep{ning2021autonomic} & UR3 Robot &   Wireless UProbe
ultrasonic device, HengTeng, China & Wireless US probe & 6  & Extra-corporeal scanning  \\
\hline


~\citep{lin2023deep} & Intra-operative US robot & -  & - & 4 &  Intra-corporeal scanning \\
\hline

~\citep{shen2023towards} & Portable US robot (Stewart-Gough platform) ~\citep{deng2021workspace} & -  & 3D US probe & 6 &  Extra-corporeal scanning \\
\hline

~\citep{chen2021learning} & - &  - & B-US probe equipped with an inertial sensor and an Force/Torque sensor & - & Extra-corporeal scanning  \\
\hline

~\citep{hu2024probe} & - & E-CUBE 12R US system (Alpinion, South Korea) & curvilinear US probe & 3 &  Extra-corporeal scanning \\
\hline

~\citep{ning2023inverse} & UR3 Robot & -  & Wireless linear array US probe & 6 & Extra-corporeal scanning  \\
\hline

~\citep{luo2024multi} & UR5 &  Wisonic US machine & - & 6 & Extra-corporeal scanning  \\
\hline

~\citep{deng2024portable} & Parallel-mechanism-based bespoke
robot (master–slave teleoperation) & VINNO Inc. (Suzhou, China) & VINNO 10  & 2 &  Extra-corporeal scanning \\
\hline

~\citep{li2024adaptive} & Ur5e Robot & Mindray DC-8 Pro  & - & 6 & Extra-corporeal scanning  \\
\hline

~\citep{hase2020ultrasound} & KUKA LBR iiwa 7 R800 (KUKA Roboter GmbH, Augsburg, Germany) &  Zonare z.one ultra sp Convertible US System (ZONARE Medical Systems, Inc., Mountain
View, California, United States) & L8-3 linear transducer & 2 &  Extra-corporeal scanning \\
\hline

~\citep{duan2024safe} & UFACTORY xArm & -   & USB US probe Sonoptek & 6 &  Extra-corporeal scanning \\
\hline

~\citep{raina2024coaching} &  Rethink Robotics Sawyer arm & -  & Micro Convex MC10-5R10S-3 probe by Telemed Medical Systems, USA & 7 & Extra-corporeal scanning  \\
\hline

~\citep{shida2024robotic} & Seated echocardiographic robot & EPIQ 7G, Philips, The Netherlands  & Matrix array sector probe (X5-1, Philips, The Netherlands) & 6 & Extra-corporeal scanning  \\
\hline

~\citep{li2024action} & Xmate ER7pro, ROKAE (Beijing) Robotics Co., Ltd., China &  MyLab SigmaD, Esaote
Co., Ltd., Genoa, Italy & Linear US probe (L4-15), Esaote Co., Ltd. (Genoa, Italy) & 3 & Extra-corporeal scanning  \\
\hline

~\citep{stevens2022accelerated} & - &  - & - & - & Intra-corporeal scanning  \\
\hline

\end{tabular}
}
\label{tab:robots}
\end{table*}
    
\begin{figure}[htbp]
    \begin{center}
    \centering
\includegraphics[width=0.6\linewidth]{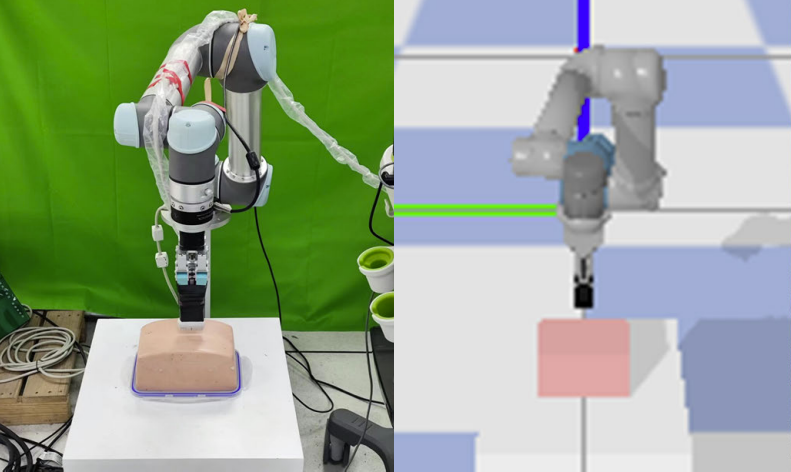}
    \caption{Simulation Environment Using the PyBullet Physics Engine ~\citep{luo2024multi}.}
    \vspace{-30pt} 
    \label{pybullet}
    \end{center}
\end{figure}

In literature, various studies use physical simulators such as Mujoco, Gazebo, and Pybullet for both the robot and the US probe ~\citep{luo2024multi}. Figure \ref{pybullet} illustrates an example of a simulation environement using Pybullet. 

\begin{figure}[htbp]
    \begin{center}
    \centering
\includegraphics[width=0.6\linewidth]{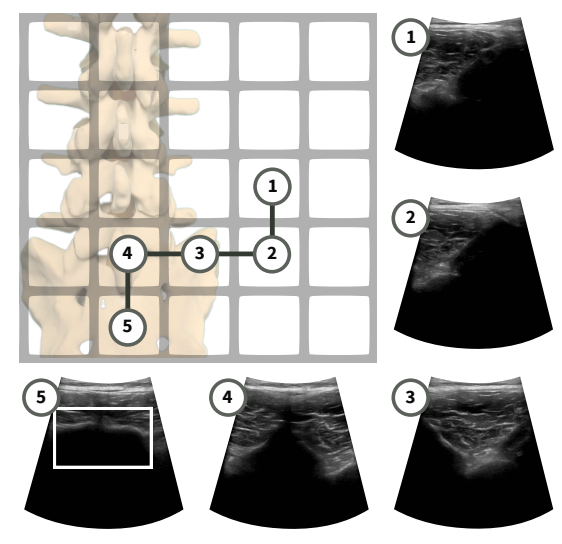}
    \caption{Grid-based Simulation Environment ~\citep{hase2020ultrasound}.}
    \vspace{-30pt} 
    \label{grid}
    \end{center}
\end{figure}

\begin{figure}[htbp]
    \begin{center}
    \centering
\includegraphics[width=0.6\linewidth]{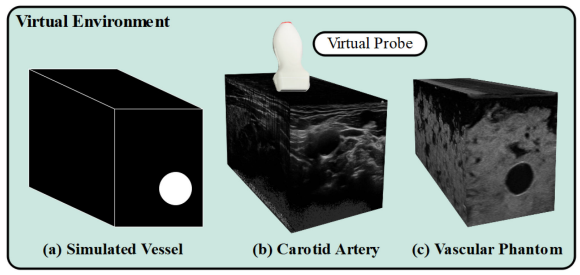}
    \caption{Binary-Based Simulation Environment ~\citep{bi2022vesnet}.}
    \vspace{-30pt} 
    \label{binary}
    \end{center}
\end{figure}

However, challenges persist in simulating the US scanned area. Due to the scarcity and restricted availability of US data, similar to other medical data, creating a realistic simulation of an organ's US environment is complex. Existing literature presents diverse approaches to address this issue. Some authors have opted to simplify the environment by constructing binary or grid-based models as shown in Figures \ref{grid}, \ref{binary} ~\citep{bi2022vesnet,hase2020ultrasound}, while others have developed simulation environments by aligning and concatenating sequences of labeled US images to create realistic panoramic environment similar to US scan area ~\citep{su2024fully}.

\begin{figure}[htbp]
    \begin{center}
    \centering
\includegraphics[width=0.6\linewidth]{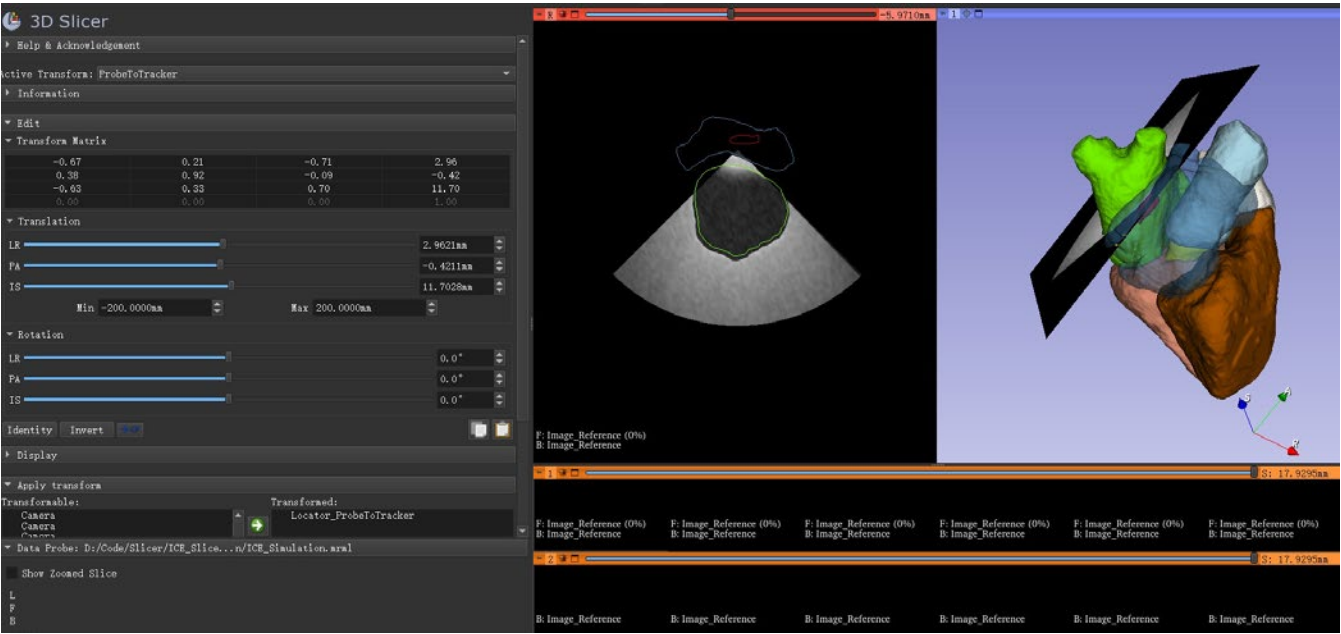}
    \caption{Simulating Ultrasound Environments Using Computed Tomography Scans ~\citep{lin2023deep}.}
       \vspace{-10pt}
    \vspace{-15pt} 
    \label{simulated}
    \end{center}
\end{figure}

Additionally, some studies have utilized US images or data from other imaging modalities, such as computed tomography (CT) scans, to generate 3D models of human organs, such as the heart, as shown in Figure \ref{simulated}. Subsequently, the US environment is simulated by generating US images and determining the corresponding positions of the US probe ~\citep{bi2024autonomous,lin2023deep,li2021autonomous}.

Despite significant progress in simulating US environments for training and validating RL models, several key challenges remain. One major issue is the scarcity of data required to create realistic simulation environments. In contrast to many other domains, medical data, including US images, is often limited due to privacy regulations, ethical considerations. This restricted availability of real US data significantly hampers the development of accurate and diverse simulation environments.
Furthermore, even when data is available, generating realistic US images remains a complex problem. Many simulation approaches rely on simplified models of the scanned area, such as binary or grid-based representations, which lack the fine details required for high-quality training ~\citep{bi2022vesnet,hase2020ultrasound}. While some methods attempt to overcome this by synthesizing panoramic views from labeled image sequences or using data from other imaging modalities (such as CT scans), these approaches often fail to capture the full richness of the tissue interactions and probe behavior seen in real US scans ~\citep{lin2023deep}.
These challenges highlight the ongoing need for more robust and realistic simulation frameworks that can address both the diversity of real-world US scenarios and the limitations of available data.

\subsection{RL Implementation and Training}

RL has shown promise in a wide range of autonomous control tasks, and its application in US image acquisition is no exception. In this context, RL algorithms enable an agent (or multiple agents) to learn optimal strategies through interactions with the environment, aiming to maximize cumulative rewards. For US applications, these rewards often correspond to achieving high-quality images or successfully navigating the probe. Since US image acquisition is inherently challenging due to factors like noise, occlusions, and the need for precise probe placement, RL-based solutions can offer significant advantages in automating and improving this process.

    
There are several RL algorithms that have been explored in the context of US image acquisition, each offering unique strengths depending on the specific task. Two of the most commonly applied algorithms are DQN ~\citep{su2024fully} and its variants, such as Double DQN ~\citep{hase2020ultrasound}, and Dueling DQN ~\citep{shen2023towards}. For a detailed background on these algorithms, please refer to section \ref{sec:rl_fundamentals}. These algorithms have been used in US-related studies due to their ability to learn optimal control policies in environments where the action space is discrete, and the state can be represented by high-dimensional inputs such as images or volumes. DQN's popularity in US image acquisition can be attributed to several factors. First, it excels in discrete action spaces, which is common in many US tasks, where the agent's actions may be limited to specific movements (e.g., move the probe left, right, or to the center, or make discrete rotations or translations along certain axes). Additionally, DQN handles high-dimensional image inputs effectively, such as labeled US images or stacked US frames, by using convolutional neural networks (CNNs) to extract relevant features. This makes DQN suitable for complex tasks like real-time probe positioning and image acquisition, where precise control over the probe's movements is needed to ensure high-quality imaging.

An alternative RL algorithm, PPO, has gained significant attention in the US field ~\citep{ning2021autonomic, li2024adaptive, luo2024multi, deng2024portable,ning2023autonomous, ning2021force, ning2023inverse}. Unlike DQN, PPO is well-suited for continuous action spaces, which are often necessary in US image acquisition tasks that involve subtle, continuous probe movements to achieve optimal image quality. PPO is favored for its stability, data efficiency, and the ability to avoid large, disruptive updates to the policy, which are common challenges in RL. In medical applications like US, where precision and reliability are critical, these properties of PPO make it a preferred choice.

Additionally, other RL methods have been explored, such as Broad RL, which uses a Broad Learning System (BLS) for more efficient and incremental Q-function estimation ~\citep{yao2024decision}. This method is particularly suitable for US, where real-time adaptability and continuous learning are often necessary due to the dynamic nature of the environment. Another notable example is the Twin Delayed Deep Deterministic Policy Gradient (TD3), which combines actor-critic architectures and has shown promise in US image-guided tasks ~\citep{chen2021learning}. TD3’s ability to leverage both a critic (for value estimation) and an actor (for action selection) can be particularly beneficial for learning optimal strategies in US, where there are multiple factors influencing image quality. The adoption of actor-critic models, including TD3, has been further extended in US applications due to their capacity to offer a robust and efficient value function. These models are well-suited for tasks where the agent needs to consider both long-term goals (e.g., high-quality images) and short-term actions (e.g., probe adjustments) simultaneously ~\citep{stevens2022accelerated,raina2024coaching, li2024action}.

Moreover, RL solutions for US image acquisition often incorporate various deep learning (DL) techniques for state representation, reward function design, and training. For example, many studies utilize segmented images rather than raw images as input to the RL model to ensure the US probe remains focused on the target. Methods such as UNet ~\citep{bi2022vesnet, li2024action}, CNN-based architectures ~\citep{su2024fully}, and hybrid CNN-based decoders with transformer-based encoders ~\citep{ning2023autonomous} are commonly employed for image segmentation. Furthermore, neural networks like CNN ~\citep{bi2024autonomous, luo2024multi}, SonoNet ~\citep{li2021autonomous}, MobileNet ~\citep{yao2024decision}, Transformer ~\citep{li2023rl}, and Autoencoders ~\citep{ning2021autonomic} are used for feature extraction from both 2D and 3D volumetric data. 

To enrich state representation, some studies define multi-state representations, with LSTM frequently applied due to its ability to preserve memory of past observations and actions and process sequential data over time ~\citep{bi2022vesnet, li2024action}. LSTM is also used in reward function design, such as capturing temporal dependencies between target and current images ~\citep{deng2024portable}. In certain cases, classifiers are employed within the reward function to determine when the goal state is achieved, such as achieving optimal image quality ~\citep{yao2024decision, hase2020ultrasound}. Additionally, Recurrent Neural Networks (RNNs) are used in some works to process sequential data ~\citep{luo2024multi}.
Image quality assessment also plays an important role in RL-based solutions, with several techniques proposed, including neural networks ~\citep{deng2024portable,raina2024coaching, shen2023towards}, Bayesian Optimization (BO) ~\citep{su2024fully}, confidence maps ~\citep{yao2024decision, duan2024safe}, and SSIM ~\citep{luo2024multi}. 
Additionally, other DL techniques are sometimes included, such as YOLO v8 for detecting heart structures in US images ~\citep{shida2024robotic}, and distributed RL, which enables training across multiple nodes for parallel processing ~\citep{bi2024autonomous}. Multi-task learning is also applied in a work ~\citep{yao2024decision} to manage tasks like lesion classification, segmentation, and image quality assessment simultaneously. In the context of manual US scanning ~\citep{macuradynamic}, an RL-based approach is used to track a needle's insertion into tissue, incorporating US imaging along with a CNN model for feature extraction.


In conclusion, while RL-based approaches have shown significant promise in US image acquisition, there are still several limitations that need to be addressed. One of the key challenges is that US scanning often requires several manipulations, such as adjusting imaging parameters and diagnosing the disease, tasks that are not always covered in most existing works. As observed, many studies focus on the scanning task solely, neglecting the simultaneous handling of both technical and clinical aspects. Additionally, RL models can be time-consuming to train and computationally intensive, which limits their applicability in real-time clinical settings.
To overcome these challenges, one potential direction is the integration of multi-agent RL, which could enable multiple agents to collaborate and optimize probe positioning, imaging tasks, and even diagnosis in more dynamic environments. Moreover, incorporating transformer-based models for feature extraction and state representation could improve the ability to capture long-range dependencies in sequential data, making RL systems more effective in tasks that require continuous adjustments over time. Finally, transfer learning could be leveraged to reduce training time and computational complexity, enabling the model to quickly adapt to new tasks or patient-specific conditions.

\subsection{RL Validation and Fine-Tuning}

Validating and fine-tuning a RL model in real-world scenarios, such as US image acquisition, is a crucial step to ensure the model can operate efficiently in practical situations. However, before reaching this stage, pretraining the RL model in an offline or simulated environment is essential. As demonstrated in the work by ~\citep{luo2024multi}, without pretraining, the agent must learn the policy from scratch, leading to slower learning rates and poor generalization with fewer training iterations.
Thus, the validation phase occurs later, after the model has been trained in a simulated environment or with offline data, to verify its performance in real-world scenarios. Based on related work, there are three primary approaches of validation: real-time human validation, real-time phantom validation, and data validation. The choice of validation type depends on the complexity of the scanning case. 
For example, in transesophageal echocardiography (TEE) or other intra-corporeal scans, human validation is particularly challenging due to the invasive nature of probe insertion, which makes it difficult to test directly on humans without extensive prior validation and safety testing ~\citep{li2023rl}. In contrast, phantom validation provides a safer and more controlled environment for testing, while data validation is often used to simulate a variety of scenarios for generalization purposes.

\textbf{Human Validation:}
Many studies validate models using real-time human interactions, ensuring that the system can function effectively in a clinical setting. For instance, in ~\citep{su2024fully}, the system maintained a stable scanning speed even when patient movements or the absence of a thyroid gland were present. Image quality was assessed using various metrics, such as confidence maps, centering errors, orientation errors, and image entropy. Human operators were more efficient, as the proposed system took 213.0 ± 85.3 seconds to perform a single thyroid lobe scan for 70 participants, whereas five doctors completed the same task in an average of 67.2 ± 27.6 seconds for 13 participants. This difference is attributed to the system’s dynamic path planning and controlled feedback for participant comfort and safety.
Similarly, ~\citep{ning2023autonomous} and ~\citep{ning2021force} conducted human validation trials for RL solutions in robotic US systems. For example, in ~\citep{ning2021force}, the model demonstrated its ability to adapt to challenging body areas like the arm, abdomen, and thighs, showing robustness in dynamic environments. The agent needed to make more adjustments when scanning sloped areas, such as the arm. In ~\citep{ning2023autonomous}, human trials involving two adults showed the system's ability to produce stable vascular images even during dynamic arm movements with minimal occlusion.

\textbf{Phantom Validation:}
Phantom models, which simulate human anatomy, are commonly used to validate RL models in a controlled setting. For example, in ~\citep{bi2022vesnet}, two vascular phantoms were used to test the model's ability to locate standard artery planes. Similarly, in ~\citep{deng2024portable}, an abdominal phantom was used to simulate patient-specific scenarios and assess how well the RL model performed under disturbances such as shaking. \cite{luo2024multi} used phantoms to reduce errors during task execution and minimize real-world training iterations. They collected real-world data and conducted an additional 10,000 iterations of interactive training to optimize the US probe scanning process.

\textbf{Data Validation:}
Some studies focus on data validation, using pre-collected datasets for model training and testing. For instance, in ~\citep{stevens2022accelerated}, different datasets, such as simulated wire targets, wire phantoms, and in-vivo data from a porcine model, were used to validate the model’s functionality. The robustness of the model was then evaluated in more realistic scenarios, including phantom wires, ultimately demonstrating its clinical potential with porcine data. Similarly, ~\citep{li2021autonomous} used data validation to evaluate the model's performance on both inter-patient and intra-patient tasks. A notable advantage of this validation is the assessment of intra-patient variability, showing how well the model adapts to variations in spinal anatomy. In the context of liver scanning, ~\citep{bi2024autonomous} used different datasets to validate the RL model, noting lower success rates in narrower intercostal spaces and when the number of targets increased. Finally, the dynamic RL model developed for needle tracking in ~\citep{macuradynamic} was validated using a dataset of frames generated from tofu and a water tank.
  
In the validation of RL models, various aspects are validated depending on the case study and the scan objectives. Typically, US image quality is validated, using commonly adopted metrics in the literature such as LPIPS Loss (Learned Perceptual Image Patch Similarity) which measures perceptual differences between images using a pre-trained deep network, Dis (Distance Metric) which quantifies spatial differences between key features in the images, while Ang (Angle Metric) evaluates the probe's angular alignment with the target and SSIM which is widely used to assess structural similarity, considering luminance, contrast, and structure. These metrics help compare the current US image with the target US image that the RL model aims to achieve, evaluating their similarities or differences. Among these, SSIM is the most widely used for qualitative evaluation, as shown in several studies such as ~\citep{stevens2022accelerated,luo2024multi,deng2024portable,li2021autonomous,ning2023autonomous, shen2023towards,li2024action}. For example, in ~\citep{luo2024multi}, if the SSIM value is greater than or equal to 0.55, the task is considered successful, indicating a high-quality standard US image. 
Other metrics such as Normalized Cross Correlation (NCC), used in ~\citep{deng2024portable} to quantify the correlation between two images, and confidence probability ~\citep{duan2024safe} are used also. Additionally, image quality is sometimes assessed qualitatively, considering factors like the force applied by the robot. For instance, in ~\citep{duan2024safe}, it was observed that the clarity of the US images increased as the applied force magnitude grew. Other validations are also sometimes conducted to check force control and the stability of the robot. For example, ~\citep{li2024adaptive} validates their solution by examining factors that influence the robot, such as stiffness and damping.


 Hyperparameter fine-tuning is an important step in developing RL models for US image acquisition, as it adjusts key parameters like batch size, learning rates, and exploration rates to improve model generalization and performance during training. A commonly used optimizer is Adam, known for its efficiency in handling high-dimensional data such as US images and sparse gradients, making it well-suited for complex tasks like US image acquisition. In terms of batch size, studies typically use values ranging from 8 to 256, depending on the model’s needs and available computational resources. For example, batch size 8 is used in ~\citep{shida2024robotic}, 16 in ~\citep{chen2021learning,li2024action}, and 32 in ~\citep{li2023rl,shen2023towards}, while larger batch sizes, such as 256, are employed in ~\citep{lin2023deep} to facilitate faster training and manage large datasets. In terms of learning rates, for example, in ~\citep{li2021autonomous}, the authors employed a learning rate schedule, starting at 0.01 for the first 40k steps and gradually decreasing it to 1e-4 during the final steps. This gradual reduction helped stabilize the model and improve its performance in the later stages of training. Similarly, in ~\citep{bi2022vesnet}, the learning rate was fine-tuned across different training phases, beginning at 0.0005 for the first 500 episodes and decreasing over time. In ~\citep{ning2023autonomous}, a learning rate of 0.01 was used, halved every 20 epochs until reaching a final value of 0.0001, optimizing the agent’s exploration-exploitation balance. Additionally, the epsilon-greedy exploration strategy is often used to balance exploration and exploitation, with epsilon decayed over time to allow for efficient RL policy learning. These hyperparameter choices are tailored to enhance the stability and generalization of the RL model in the context of US image acquisition.

In conclusion, validating RL-based solutions for US image acquisition remains a challenging task, particularly when it comes to real-world validation. While human validation is difficult in some cases, even non-invasive scenarios can pose challenges due to the complexity of ensuring that robotic systems work effectively in dynamic clinical environments. Phantom models and data validation offer controlled testing conditions but may not capture all the real-time variables encountered in actual scanning procedures. Therefore, thorough and rigorous testing, along with fine-tuning of hyperparameters, is essential to ensure that RL models can generalize well to real-world applications. Future research should focus on refining validation techniques, improving hyperparameter optimization, and addressing the complexities involved in real-time autonomous US scanning to bridge the gap between simulated training and practical deployment.

\section{RL for US Image Enhancement}

Following US image acquisition, enhancement plays a pivotal role in improving US images' visual quality and diagnostic interpretability. Due to the reliance on high-frequency sound waves, US images are prone to artifacts such as speckle noise, low contrast, blurring, and shadowing artifacts, which can obscure critical anatomical details and hinder accurate clinical interpretation ~\citep{image_enhencement_00}. These challenges necessitate specific processing steps to ensure that US images meet the high standards required for diagnostic accuracy. Effective image enhancement improves visibility and aids in image analysis tasks such as segmentation, classification, feature extraction, and so on, thereby reducing diagnostic errors and improving clinical outcomes ~\citep{image_enhencement_01}.

In US image enhancement, conventional image enhancement methods often rely on filtering techniques, such as median filtering or wavelet transforms. Additionally, machine learning methods like CNNs, autoencoders, and generative adversarial networks (GANs) are commonly employed to denoise, improve resolution, and enhance contrast of US images ~\citep{Image_enhence_RL_03}. However, these methods face several limitations. First, conventional techniques often require extensive manual tuning of parameters, which can be inefficient and prone to errors across diverse imaging conditions. They also rely heavily on labeled training data, which is labor-intensive and challenging to obtain in medical imaging due to privacy and annotation constraints ~\citep{Image_enhence_RL_02}. Additionally, these models typically work in a static framework, optimizing enhancement for predefined conditions, which limits their adaptability to real-time variations, such as probe pressure or patient-specific anatomical differences. Lastly, many traditional methods lack the ability to dynamically learn from their errors, resulting in suboptimal performance when faced with unforeseen imaging conditions or noise levels ~\citep{Image_enhence_RL_01}.

However, RL approaches offer unique advantages by addressing challenges that traditional computer vision methods may struggle with. Unlike static models, RL can handle sequential decision-making processes, such as step-by-step denoising or edge sharpening, while maintaining global context within an image. RL agents also eliminate the need for extensive labeled datasets by using self-supervised feedback loops, where the reward signal itself guides the enhancement process ~\citep{image_enhencement_05}. Furthermore, RL's ability to learn from past experiences allows it to generalize better across diverse datasets, making it more robust against artifacts and noise. By leveraging policies trained on a wide range of conditions, RL methods provide a scalable and adaptable solution for enhancing US images, paving the way for improved diagnostic accuracy and reduced operator dependency. This adaptability makes RL particularly useful in the US, where heterogeneity and real-time requirements are critical ~\citep{image_enhencement_06}. In this section, we will explore the use of RL in US image enhancement. In this section, we will explore deeper into studies investigating how RL has been used for image enhancement. With recent breakthroughs, RL has emerged as a transformative approach to improving the efficiency, precision, and reliability of US image enhancement. We will examine these advancements in detail, highlighting the innovative ways RL is reshaping image enhancement, as well as examine the stages involved in implementing RL models, including data acquisition, preprocessing, model training, validation, and optimization, providing a comprehensive view of this cutting-edge methodology.

\subsection{RL Data Preparation and Processing}

The success of RL in US image enhancement heavily relies on the quality and preprocessing of the data used during model development. Preparing data for RL involves critical steps such as selecting appropriate datasets, defining sources, and employing advanced preprocessing techniques like data normalization, augmentation, and resolution standardization. This section highlights the methods used to process and curate data for RL applications in enhancing US images ~\citep{image_enhence_data_prep}. Particular emphasis is placed on tasks such as noise reduction, US quality assessment, and so on.

\begin{table*}[!h]
  \centering
  \caption{Overview of Data Sources of RL Model Development for US Image Enhancement}
  \resizebox{\textwidth}{!}{
  \begin{tabular}{|p{1.5cm}|p{2.0cm}|p{2.5cm}| p{2.0cm}| p{7.5cm}|}
\hline
    Reference & Scanned Organ &  Dataset Type & Data Source &  Description \\ \hline

~\citep{enhence_quality_data_00} & Fetal brain & Private dataset & Human & Private 878 US video recordings from fetal screening from a Shenzhen hospital \\
\hline

~\citep{enhence_quality_data_01} & Prostate & Private dataset & Human & 6644 2D US images from 249 prostate cancer patients \\

\hline

~\citep{enhence_quality_data_02} & Prostate &  Private dataset & Human & using transrectal 6712 US images from 259 prostate cancer patients\\ 

\hline

~\citep{enhence_denoise_data_00} & Breast & Private dataset & Simulation & OASBUD US images by artificially corrupting them with Gaussian noise\\
\hline

~\citep{enhence_denoise_data_01} & Fetal head & Private dataset & Human & Data size: 17,620 volumes from 122 subjects \\
\hline

~\citep{enhence_data_01} & Lung,  prostate & Public dataset & Human & Gaussian noise was added varying between 0.0 and 0.8 with 1.0 meaning all pixels in the image are corrupted by random noise and 0.0 means that no pixels are corrupted\\
\hline
 
~\citep{enhence_data_02} & Not a specific organ & Private dataset & Human  & Data Consists of 100 samples \\
\hline

\end{tabular}
}
\label{tab: Enhecement_data}
\end{table*}

For RL applications in US image enhancement, the data is structured around the components of state, action, and environment to guide the agent's learning and decision-making process effectively. The state represents the current quality and features of the US image, such as its raw pixel data, noise level, contrast, resolution, or specific extracted features like texture or edges, which characterize the image's condition before enhancement. The action involves the potential operations the RL agent can apply to improve the image, such as noise reduction, contrast adjustment, sharpening, or region-specific enhancements ~\citep{image_enhence_data_prep_}. The updated environment refers to the updated image and its characteristics after the selected action is performed, reflecting improvements in quality metrics like noise reduction, structural similarity, or perceptual clarity. Table \ref{tab: Enhecement_data} provides a comprehensive overview of the datasets used in various studies, detailing the organ types and imaging characteristics considered in RL-driven US enhancement.


The table shows that existing RL approaches US image enhancement often rely on private datasets due to ethical constraints in sharing patient or volunteer data. However, the availability of datasets is a critical factor for the continued development and evaluation of RL-based approaches for US image enhancement. So, It highlights the urgent need for publicly accessible datasets to facilitate reproducibility and broader experimentation in this domain. Furthermore, the generation of simulated US images and their availability for public use can play a pivotal role in overcoming current data limitations. Whereas, some studies utilized hospital patient data ~\citep{enhence_quality_data_00, enhence_quality_data_01, enhence_quality_data_02}, while others relied on data from human volunteers ~\citep{enhence_denoise_data_01}. Simulated datasets also played a significant role in experimental studies, proving valuable when real-world data was unavailable ~\citep{enhence_denoise_data_00}. This highlights the importance of simulated data for advancing research and the need for publicly available datasets to foster innovation. Notably, the datasets used in these studies span various organs, reflecting the versatility of RL approaches across different imaging contexts. For example, medical organs were used during the experiments such as the fetal brain ~\citep{enhence_quality_data_00}, breast ~\citep{enhence_denoise_data_00}, prostate ~\citep{enhence_quality_data_01, enhence_quality_data_02}, and lung ~\citep{enhence_data_01}. However, certain studies, such as ~\citep{enhence_data_02}, did not specify the organs utilized during their experiments, indicating variability in reporting standards. The diversity of organs included in these datasets further emphasizes the need for a standardized and publicly available resource to support future RL-driven advancements in US image enhancement.


The quantity and diversity of data are critical factors in US image enhancement, as they directly impact the effectiveness of noise reduction and model robustness. All of the approaches utilized a wide range of datasets, often augmented through techniques such as geometric transformations (e.g., rotation, flipping, and scaling), brightness and contrast adjustments, and image cropping and padding. These augmentation strategies enhance variability, allowing models to generalize effectively across diverse imaging conditions. For instance, ~\citep{enhence_denoise_data_00} specifically focused on breast images artificially corrupted with Gaussian noise to replicate real-world noise challenges, further strengthening model robustness. By integrating diverse datasets, noise simulation, and augmentation techniques, RL models facilitate noise reduction but also improve image quality assessment, alignment, and the preservation of anatomical details in US imaging.

\subsection{RL Problem Formulation and Algorithms}

The application of RL in US imaging has gained significant traction, offering innovative solutions for enhancing image quality. Different US imaging tasks, such as image denoising, quality assessment, and alignment, present unique challenges that are addressed through tailored RL frameworks. Each task defines its RL formulation with carefully designed state spaces, action sets, and reward mechanisms to achieve task-specific optimization. This section delves into the diverse applications of RL in US image enhancement, highlighting the strategies and methodologies employed to tackle these critical challenges effectively. The methods focus on optimizing key parameters such as quality assessment, pose correction, and denoising to achieve superior diagnostic outcomes. Table \ref{tab: image_enhencement_RL_prob_formulation} summarized the RL problem formulation and algorithm used for the image enhancement.

In the context of US image denoising, the RL problem is formulated by the existing work to optimize the removal of noise while preserving critical anatomical details. The state space represents the current condition of the US image, including pixel intensity distributions and noise patterns. Actions correspond to applying specific denoising operations, such as filtering or adjusting pixel values in localized regions ~\citep{Image_enhence_RL_prob_denoise}. The agent learns to select these actions iteratively to enhance image clarity. The reward function is designed to balance noise reduction with detail preservation, using metrics such as SSIM by \cite{enhence_denoise_data_00} or peak signal-to-noise ratio (PSNR) by \cite{enhence_denoise_data_01} as feedback. By iteratively interacting with the noisy image environment, the RL agent refines its denoising strategy, ensuring high-quality output tailored to diverse imaging conditions. The diversity of algorithms used in US image denoising further illustrates the adaptability of RL to various scenarios. For instance, Jarosik et al. leverage the Asynchronous Advantage Actor-Critic (A3C) algorithm for 2D images, where states represent the current image, actions involve filtering and pixel adjustments, and rewards minimize denoising error ~\citep{enhence_denoise_data_00}. Similarly, Wright et al. apply a deterministic policy tailored for 3D US imaging, optimizing actions and rewards to handle the added complexity of three-dimensional data ~\citep{enhence_denoise_data_01}. These algorithmic approaches showcase the flexibility of RL in addressing the unique challenges of denoising across different imaging dimensions.

In the context of US image quality assessment, RL is employed to automate the evaluation of image quality based on predefined clinical criteria, such as clarity, contrast, and anatomical visibility. The state space represents the current quality attributes of the US image, which may include metrics derived from texture, intensity histograms, or edge sharpness. Actions involve selecting specific image enhancement techniques or parameter adjustments, such as modifying gain, contrast levels, or probe positioning. The reward function is carefully designed to reflect improvements in image quality metrics, ensuring alignment with clinical standards and practitioner preferences. By iteratively interacting with the imaging environment, the RL agent learns to identify and address quality deficiencies, producing diagnostically relevant images ~\citep{Image_enhence_RL_prob_quality}. On the other hand, RL algorithms showcase their capacity to optimize decision-making at multiple levels of complexity. For instance, hierarchical reinforcement learning (HRL) approaches integrate frame-level and video-level evaluations, with specialized agents like Bi-LSTM and 3D CNN working in tandem to assess and classify image quality ~\citep{enhence_quality_data_00}. Meanwhile, PPO models, enhanced with RNN controllers, adapt dynamically to varying label distributions, focusing on metrics such as accuracy and Dice score to refine image selection processes ~\citep{enhence_quality_data_01}. Additionally, Deep Deterministic Policy Gradient (DDPG) methods prioritize task-specific improvements, rewarding actions that enhance downstream task performance, such as segmentation or classification accuracy ~\citep{enhence_quality_data_02}. These tailored RL formulations ensure that quality assessment systems can adapt seamlessly to diverse imaging scenarios, promoting the generation of diagnostically relevant and high-quality images across applications.


US image registration is a critical task in image enhancement, where the RL problem is formulated to optimize the spatial arrangement of images for accurate alignment. The state space encapsulates the current alignment parameters, including relative positions, orientations, and overlap between image segments. Actions correspond to transformations such as translations, rotations, or scaling operations applied to the images ~\citep{Image_enhence_RL_prob_registration}. However, the reward function is designed to assess alignment quality, often using metrics like mutual information, normalized cross-correlation (NCC), or reflect improvements in accuracy and Dice score ~\citep{enhence_data_01}. In certain cases, a combination of multiple metrics like rewards tied to frequency matching accuracy and vibration stability has been utilized ~\citep{enhence_data_02}. By iteratively adjusting the alignment parameters through interactions with the image environment, the RL agent learns an optimal alignment strategy that ensures seamless integration of images while preserving anatomical continuity. Despite the promising potential of RL for image alignment, the number of studies applying RL to this domain remains notably limited.

The formulation of RL problems for US image enhancement presents unique challenges, particularly in designing effective reward functions that can generalize across diverse imaging conditions and anatomical structures. A significant challenge lies in crafting rewards that balance competing objectives, such as noise reduction and preservation of anatomical details. While some approaches have incorporated perceptual metrics like SSIM or PSNR into the reward structure, these metrics alone may not capture clinical relevance. For example, methods that focus on denoising often prioritize minimizing pixel-level errors, which can lead to over-smoothing or loss of critical diagnostic features. Moreover, the effectiveness of these reward functions is often tested on specific datasets, such as fetal brain or breast images, raising questions about their applicability to other organs or imaging scenarios. This highlights the need for task-specific reward designs that integrate clinical input and consider broader generalizability to ensure robust and meaningful enhancement across varied US imaging applications.

\begin{figure}[htbp]
    \begin{center}
    \centering
\includegraphics[width=1.0\linewidth]{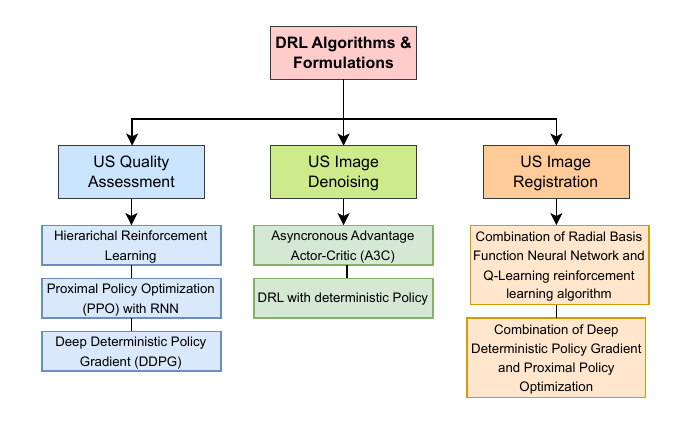}
   \caption{ Deep Reinforcement Learning (DRL) Algorithms for Ultrasound (US) Image Enhancement Tasks. The Abbreviation RNN Refers to Recurrent Neural Networks.}
      \vspace{-40pt}
    \label{fig: DRL_enhecement_}
    \end{center}
    \vspace{10pt}
\end{figure}

\begin{table*}[!t]
  \centering
    \caption{Overview of Related Work about US Image Enhancement}
    \resizebox{\textwidth}{!}{
  \begin{tabular}{|p{1.5cm}|p{2.0cm}| p{2.0cm}| p{4.0cm}| p{7.5cm}|}
\hline
    Reference & Organ & Reward Function & RL Algorithm & Metrics \\ \hline

 ~\citep{enhence_denoise_data_00} & Brain  & Dense  & A3C & Average reward \& PSNR and SSIM \\
\hline

 ~\citep{enhence_denoise_data_01} & Prostate & Dense  & Q-learning with a deterministic policy & Qualitative feedback from expert sonographers \\
\hline

~\citep{enhence_quality_data_00} &  Prostate  &  Dense & Agentsub and Agentsup & standard accuracy, precision, specificity, sensitivity, and F1-score \\
\hline

 ~\citep{enhence_quality_data_01} & Breast  & Dense & PPO with an RNN-based controller & Mean classification accuracy, Dice score and t-tests \\
\hline

~\citep{enhence_quality_data_02} & Head & Dense  & DDPG & Accuracy, Dice score and Cohen’s kappa  \\
\hline

~\citep{enhence_data_01} & Lung  &  Dense  & DDPG  & Accuracy and Dice score \\
\hline

~\citep{enhence_data_02} & Not a specific organ  & Sparse  & RBFNN & -  \\
\hline

\end{tabular}
}
\label{tab: image_enhencement_RL_prob_formulation}
\end{table*}

\subsection{RL Simulation Environment}

Simulation environments play a critical role in the development and evaluation of RL models for US image enhancement tasks. These environments allow for training, validation, and fine-tuning of RL algorithms without the constraints and risks associated with real-world scenarios, including ethical concerns and patient safety. By simulating the enhancement process, RL agents can explore diverse scenarios, refine their decision-making strategies, and optimize performance under controlled conditions ~\citep{Image_simulation_Environment}. In the denoising stage, the environment simulates the presence of noise artifacts commonly found in US images, such as speckle noise, motion blur, and electronic interference. A variety of noise is usually present when experimental data (US image) is acquired through real patients at hospitals, which resembles to actual RL environment in the images ~\citep{enhence_denoise_data_01}. However, when data is generated or acquired through the simulation or volunteer in the hospital environment, then different noise patterns are generated synthetically using statistical models like Gaussian, Poisson, or Rayleigh distributions, based on real-world observations. For example, in the work by ~\citep{enhence_denoise_data_00}, breast US images were corrupted with simulated noise to create training environments that reflect real-world challenges. The environment also includes tools for simulating the effect of various denoising filters (e.g., median, Gaussian, or wavelet-based) as part of the action space. By iterating within this simulated setup, the agent learns to select optimal filters or apply pixel-wise adjustments, ensuring minimal loss of critical anatomical details.

\begin{figure}[htbp]
    \begin{center}
    \centering
\includegraphics[width=1.0\linewidth]{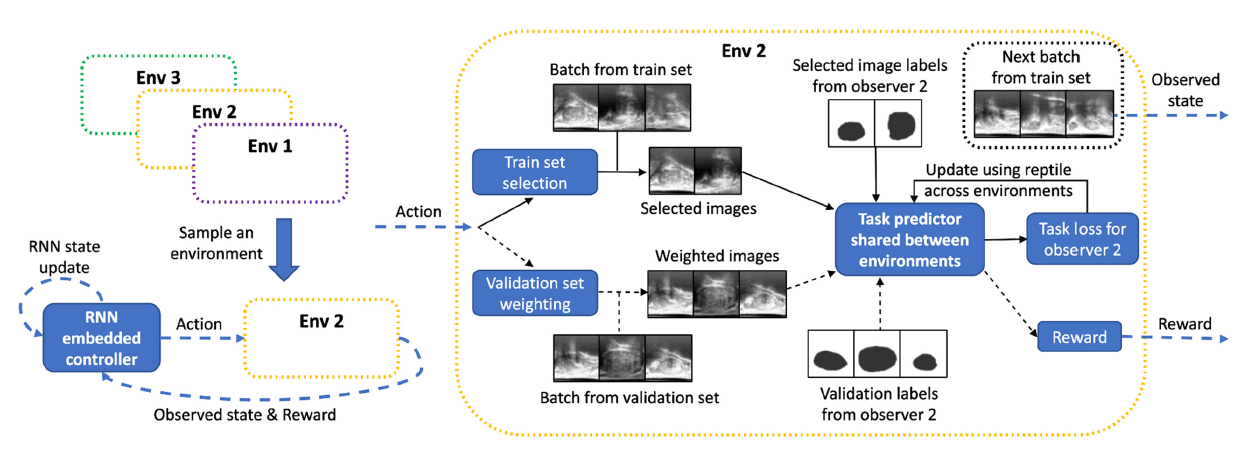}
   \caption{ An Overview of the Proposed Meta-RL Framework for Training the Task Predictor and the Recurrent Neural Networks (RNN)-Embedded Controller ~\citep{enhence_quality_data_01}.}
      \vspace{-30pt}
    \label{fig: image_quality_assessment}
    \end{center}
    \vspace{-2em}
\end{figure}

The RL environment for US image quality assessment tasks is crafted to replicate real-world conditions where agents evaluate and optimize image quality. This environment provides a structured framework for the agent to interact with data, assess image quality, and optimize decision-making strategies to ensure diagnostic reliability and consistency. The simulation environment enables structured interaction with data, simulating challenges like low contrast, uneven brightness, and anatomical visualization issues. Whereas, synthetic datasets with artificially induced degradations allow agents to learn from diverse scenarios ~\citep{Image_simulation_Environment}. Moreover, the state space typically includes quality features such as resolution, contrast, noise levels, and anatomical clarity. Hierarchical RL approaches enhance this framework by dividing the state space into frame-level and overall video-level assessments, ensuring comprehensive evaluations ~\citep{enhence_quality_data_00}. In another approach, meta-RL methods adapt to diverse datasets by modeling multiple MDPs for improved generalization ~\citep{enhence_quality_data_01}. Figure \ref{fig: image_quality_assessment} depicts an overview of the sample environment for the proposed framework, where a sample environment is chosen for the experiment from several environments. Actions in this setup involve applying quality metrics, modifying imaging parameters, or selecting images for diagnostic tasks. For instance, in prostate US imaging, task predictor models guide agents in selecting optimal frames, improving both predictor performance and image quality evaluation ~\citep{enhence_quality_data_02}. These environments provide a scalable and safe training platform and allow RL agents to explore and address diverse quality scenarios, including rare and edge cases, leading to robust strategies for real-world diagnostic improvements.

The RL environment for US image registration or alignment tasks is designed to simulate scenarios where agents learn to align multiple US images accurately, ensuring spatial consistency and anatomical coherence. This environment models the complexities of image alignment by incorporating factors such as tissue deformation, probe orientation variability, and imaging artifacts. States in this environment encode features representing the spatial relationships between images, such as pixel intensity mappings, transformation parameters, and similarity metrics like mutual information ~\citep{Image_simulation_Environment}. For task-specific adaptations, some frameworks model the environment as a MDP, where the state transitions reflect changes in alignment accuracy or task predictor performance. Here, the agent interacts with labeled image batches and a task predictor model to iteratively refine alignment parameters. The actions involve selecting transformation operations, such as translation, rotation, or scaling, to minimize misalignment between images ~\citep{enhence_data_01}. In another case, an equivalent circuit model simulates the ultrasonic scalpel transducer, linking electrical parameters to mechanical variables like displacement and force, enabling precise optimization of alignment and image quality ~\citep{enhence_data_02}. By leveraging these simulation environments, RL agents can explore and fine-tune alignment strategies in a controlled and scalable setting, overcoming real-world challenges such as safety concerns, variability in imaging conditions, and limited annotated data. This facilitates the development of robust and efficient alignment techniques that enhance the diagnostic utility of US imaging.

\subsection{RL Implementation and Training}
The implementation of RL in image enhancement tasks varies significantly depending on the nature of the input data and the specific challenges associated with each application. This section delves into the application of RL for key image enhancement tasks in US imaging, including image denoising, quality assessment, and registration or alignment. We will examine the methods and technologies employed in various studies, highlighting RL's adaptability in addressing each task's unique requirements and its role in improving the overall quality and utility of US images for diagnostic and clinical purposes.

\begin{figure}[htbp]
    \begin{center}
    \centering
\includegraphics[width=0.9\linewidth]{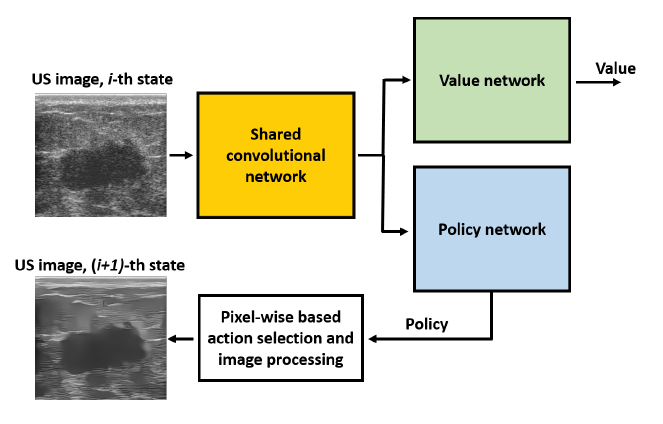}
   \caption{ Overview of Scheme Presenting a Single Iteration of the PixelRL Technique ~\citep{enhence_denoise_data_00}.}
      \vspace{-50pt}
    \label{fig: Image_simulation_Environment}
    \end{center}
\end{figure}

The image-denoising task aims to reduce noise while preserving critical anatomical details. RL-based approaches for denoising leverage iterative learning to adapt to diverse noise patterns and enhance image quality, ensuring diagnostic reliability. The method by Robert et. al. incorporates a spatial transformer network (STN) within an RL framework to address both denoising and pose correction. Implemented in PyTorch, this approach utilizes data augmentation techniques, such as rotations and intensity variations, to improve robustness and generalization. By combining noise reduction and spatial alignment, the framework produces cleaner, well-aligned images suitable for diagnostic tasks ~\citep{enhence_denoise_data_01}. Another notable approach trains the PixelRL model on grayscale images from the BSD68 and Waterloo Exploration datasets, modified to resemble US images with added artificial noise. The RL agent learns to perform pixel-level adjustments iteratively, optimizing metrics like SSIM and PSNR to achieve high-quality outputs ~\citep{enhence_denoise_data_00}. Figure \ref{fig: Image_simulation_Environment} shows how the model has been trained with noisy data by passing through the shared convolution network connected to the value and policy network, where the policy network determines the policy for pixel-based action selection.

The structured training setups for US image quality assessment provide valuable insights for US image denoising tasks. Liu et al.'s approach, with a 5-episode training scheme using an SGD optimizer, emphasizes stability and gradual refinement, which can help RL agents improve denoising iteratively while preserving anatomical details ~\citep{enhence_quality_data_00}. The integration of a Task Predictor with AlexNet and U-Net, and a controller using PPO, can capture spatial and temporal dependencies, crucial for handling complex noise patterns. This setup enables the agent to balance noise reduction with detail preservation ~\citep{enhence_quality_data_01}. Furthermore, Saeed et al. propose a system where a task predictor and controller, trained using DDPG, evaluate and rank images based on task amenability. Implemented on an Nvidia Quadro P5000 GPU, this model follows standard DDPG configurations for hyperparameter tuning ~\citep{enhence_quality_data_02}. These implementations highlight the versatility of RL in addressing the challenges of US image quality assessment, ensuring accurate and reliable evaluations across diverse clinical applications.

In the context of image registration tasks for US imaging, RL implementation, and training approaches are tailored to select the most relevant image samples and optimize task performance. For instance, neural networks like AlexNet and U-Net are used as task predictors for tasks such as prostate detection and segmentation. The RL agent, using DDPG or PPO controllers, learns to select the most task-relevant images for processing, enhancing the performance of segmentation or registration tasks ~\citep{enhence_data_01}. These networks enable the model to focus on the most critical features of the images, facilitating more accurate registration outcomes. Additionally, an innovative approach uses an equivalent circuit model combined with Radial Basis Function Neural Networks (RBFNN) and Q-learning to optimize the piezoelectric transducer behavior in real-time, adjusting the resonant frequency dynamically. Although focused on the mechanical aspects of US transducers, this framework provides insights into how RL can fine-tune system parameters for better image quality, and by extension, more accurate image registration. The combination of real-time optimization with Q-learning highlights the flexibility of RL in adapting to various imaging conditions and registration challenges ~\citep{enhence_data_02}. These diverse strategies exemplify how RL can be leveraged to address complex image registration tasks, improving both the accuracy and efficiency of US image processing systems.

\hspace{1.2pt}

\subsection{RL Validation and Fine-Tuning}
RL validation and fine-tuning are essential steps for ensuring model performance and adaptability in diverse US imaging tasks. These processes involve rigorous evaluation using real-world datasets and expert-labeled ground truth, alongside iterative refinement to optimize model accuracy and robustness. The US quality assessment framework by Liu et al. was validated using an 878-video fetal brain US dataset with expert annotations, achieving superior accuracy, precision, and F1 scores compared to state-of-the-art models ~\citep{enhence_quality_data_00}. And, fine-tuning involved pre-training for frame-level quality and collaborative learning for frame- and video-level optimization. In certain cases, different sets of images have been used to validate to test the adaptability of the RL system. One of the prominent approaches uses prostate cancer datasets (6,644 and 6,712 images) to validate the model ~\citep{enhence_quality_data_01}. However, utilization of a range of metrics for fine-tuning the model has been observed. Another US image quality assessment approach uses several metrics included classification accuracy, Dice score, and Cohen’s kappa, with fine-tuning incorporation on ng expert-labeled data, meta-learning, and refined reward definitions ~\citep{enhence_quality_data_02}. This helps to ensure more robust RL systems to improve suitability in real-life contexts.

RL validation and fine-tuning are for denoising tasks to ensure their effectiveness in handling complex and diverse noise patterns. For example, Jarosik et al. validated their denoising model on OASBUD images with Gaussian noise, demonstrating improvements in PSNR and SSIM scores ~\citep{enhence_denoise_data_00}. However, the model's performance needs further fine-tuning to address more intricate and variable noise types found in real-world scenarios. Wright et al. employed direct policy optimization with data augmentation to optimize pose correction in US imaging, validating their approach on 17,620 US volumes using image sharpness, consistency, and sonographer feedback as metrics ~\citep{enhence_denoise_data_01}. While these validations provided valuable insights into the model’s performance, broader evaluation is still required to ensure robustness across a wider range of noise conditions and imaging challenges. Fine-tuning through data augmentation, expert feedback, and targeted reward refinements can further enhance the RL model’s capability to handle diverse noise patterns, leading to improved denoising outcomes in practical clinical settings.

Finally, US image registration and alignment tasks is required in-depth validation and fine-tuning for the diverse datasets in real-world clinical conditions. The method proposed by Saeed et al. validated its task predictor model using classification accuracy and Dice score, fine-tuning across various environments with meta-RL techniques like the Reptile algorithm ~\citep{enhence_data_01}. This approach ensures that the model adapts effectively to different imaging conditions. Another RL-based method used Q-learning to adjust RBFNN weights for dynamically optimizing piezoelectric transducer behavior in response to environmental factors, demonstrating real-time optimization ~\citep{enhence_data_02}. While these approaches showed promising results, further validation across more diverse datasets and clinical scenarios is needed to evaluate their generalization capabilities and robustness in practical image registration and alignment tasks.

\section{RL for US Image Analysis}

US image analysis involves interpreting US images to aid medical diagnosis and treatment planning. It includes feature extraction, segmentation, classification, and quantitative analysis to identify structures, isolate regions of interest, and measure parameters like size and motion. Advanced imaging technologies like 2D and 3D imaging allow for dynamic assessments. Further involvement of intelligent systems can enhance automation and accuracy in applications like fetal monitoring, cancer detection, and cardiovascular evaluations ~\citep{US_image_analysis_00}. This task is challenging due to noise, low contrast, and variability in imaging conditions. However, RL is promising to enhance US image analysis, including segmentation, landmark detection, feature extraction, and 3D localization, while the adaptability and iterative learning of RL has the potential to make it a transformative tool for improving automation, precision, and efficiency in medical US imaging.

\subsection{RL Data Preparation and Processing:}

RL for US image analysis requires meticulous data preparation and processing to ensure effective training and performance. This involves curating diverse and representative datasets, preprocessing images to standardize quality, and defining state representations that capture critical features like texture, intensity, or anatomical landmarks. Additionally, simulation environments are often created with the help of data to mimic real-world conditions, allowing RL agents to iteratively learn and adapt to complex, noisy, and variable US imaging scenarios ~\citep{image_enhence_data_prep}. Researchers rely on diverse datasets for US image analysis tasks using RL to ensure the model's ability to perform robustly across various medical domains. Table \ref{tab: data_image_analysis} provides an overview of data sources used for US image analysis, where scanned organ, dataset accessibility, data source, and description have been described for each of the existing RL approaches.

Studies have shown that combining private and public datasets spanning various imaging modalities, anatomical regions, and clinical conditions enhances the generalizability and robustness of RL models. For landmark detection tasks, diverse private datasets, such as freehand US images of the spine ~\citep{analysis_landmark_00} and 3D fetal head scans from the iFIND project ~\citep{analysis_landmark_01}, have been used effectively. Additional data from brain and cardiac magnetic resonance imaging (MRI) ~\citep{analysis_landmark_02}, as well as 3D+t TEE sequences of the heart ~\citep{analysis_landmark_03}, further improve the variability in the training process. Whole-body DXA, cardiac MRI, CT scans, and cephalometric X-rays ~\citep{analysis_landmark_04, analysis_landmark_05} have also been employed, showcasing the importance of diverse data sources in ensuring RL models are equipped to handle complex anatomical structures and varied imaging conditions. Moreover, data augmentation techniques, such as rotation and flipping, have been widely applied, enabling models to adapt better to different imaging angles and conditions, ultimately improving landmark detection accuracy.


\begin{table*}[!t]
  \centering
  \caption{Overview of Data Processing for RL Image Analysis }
  \resizebox{\textwidth}{!}{
  \begin{tabular}{|p{2.2cm}|p{2.5cm}|p{2.0cm}| p{1.8cm}| p{8.5cm}|}
\hline
    Reference & Scanned Organ &  Dataset Type & Data Source &  Description \\ \hline

 ~\citep{analysis_landmark_00} & Spine & private dataset & Human & A dataset with 97 subjects using freehand US images\\
\hline

 ~\citep{analysis_landmark_01} & Brain & Private datasets & Human & Two datasets: (i) 832 T1-weighted 1.5T MRI brain scans from the Alzheimer's Disease Neuroimaging Initiative (ADNI)
(ii) 72, 3D fetal head US scans from the iFIND project \\
\hline

~\citep{analysis_landmark_02} & Brain \& cardiac & Private dataset & Human & 3 datasets were used: 455 cardiac MRI scans with six landmark locations, 832 adult brain MRI scans with fifteen landmarks, and 72 fetal head US with thirteen labeled points each \\
\hline

 ~\citep{analysis_landmark_03} & Heart & Private Data & Human & 795 3D+t TEE sequences from 659 patients\\
\hline
 ~\citep{analysis_landmark_04} & Whole-body, heart, and head & Private dataset & Human & Four datasets include (1) the whole-body dual-energy X-ray absorptiometry (DXA) images, (2) the cardiac MRI images, (3) the whole-head CT images and (4) the half-head CT images\\
\hline
 ~\citep{analysis_landmark_05} & Head & Public Dataset & human &  Publicly available ISBI 2015 cephalometric X-ray dataset\\
\hline
 ~\citep{analysis_landmark_06} & Uterus \& fetal brain &  Private dataset & Human & 683 volumes obtained from 476 patients. The average size of the volumes is 261X175X277 with isotropic voxel of size 0.5X0.5X05 mm 
 \\
\hline

~\citep{analysis_segment_00} & Breast & Private dataset & Human & Two datasets with 1000 US images\\
\hline

 ~\citep{analysis_segment_01} & Breast & private dataset & Human & Use US and mammography images \\
\hline

~\citep{analysis_segment_02} & Lung \& Brain &  Public Datasets & Human & Utilizing multimodal CT scans: LIDC-IDRI (lung CT), SCR (chest radiographs), and DeepLesion (CT images)\\
\hline

 ~\citep{analysis_segment_03} & Heart & Private Data & Human & Data with 3D CMR volumes\\
\hline
 ~\citep{analysis_segment_04} & Heart & Private dataset & Human & images were re-sampled to have a volume size of 160X160X160 voxels with padding applied, which leads to a voxel size ranging from 0.3 - 0.9 mm\\
\hline
 ~\citep{analysis_segment_05} & Fetal face & Private dataset & Human &   3D US Datasets for different standard planes: MSP (440 cases), AP (351 cases), and FCP (459 cases)\\
\hline
 ~\citep{analysis_segment_06} & Kidney, pancreas, and breast &  Private dataset & Human & 3D US datasets, including the fetal brain, fetal abdomen, and uterus\\
\hline
 ~\citep{analysis_segment_07} & Nuchal translucency area & private dataset & Human &  behind the fetus’s neck during the first trimester\\
\hline
 ~\citep{analysis_segment_08} & Neck &  Private dataset & Human & 280 sequences of carotid US imaging and radial and longitudinal pixel size for image acquisition was 19.2 pixels/mm\\
\hline

 ~\citep{Analysis_data_F_extract_00} & Brain \& breast & private datasets & Human & 653 patients at the following hospital with 2606 videos (height of 400 and a width of 600) of breast lesions\\
\hline

 ~\citep{Analysis_data_F_extract_01} & Breast & private dataset & Human &  780 single-channel grayscale US images categorized into 133 normal, and 647 cancer cases \\
\hline

~\citep{Analysis_data_F_extract_02} & Uterus \& fetal brain & private dataset & human & normal and abnormal anatomical variations in the uterus and fetal brain\\
\hline

~\citep{Analysis_data_auto_local_00} & Uterus \& Fetal brain & private dataset & Human & 2D echocardiographic images\\
\hline

 ~\citep{Analysis_data_auto_local_01} & Uterus \& Fetal brain &  Private dataset & Human & normal and abnormal anatomical variations in the uterus and fetal brain \\
\hline

~\citep{Analysis_data_auto_local_02} & Prenatal (brain, fetal abdomen, and uterus) &  Private dataset & Human & Data with 1,635 3D US volumes with three plane localization tasks each for the fetal brain, abdomen, and uterus\\
\hline

\end{tabular}
}
\label{tab: data_image_analysis}
\end{table*}

For segmentation tasks, the use of task-specific datasets demonstrates a targeted approach to training RL models for diagnostic imaging. Private datasets, such as 3D US images of the fetal face comprising MSP, AP, and FCP standard planes ~\citep{analysis_segment_05}, and nuchal translucency area scans captured during the first trimester ~\citep{analysis_segment_07}, cater to specialized clinical needs. Complementing these are public multimodal datasets like LIDC-IDRI for lung CT, SCR for chest radiographs, and DeepLesion for CT images ~\citep{analysis_segment_02}, which provide broader perspectives for training. Furthermore, datasets specific to breast cancer imaging, encompassing both US and mammography modalities ~\citep{analysis_segment_01}, along with 3D US datasets of the kidney, pancreas, and common carotid artery (CCA) ~\citep{analysis_segment_06, analysis_segment_08}, contribute to developing segmentation models capable of handling diverse anatomical structures. The inclusion of such varied datasets highlights the effort to improve model performance in identifying and segmenting regions of clinical interest across multiple diagnostic domains.


Localization and feature extraction tasks further emphasize the importance of diverse datasets in RL model development. Examples include private datasets of 2D echocardiographic images of the uterus and fetal brain ~\citep{Analysis_data_auto_local_00} and 1,635 3D US volumes covering the fetal brain, abdomen, and uterus localization tasks ~\citep{Analysis_data_auto_local_02}. Additionally, datasets containing normal and abnormal anatomical variations in the uterus and fetal brain ~\citep{Analysis_data_F_extract_02, Analysis_data_auto_local_01} allow models to learn from a wide spectrum of physiological and pathological conditions. Similarly, breast imaging studies incorporate private datasets for feature extraction and analysis ~\citep{Analysis_data_F_extract_00, Analysis_data_F_extract_01}, further underscoring the value of tailored data sources in RL training.

In summary, the integration of diverse datasets, coupled with advanced data augmentation techniques, equips RL models to generalize effectively across different anatomical regions, imaging modalities, and clinical scenarios. Observations from these studies highlight the need for both private and public datasets to ensure robust training, as well as the role of task-specific data preparation in addressing unique clinical challenges. However, the reliance on private datasets, often lacking diversity, limits model generalizability, while the scarcity of large-scale annotated datasets and task-specific designs further constrains applicability. Data augmentation improves variability but may not fully capture real-world complexities. Addressing these limitations through the strategic use of multimodal data, standardized benchmarks, and real-world validation can enhance the adaptability of RL models, ensuring reliability and accuracy in clinical applications.




\subsection{RL Problem Formulation and Algorithms}
\begin{figure}[htbp]
    \begin{center}
    \centering
\includegraphics[width=1.0\linewidth]{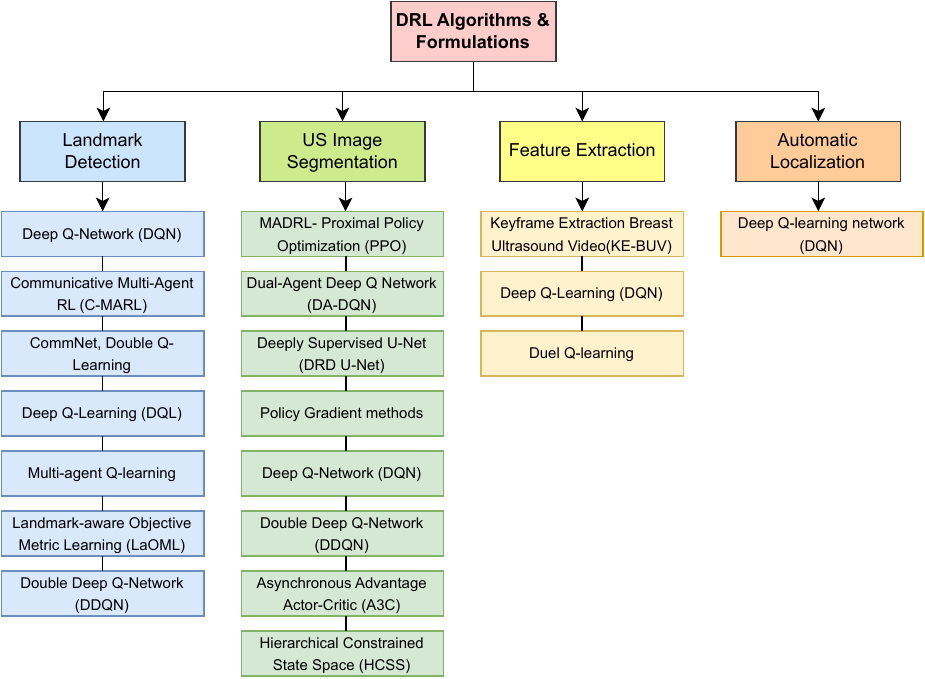}
   \caption{ Deep Reinforcement Learning (DRL) Algorithms for Ultrasound (US) Image Analysis Tasks.}
    \label{fig: image_analysis_algorithm}
    \end{center}
    \vspace{-2em}
\end{figure}

The application of RL in US image analysis has emerged as a transformative approach, providing advanced methodologies for tasks such as landmark detection, image segmentation, feature extraction, and automatic localization. These tasks pose distinct challenges that are addressed through customized RL problem formulations, incorporating task-specific state representations, action spaces, and reward functions. By leveraging these tailored frameworks, RL models are designed to optimize performance metrics such as anatomical accuracy, segmentation quality, and localization precision. Figure \ref{fig: image_analysis_algorithm} depicts the RL algorithms used across these critical tasks in US image analysis. This section explores the diverse strategies employed in RL-based US image analysis, emphasizing the innovations in state space design, reward structures, and algorithm selection to address complex clinical scenarios. Advanced algorithms and techniques are employed to improve adaptability, efficiency, and diagnostic accuracy. Table \ref{tab:image_analysis_algo} provides a comprehensive summary of RL problem formulations and algorithms implemented across these critical tasks including reward function and metrics in US image analysis.

\begin{table*}[!t]
  \centering
    \caption{Overview of Related Work about US Image Analysis}
    \resizebox{\textwidth}{!}{
  \begin{tabular}{|p{2.2cm}|p{2.7cm}| p{3.0cm}| p{4.0cm}| p{6.3cm}|}
\hline
    Reference & Task & Reward Function & RL Algorithm & Metrics \\ \hline

 ~\citep{analysis_landmark_00} & Landmark Detection & Euclidean distance &  DQN & Euclidean distance error \\
\hline

~\citep{analysis_landmark_01} & Landmark Detection & Euclidean distances &  Communicative Multi-Agent Reinforcement Learning (C-MARL) & Euclidean distance error\\
\hline

~\citep{analysis_landmark_02} &  Landmark Detection & Euclidean distances & Double Q-Learning with CommNet, a communication network model with multiple agents & Distance error\\
\hline

~\citep{analysis_landmark_03} &  Landmark Detection & Euclidean distance & DQN & Euclidean distance errors, view angle prediction errors, annulus curve-to-curve errors, median errors \\
\hline

~\citep{analysis_landmark_04} &  Landmark Detection & Euclidean distance & DeepMaQ network with multi-agent Q-learning & Dice score, Hausdorff distance\\
\hline

~\citep{analysis_landmark_05} & Landmark Detection & - & LaOML (Landmark-aware Objective Metric Learning) & MSE\\
\hline

~\citep{analysis_landmark_06} & Landmark Detection & - & MARL with DDQN & Angular difference, euclidean distance, SSIM \\
\hline

~\citep{analysis_segment_00} & US Segmentation & Gestalt laws & MADRL, PPO & Hausdorff distance, Dice coefficient, Jaccard Index\\
\hline

~\citep{analysis_segment_01} & US Segmentation & segmentation accuracy & Dual-Agent DQN (DA-DQN) & Accuracy, precision, recall, F-measure, computation time\\
\hline

~\citep{analysis_segment_02} & US Segmentation & Relative overlap & Deeply Supervised U-Net (DRD U-Net) with DRL agent & SSIM, signal-to-noise ratio (SNR), training loss, relative overlap, accuracy \\
\hline

~\citep{analysis_segment_03} & US Segmentation & Structural similarity & DQN or Policy Gradient methods & Accuracy\\
\hline

~\citep{analysis_segment_04} & US Segmentation & Euclidean distance & DQN, leveraging Q-learning & Dice Score, jaccard score\\
\hline

~\citep{analysis_segment_05} & US Segmentation & Spatial Location Reward, Anatomical Structure Reward & DDQN & Angular deviation, euclidean distance, SSIM, success rate, improvement rate\\
\hline

~\citep{analysis_segment_06} & US Segmentation & - & A3C & Accuracy\\
\hline

~\citep{analysis_segment_07} & US Segmentation & - & RL is indirectly applied through model tuning & Accuracy \\
\hline

~\citep{analysis_segment_08} & US Segmentation & Intersection-over-Union (IoU) & Hierarchical Constrained State Space (HCSS) & Radial and longitudinal motion error, IoU, Dice similarity coefficient (DSC), recall, precision, Hausdorff distance\\
\hline

~\citep{Analysis_data_F_extract_00} & Feature Extraction & Custom Reward function & KE-BUV 
(Keyframe Extraction for Breast US Video) & - \\
\hline

~\citep{Analysis_data_F_extract_01} & Feature Extraction & Prediction model's loss after reconstructing the masked region & DQN & Accuracy, macro F1 score, ROC\\
\hline

~\citep{Analysis_data_F_extract_02} & Feature Extraction & Combine spatial location and anatomical structure & Dueling Q-learning &  Angle and distance between planes, SSIM, NCC\\
\hline

~\citep{Analysis_data_auto_local_00} & Automatic Localization & Reward based on DSC function  & DQN & DSC, Hausdorff distance, mean absolute distance, accuracy\\
\hline

~\citep{Analysis_data_auto_local_01} & Automatic Localization & Combine spatial location and anatomical structure & Dueling Q-learning. & Angle and distance between planes, SSIM, NCC\\
\hline

~\citep{Analysis_data_F_extract_02} & Automatic Localization & Difference between plane parameters and the ground truth & Q-learning & Accuracy, SSIM \\
\hline

\end{tabular}
}
\label{tab:image_analysis_algo}
\end{table*}

The RL formulation for US landmark detection involves training agents to navigate and identify anatomical landmarks within US images. The state space is typically defined by local regions around the agent’s current position. In 2D US, a square region around the agent is often used to balance computational efficiency and spatial accuracy ~\citep{analysis_landmark_00}. However, for 3D US images, regions of interest (RoIs) or voxel-based sub-volumes are used to capture more complex spatial relationships ~\citep{analysis_landmark_01}. The agent’s state representation must carefully balance resolution and context to ensure effective landmark localization. The action space usually involves pixel-wise or voxel-wise movements within the image. In 2D, pixel-wise movements are employed for fine-grained exploration ~\citep{analysis_landmark_00}, while in 3D tasks, actions often include movement along Cartesian coordinates (x, y, z) ~\citep{analysis_landmark_01}. These formulations indicate the importance of both coarse and fine exploration strategies in RL-based landmark detection, especially when high precision is required. Furthermore, the use of hierarchical action strategies can enhance exploration efficiency in complex 3D volumes, as demonstrated in multi-scale search approaches ~\citep{analysis_landmark_01}. Reward functions in RL for landmark detection are designed to guide the agent toward accurate landmark localization. One of the approaches is to define rewards based on the Euclidean distance to the target landmark, with positive rewards for movement toward the target and penalties for deviation ~\citep{analysis_landmark_00}. However, more sophisticated reward functions are also employed, incorporating anatomical relevance or cooperative behaviors. In multi-agent setups, for instance, agents are encouraged to collaborate and explore different regions ~\citep{analysis_landmark_02}, improving the overall efficiency of landmark detection. Other approaches, such as using Gaussian distributions for dynamic landmark adjustments, offer more flexible reward functions that adapt to varying landmark shapes ~\citep{analysis_landmark_05}. These reward mechanisms highlight the need for nuanced supervision that aligns with clinical goals. Regarding algorithms, DQN has been widely applied for 2D tasks, enabling pixel-wise exploration and precise localization ~\citep{analysis_landmark_00}. For 3D tasks, multi-agent RL frameworks like Communicative Multi-Agent RL (C-MARL) ~\citep{analysis_landmark_01} and Double Q-Learning with communication networks (CommNet) ~\citep{analysis_landmark_02} allow agents to collaborate, optimizing exploration and improving performance in high-dimensional spaces. These algorithms reflect the growing trend toward cooperative, multi-agent systems that tackle complex tasks by enhancing exploration efficiency. RL-based US landmark detection has evolved to address the balance between computational efficiency, spatial accuracy, and clinical relevance.

US image segmentation tasks framed as RL problems involve creating environments where agents learn to delineate anatomical structures or regions of interest within US images. The state space in this context is designed to represent the agent’s understanding of the image, incorporating pixel intensities, spatial gradients, and contextual features like anatomical landmarks. Previous works have defined states using features such as superpixel decomposition-based attention masks for tumor boundaries ~\citep{analysis_segment_00}, tumor-specific attributes like density and size ~\citep{analysis_segment_01}, and multi-scale context for refined segmentation ~\citep{analysis_segment_02}. For 3D segmentation tasks, states often rely on voxel-based observations that track anatomical structures across multiple slices ~\citep{analysis_segment_04}. In terms of the action space, RL approaches in segmentation often include pixel-wise movements, region-specific adjustments, and hierarchical refinements. Pixel-wise navigation offers fine-grained exploration, especially in 2D segmentation ~\citep{analysis_landmark_00}, while multi-scale approaches support coarse-to-fine adjustments, especially in 3D ~\citep{analysis_landmark_01}. The flexibility in action spaces is important for dealing with complex image structures that require detailed localization. The reward function in RL for image segmentation typically aims to guide the agent toward accurately identifying boundaries or specific anatomical structures. Rewards are often based on proximity to true boundaries or anatomical structure similarity ~\citep{analysis_landmark_00, analysis_landmark_03}, ensuring that the agent's actions are aligned with clinical goals. For 3D segmentation, hierarchical or multi-agent reward functions facilitate cooperative refinement of the segmentation process, leading to improved performance ~\citep{analysis_landmark_01}. Algorithmically, DQN is commonly used in 2D segmentation tasks where pixel-wise or small region-based actions guide the agent’s decisions. The DQN framework allows the agent to learn the value of state-action pairs, optimizing decisions for boundary accuracy ~\citep{analysis_landmark_00}. For more complex 3D segmentation tasks, Double Q-Learning is often employed to mitigate overestimation bias inherent in Q-value learning, improving stability and performance ~\citep{analysis_landmark_02}. Multi-agent systems, including approaches like DeepMaQ ~\citep{analysis_landmark_04}, further enhance segmentation by enabling collaborative exploration and segmentation across multiple agents, particularly useful in high-dimensional spaces. In conclusion, RL-based US image segmentation tasks leverage well-structured state spaces, dynamic action spaces, and reward functions tailored to segmentation accuracy. The combination of algorithms such as DQN, Double Q-Learning, and multi-agent frameworks enables scalable and precise segmentation, providing powerful tools for a range of clinical applications.

US feature extraction focuses on identifying and isolating critical information from US data to support downstream tasks such as diagnosis, segmentation, and localization. The task involves selecting and refining key features or regions within US images or videos that contain relevant diagnostic or structural information. RL-based approaches optimize this process by framing it as a decision-making problem where agents iteratively identify and extract salient features guided by task-specific objectives ~\citep{US_feature_extraction}. The state space in these tasks represents the agent's contextual understanding of the image or video data. For instance, CNN-encoded frames in KE-BUV focus on diagnostically relevant video sequences ~\citep{Analysis_data_F_extract_00}, while 2D slices reconstructed from 3D volumes enhance feature detection with anatomical context ~\citep{Analysis_data_F_extract_02}. These state designs ensure that agents effectively capture both spatial and temporal information, with the formulation tailored to the dimensionality and modality of the task. The action space defines the decisions available to the agent for extracting features. In some tasks, such as KE-BUV, the action space allows for binary selection of keyframes to isolate relevant diagnostic moments ~\citep{Analysis_data_F_extract_00}. In other tasks, such as 3D feature extraction, the action space includes adjustments to the 3D coordinates to ensure anatomical alignment with the target region ~\citep{Analysis_data_F_extract_02}. These diverse action spaces highlight the adaptability of RL to varied imaging tasks, balancing complexity with task-specific requirements. The reward function serves as the optimization driver, aligning the agent's decisions with clinical objectives. For example, KE-BUV rewards agents based on diagnostic attributes, encouraging the selection of keyframes that contain critical features ~\citep{Analysis_data_F_extract_00}. In 3D feature extraction, hybrid rewards combine spatial alignment and anatomical relevance, ensuring precise feature extraction ~\citep{Analysis_data_F_extract_02}. These reward strategies guide the agent toward achieving clinically relevant outputs, improving the accuracy and efficiency of the feature extraction process. RL algorithms play a crucial role in optimizing the feature extraction process. For example, KE-BUV uses DQN, where the agent learns the value of selecting certain frames based on the reward structure ~\citep{Analysis_data_F_extract_00}. DQN has proven effective in balancing exploration and exploitation, allowing the agent to learn which frames or regions contribute most to the task. For 3D feature extraction tasks, algorithms like Dueling Q-Learning and dueling networks are often employed ~\citep{Analysis_data_F_extract_02}. These algorithms incorporate two value streams—one for the state value and one for the action advantage—allowing the agent to handle the complexity of 3D spatial configurations and anatomical structures more effectively. By incorporating algorithms, the approach efficiently addresses the challenges of varied imaging modalities, ensuring accurate and reliable feature extraction.

Automatic localization in US imaging involves using RL to precisely identify anatomical structures within US images. The state space typically includes image features extracted via CNNs or 2D slices from 3D volumes, which provide contextual and anatomical information to guide the agent's localization ~\citep{Analysis_data_auto_local_00, Analysis_data_auto_local_01}. Temporal and spatial aspects are often incorporated to improve the agent’s understanding over iterations. The action space enables decisions like pixel-wise mask expansions, shrinking, or 3D coordinate adjustments ~\citep{Analysis_data_auto_local_00, Analysis_data_auto_local_01}. These actions allow the agent to refine its localization progressively. The reward function generally encourages accurate localization by using metrics like the Dice similarity coefficient (DSC) or anatomical heatmaps around key landmarks ~\citep{Analysis_data_auto_local_00, Analysis_data_auto_local_01}. On the contrary, for automatic localization tasks, algorithms like DQN and Dueling Q-Learning had been applied, with DQN used for simpler 2D tasks and Dueling Q-Learning offering enhanced performance for more complex 3D localization ~\citep{Analysis_data_auto_local_00, Analysis_data_auto_local_01}. Overall, RL-based approaches in US localization demonstrate how tailored state-action designs and reward functions can improve localization accuracy, making them effective tools in clinical applications.

\subsection{RL Simulation Environment}
RL simulation environments for US image analysis play a crucial role in training and testing RL agents across image analysis tasks. These environments define a state space, action space, and reward functions to guide the agent's learning process. Using real or synthetic datasets, these setups simulate realistic scenarios, allowing agents to iteratively refine their outputs. By simulating realistic scenarios, these environments allow agents to iteratively improve their performance and tackle complex challenges in medical imaging ~\citep{us_RL_env_simulation}. In this section, we will explore the RL simulation environments for different US image analysis tasks such as landmark detection, segmentation, feature extraction, and so on.

\begin{figure}[htbp]
    \begin{center}
    \centering
\includegraphics[width=.70\linewidth]{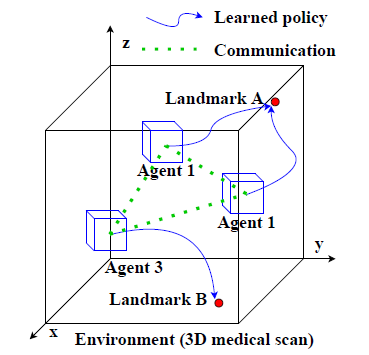}
   \caption{ Reinforcement Learning Simulation Environment for Ultrasound Image Landmark Detection Tasks ~\citep{analysis_landmark_01}.}
      \vspace{-20pt}
    \label{fig: analysis_landmark_01}
    \end{center}
    \vspace{-1em}
\end{figure}

The RL environment simulation for US landmark detection lies in its ability to model and adapt to the unique challenges of US imaging, such as low image resolution, noise, and variability in anatomical structures. These simulations create environments where RL agents can learn to effectively navigate and process the complex 2D or 3D spatial information typical of US scans. Additionally, RL environments allow agents to iteratively refine their detection strategies by receiving feedback through rewards, enabling them to optimize for accuracy in detecting landmarks under varying clinical scenarios ~\citep{us_RL_env_simulation_landmark}. For instance, some simulation environments involve agents navigating 3D US scans to detect anatomical landmarks. Ran et al. used a framework involving a MDP in a partially observable environment to locate spinous processes in the spine ~\citep{analysis_landmark_00}, while several other methods simulate multi-agent systems to detect landmarks in 3D medical scans like US or MRI ~\citep{analysis_landmark_01, analysis_landmark_02}. In addition, Leroy et. al. use a simulation environment for landmark detection tasks that is suitable for multi-agents, Figure \ref{fig: analysis_landmark_01}. Where different agents focus on detecting different landmarks and this is how the agents learned the policy. Moreover, a specific example is the use of 3D TEE for landmark detection based on defined states and rewards ~\citep{analysis_landmark_03}. In addition, shape-guided Multi-Agent RL (SGMaRL) approaches by Wan et. al. integrate statistical shape models to improve landmark localization ~\citep{analysis_landmark_04}. Further, for fetal imaging, agents navigate within 3D US volumes to identify planes like sagittal, transverse, and coronal for the uterus, or trans-thalamic and trans-cerebellar for the brain ~\citep{analysis_landmark_06}. A common observation across existing work is the flexibility in action spaces and the ability to incorporate multi-agent systems for more complex tasks like anatomical landmark detection. These setups help simulate real-world challenges, paving the way for RL approaches to improve landmark detection.

For segmentation tasks, simulation environments provide controlled settings where models can learn to accurately delineate anatomical structures. These simulations are designed to address the unique challenges of US imaging, such as noise, low resolution, and complex boundaries, by simulating realistic environments where agents can iteratively refine their segmentation strategies ~\citep{us_RL_env_simulation_segment}. RL environments offer dynamic learning opportunities, driving improvements in segmentation accuracy for various clinical applications. For instance, in Karunanayake et al.'s work, this environment helps agents handle the ambiguity in tumor boundary detection within noisy US edge maps ~\citep{analysis_segment_00}. Moreover, the use of dual-agent architectures, as seen in DA-DQN approaches, suggests that collaborative learning among multiple agents can improve segmentation performance by focusing on different regions of interest, such as fused mammography and US images ~\citep{analysis_segment_01}. The inclusion of noise in simulation environments, like Gao et al.'s Deeply Supervised U-Net, reflects a significant insight: by incorporating high- and low-frequency noise, these environments force agents to adapt and generalize their learning strategies, making them more robust in real-world scenarios ~\citep{analysis_segment_02}. Additionally, simulations for specialized tasks such as cardiac imaging and catheter localization underscore the adaptability of RL environments to 3D datasets, highlighting the ability to refine segmentation strategies in both volumetric and multi-modal settings ~\citep{analysis_segment_03, analysis_segment_04}. ~\citep{analysis_segment_04} shows the RL simulation environment for 3D US segmentation, where its been patch-based segmentation had been applied. In addition, the closed-loop frameworks, such as Yang et al.'s integration of multiple networks to adapt images based on segmentation accuracy, reveal a crucial advantage of these environments: the capacity for real-time feedback and iterative improvement, allowing agents to continuously refine their decision-making processes ~\citep{analysis_segment_06}. Collectively, these approaches demonstrate that RL simulation environments are not just useful for segmentation but essential for overcoming the inherent challenges of US imaging, such as noise, resolution limitations, and complex anatomical structures.

\begin{figure}[htbp]
    \begin{center}
    \centering
\includegraphics[width=1.0\linewidth]{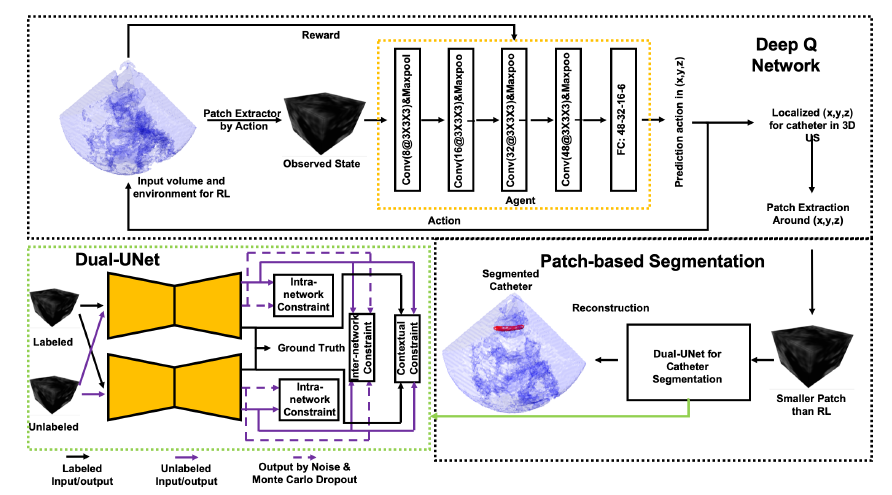}
   \caption{ Reinforcement Learning Simulation Environment for Ultrasound Image Segmentation Tasks ~\citep{analysis_segment_04}.}
      \vspace{-20pt}
    \label{Fig: analysis_segment_04}
    \end{center}
\end{figure}
\hspace{10pt}

Simulation environments for feature extraction in US image analysis play a critical role in improving task efficiency, such as lesion identification and region selection. Huang et al. simulate a sonographer’s process of identifying keyframes of lesions in breast US videos, creating a realistic environment for training ~\citep{Analysis_data_F_extract_00}. This approach is particularly effective for replicating real-world scenarios in lesion detection, which is crucial for developing clinically viable models. Bahrami’s framework, using MRI and US datasets, employs a masking agent to focus on key regions, streamlining feature extraction ~\citep{Analysis_data_F_extract_01}. This method is vital for accurately selecting regions of interest, which reduces computational cost and enhances the agent’s focus. Furthermore, hybrid approaches, like Jiang et al.’s combination of dueling Q-learning with imitation learning, enhance landmark detection by accelerating the learning process and improving accuracy ~\citep{Analysis_data_F_extract_02}. The integration of imitation learning with Q-learning is crucial as it combines exploration with prior knowledge, enabling faster convergence in complex environments. In US localization task, Yang et al.’s simulations refine LV segmentation masks and iteratively adjust 3D plane parameters for precise localization, demonstrating the effectiveness of step-by-step adjustments in improving model performance ~\citep{Analysis_data_auto_local_00, Analysis_data_auto_local_02}. This iterative adjustment process is particularly important for ensuring high precision in localization tasks, especially when working with dynamic US data.

\subsection{RL Implementation and Training}

RL implementation and training for US image analysis involves optimizing specialized models using task-specific environments, with training often done in frameworks like PyTorch or TensorFlow. Data augmentation, such as random rotations, helps improve robustness but may reduce generalization across unseen data, which can be a concern if real-world dataset variability isn't considered ~\citep{analysis_landmark_00}. This highlights the challenge of balancing robustness and generalization when augmenting data and emphasizes the need for task-specific strategies that avoid overfitting. Leroy et al.'s use of shared CNNs combined with E-greedy policies addresses multi-agent coordination, enhancing communication between agents and improving landmark detection performance ~\citep{analysis_landmark_01}. The choice of E-greedy policies in this context is significant as it ensures agents can explore new possibilities while also exploiting learned strategies, and optimizing performance in complex landmark localization tasks. Furthermore, the incorporation of shape priors and incomplete data handling in systems like DeepMaQ-Net and DeepSSM shows how domain-specific knowledge can be leveraged to improve RL models. Shape priors guide the agent to focus on anatomically relevant regions, making the system more effective even in the presence of missing or unclear data ~\citep{analysis_landmark_04}. This approach exemplifies how integrating domain knowledge can significantly enhance model robustness, ensuring more accurate and reliable outputs in US imaging tasks. These observations underline the importance of carefully designing RL training strategies, including the right balance of data augmentation, multi-agent coordination, and domain-specific knowledge, to improve performance in US image analysis tasks.

In US image segmentation, RL implementations utilize different strategies to optimize performance during training. For instance, Karunanayake et al.'s use of PPO with hidden layers to segment US data from multiple hospitals is critical in adapting the model to diverse imaging datasets and protocols. This flexibility helps the model generalize better across varied conditions, which is essential in clinical practice where data consistency is often a challenge ~\citep{analysis_segment_00}. Similarly, DA-DQN combined with Enhanced U-Net for segmenting fused mammography and US images leverages the power of both RL and advanced architectures, showcasing the necessity of combining complementary techniques to handle multimodal data ~\citep{analysis_segment_01}. Moreover, Gao et al.’s DRD U-Net model, designed for capturing both local and global features in 3D US segmentation, highlights the importance of addressing the complexity of 3D data. By efficiently processing spatial and contextual information, the model can tackle challenges like occlusion and anatomical variation ~\citep{analysis_segment_02}. Furthermore, Dual-UNet's dual-path structure, used in catheter localization and patch-based segmentation tasks, optimizes both segmentation and feature extraction, making it well-suited for high-accuracy tasks like catheter placement ~\citep{analysis_segment_04}. On the other hand, multi-step RL models combining CNN and Transformer layers, as seen in certain advanced models, underline the growing importance of integrating local feature extraction with global context to improve segmentation accuracy across complex US images ~\citep{analysis_segment_06}. These approaches illustrate how RL methods can be trained for specific segmentation tasks, balancing data versatility, computational efficiency, and model accuracy.

RL implementation and training for US image feature extraction focus on designing models capable of capturing relevant features while addressing task-specific challenges. One of the frameworks employs a CNN encoder to extract features from video frames, integrating a detection-based nodule filtering (DNF) module to filter redundant information and focus on lesion regions. An attribute classification network (ACN) with group-aware focal loss is included to address class imbalance in lesion attribute recognition, with the entire framework trained end-to-end for video summarization tasks ~\citep{Analysis_data_F_extract_00}. Another approach utilizes an encoder-decoder architecture built with deep neural networks like Auto-Encoder and UNet for feature prediction, supplemented by a DQN masking network for precise feature localization ~\citep{Analysis_data_F_extract_01}. In a separate study, a ResNet-based Q-network evaluates actions, while auxiliary tasks predict state-content similarity, aiding the agent in distinguishing target SPs from non-SPs. The inclusion of imitation learning enhances early-stage training, accelerating the learning process and improving overall performance ~\citep{Analysis_data_F_extract_02}. On the contrary, for the feature extraction tasks, utilization of CNN can be observed which has combined with a DQN to refine segmentation policies. It incorporates a novel reward function based on DSC improvement and leverages techniques like experience replay and target networks to stabilize training. A multi-scale strategy is applied to handle varying image resolutions ~\citep{Analysis_data_auto_local_00} effectively. Another method employs a ResNet-based Q-network to evaluate actions, with auxiliary tasks predicting state-content similarity to help the agent distinguish between target SPs and non-SPs. Early-stage performance is enhanced through imitation learning, allowing the agent to adapt efficiently during initial training phases ~\citep{Analysis_data_auto_local_01}. Additionally, a PyTorch-implemented system integrates a pre-trained VGG model as the backbone and utilizes a replay buffer for efficient learning. This setup, executed on NVIDIA Titan XP GPUs, ensures improved localization and computational efficiency ~\citep{Analysis_data_auto_local_02}.



\subsection{RL Validation and Fine-Tuning}

RL validation and fine-tuning for US image analysis ensure models are accurate, reliable, and clinically applicable. Validation evaluates performance using diverse datasets and task-specific metrics while fine-tuning adapts pre-trained models to specific tasks by refining hyperparameters, reward functions, or re-training on tailored datasets ~\citep{US_analysis_RL_valid_tune}. For US landmark detection, methods such as adjusting SSIM-based thresholds and iterative curve fitting for 3D spine curve generation demonstrate the importance of domain-specific fine-tuning, achieving high correlation (R=0.86) between automated and manual measurements ~\citep{analysis_landmark_00}. And, euclidean distance metrics provide a robust standard for evaluating localization accuracy across brain imaging datasets, such as T1-weighted MRIs and fetal head ultrasounds, demonstrating the adaptability of RL models to different scanning modalities ~\citep{analysis_landmark_01}. Fine-tuning strategies such as epsilon-greedy policies improve exploitation, while optimized reward functions and step sizes balance precision and computational efficiency ~\citep{analysis_landmark_02}. In interventional imaging, fine-tuning with simulated TEE sequences demonstrates resilience to artifacts like device shadows, achieving millimeter-level accuracy ~\citep{analysis_landmark_03}. Statistical shape models (SSMs) further refine landmark predictions, particularly in incomplete images, improving both Dice scores and anatomical alignment ~\citep{analysis_landmark_04}.

In the US segmentation task, RL fine-tuning and validation focuses on adapting models to diverse imaging conditions. Progressive fine-tuning, such as transfer learning schemes, handles increasing image complexities and achieves superior segmentation metrics like Dice coefficient and Hausdorff distance, validating performance across large datasets ~\citep{analysis_segment_00}. Iterative adjustments to models like DA-DQN and DRD U-Net highlight the importance of aligning action spaces and reward functions with imaging characteristics, ensuring robustness to noise and feature variability ~\citep{analysis_segment_01, analysis_segment_02}. Techniques such as uncertainty constraints and noise injections improve model adaptability, validated by robust performance on multiple datasets ~\citep{analysis_segment_04}. Specialized frameworks optimize action space and boundary clarity, emphasizing precision in localization and segmentation tasks, with near-human-level performance metrics ~\citep{analysis_segment_05, analysis_segment_07}. All of the approaches underscore the need for task-specific tuning, rigorous metric evaluation, and flexible state-action design to generalize across different US datasets and clinical applications.

For US feature extraction tasks, some methods omit explicit fine-tuning, their iterative frameworks and extensive validation demonstrate capability in extracting meaningful features, though specific metrics may not always be detailed ~\citep{Analysis_data_F_extract_00}. Other approaches employ techniques like Q-value updates in DQN agents for self-supervised region masking, validated using accuracy, macro F1 scores, and AUROC on tasks like glioma and breast cancer detection ~\citep{Analysis_data_F_extract_01}. Additionally, fine-tuning with dynamic e-greedy exploration and multi-stage action adjustments ensures adaptability to anatomical variability, validated with angular and distance deviations, SSIM, and NCC metrics on datasets featuring fetal brain and uterine imaging ~\citep{Analysis_data_F_extract_02}. On the other hand, for US automatic localization tasks, iterative RL approaches inherently refine outputs, such as segmentation masks, validated through metrics like DSC, Hausdorff distance, and accuracy, demonstrating effectiveness over baseline methods like U-Net and FCN ~\citep{Analysis_data_auto_local_00}. Strategies like E-greedy exploration adjust action steps dynamically, enabling adaptability to anatomical variations, with validation using angular and distance metrics on datasets featuring fetal brain and uterus scans ~\citep{Analysis_data_auto_local_01}. Moreover, advanced methods incorporate modules like landmark-aware alignment and adaptive dynamic termination for optimized searches, validated across thousands of 3D US volumes with structural similarity and localization accuracy metrics ~\citep{Analysis_data_auto_local_02}. These approaches highlight the importance of flexible fine-tuning and comprehensive validation for robust feature extraction in complex US imaging scenarios.

\section{RL for US Image-Based Decision-Making and Diagnosis}
\label{Sec: RL Diagnosis}
\begin{figure}
    \begin{center}
    \centering
\includegraphics[width=0.9\linewidth]{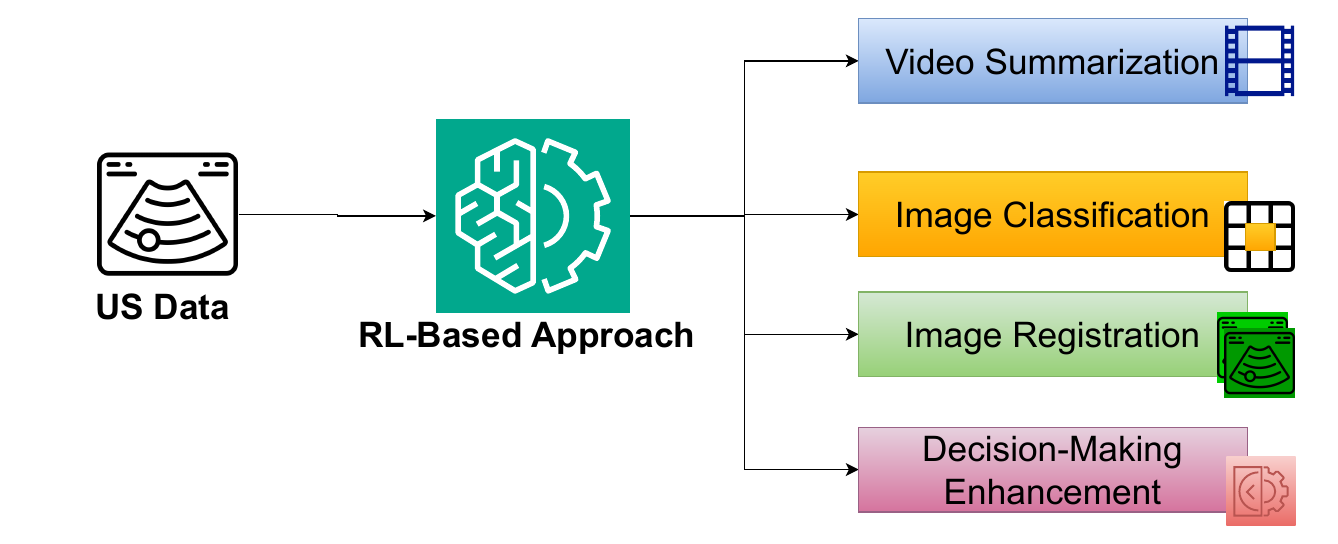}
    \caption{Overview of Current Literature on Ultrasound (US) Image-based Decision-Making and Diagnosis Using Reinforcement Learning (RL).}
       \vspace{-30pt}
    \label{diagnosis}
    \end{center}
\end{figure}

In the context of medical diagnosis using US imaging, the process typically begins with image acquisition, which can be either manual or automated, followed by image enhancement and analysis, leading to the important decision-making phase. During this phase, medical experts interpret the US images to make diagnostic decisions,  relying on their expertise and clinical experience. At this stage, the integration of advanced techniques like DL and RL can significantly assist in speeding up the diagnostic process, improving accuracy, and providing more intuitive support for clinicians. This section explores the application of RL within the decision-making and diagnostic process for US imaging. As shown in Figure \ref{diagnosis}, the existing literature indicates that current research has primarily concentrated on areas such as video summarization, image classification, image registration, and disease diagnosis.  The related works on decision-making and diagnosis are summarized in Table \ref{tab:diagnosis}, which organizes the studies based on task type. It outlines the organ examined in each study, the reward function category used, the RL algorithm employed, and the evaluation metrics. Each study will be discussed in detail in the following sections. It is worth noting that this section does not include a discussion of RL simulation environments, as all the reviewed studies used mostly private datasets for offline testing of their RL-based solutions, and no specific descriptions of simulation environments were provided.

\begin{table*}[ht]
    \centering
    \caption{Overview of Related Work about Decision-Making \& Diagnosis}
    \label{tab:diagnosis}
    \resizebox{\textwidth}{!}{
    \begin{threeparttable}
            \begin{tabular}{p{3cm} p{1.4cm} p{2cm} p{2cm} p{2.5cm} p{4cm}}
                \hline
                Task & Reference & Organ & Reward Function & RL Algorithm & Metrics \\
                \hline
                \multirow{5}{*}{\parbox{1.8cm}{\textbf{Video \newline Summarization}}} 
                & ~\citep{liu2022video} & Fetal & Dense & MDP & F1 score, summary length \\
                 \cmidrule(r){2-6}
                & ~\citep{liu2020ultrasound} & Fetal & Dense & MDP & Precision, recall, F1-score \\
                \cmidrule(r){2-6}
                & ~\citep{mathews2022unsupervised} & Lung & Dense & Unsupervised RL & Precision, recall, F1 score \\
                \cmidrule(r){2-6}
                & ~\citep{mathews2021vid} & Lung & Dense & MDP & Precision, recall, F1 score \\
                \hline
                \multirow{9}{*}{\parbox{1.8cm}{\textbf{Image \newline Classification}}} 
                & ~\citep{narmatha2023ovarian} & Ovary & Sparse & DQN & Recall, accuracy, IoU, execution Time, precision, F-measure \\ 
                \cmidrule(r){2-6}
                & ~\citep{wang2020auto} & Breast & Dense & MDP & Accuracy \\ 
                \cmidrule(r){2-6}
                & ~\citep{huang2021aw3m} & Breast & Sparse & REINFORCE & Accuracy, sensitivity, precision, specificity and F1-score \\
                \cmidrule(r){2-6}
                & ~\citep{zhang2023detection} & Thyroid & Sparse & MDP &  Confusion matrix, accuracy rate, precision and recall rate \\ 
                \hline
                \multirow{2}{*}{\parbox{1.8cm}{\textbf{Image \newline Registration}}}
                & ~\citep{zeng2020hierarchical} & Musculo-skeletal System & Dense & Q-Learning, HRL & TRE, success rate, average time \\ 
                \cmidrule(r){2-6}
                & ~\citep{shanmuganathan2024two}  & Heart & Dense & DQN, MADRL & Dice score, Hausdorff distance, mean squares, mattes mutual information \\ 
                \hline
                \multirow{2}{*}{\parbox{1.8cm}{\textbf{Disease \newline Diagnosis}}} 
                & ~\citep{xie2023reinforced} & Thyroid & Sparse  & Q-learning & Accuracy, sensitivity, specificity, classification error rates, majority rule voting \\ 
                \cmidrule(r){2-6}
                & ~\citep{li2020coronary} & Heart & Dense & A3C & Accuracy, F1 score \\ 
                \hline
            \end{tabular}
    \end{threeparttable}
    }
    \vspace{-10pt}
\end{table*}

\subsection{RL Data Preparation and Processing}

The effective application of RL in medical imaging, particularly in US imaging, is closely dependent on the quality and preparation of data. 
Data preparation involves several key aspects, including the type and source of the data, as well as the application of processing techniques such as data augmentation and resizing.  
This section focuses on the data preparation and processing methods used in RL models for assisting medical experts with diagnostic tasks, specifically focusing on tasks such as video summarization, image classification, and image registration. The table \ref{tab:data_diagnosis} details the data acquired for each study in developing the RL model, including the type of organ scanned.

Video summarization is an efficient process that enables the identification of key frames within a video without the need to view the entire content ~\citep{apostolidis2021video}. This method is particularly valuable in the medical field, where clinicians look for frames that represent specific anatomical or biological landmarks essential for diagnosis and decision-making ~\citep{mathews2022unsupervised}. Numerous DL solutions have been developed for video summarization, but the necessity for annotated frames complicates the process ~\citep{mathews2022unsupervised}. In fact, manually annotating frames in videos and US particularly is a labor-intensive and time-consuming task due to the large number of frames involved. To simplify this work, the concept of video summarization can utilize an unsupervised learning approach, where the model learns to identify labels for the frames without prior exposure. RL is one of the proposed strategies to achieve this objective. In the case of US video summarization, the collected data consists of US videos. The RL state in this context typically includes features of each video frame, such as spatial and temporal information. The RL action corresponds to the agent’s decision to either keep or reject a frame based on its importance for the summary. The RL next state reflects the updated summary after the action is taken.
There are not many works on US video summarization using RL; most use private US video datasets for training and validating their RL-based solutions. Some studies have used US video datasets focused on fetal imaging ~\citep{liu2020ultrasound,liu2022video}, while others have concentrated on lung imaging ~\citep{mathews2022unsupervised,mathews2021vid}, employing either the same US machine or videos from multiple machines to introduce variations. 

Medical image classification represents an important step in the diagnosis of diseases and the subsequent decision-making processes. The existing literature includes numerous examples of studies employing DL techniques, particularly deep neural networks, for image classification ~\citep{patricio2023explainable,litjens2017survey}. However, the emergence of RL has led to an increasing application of RL for image classification tasks, especially in medical contexts, although applications of RL in US imaging remain relatively scarce.

Existing research on RL in image classification aims to improve the US image classification process, particularly by enhancing dataset annotation and managing multi-modal classification. RL is not necessarily used as the primary classifier, but rather to support and enhance the overall classification process. From an RL perspective, the image classification task is defined as a decision-making problem, where an agent makes decisions based on the features extracted from images. In image classification, the state can consist of features extracted from US images, providing the agent with the necessary context to make informed decisions ~\citep{narmatha2023ovarian, zhang2023detection}, or, in the case of multi-modal US images, the state can be represented by the weights of the classifier model, which are optimized to enhance the classification process ~\citep{wang2020auto, huang2021aw3m}. The action refers to the agent's decision or move within the environment, such as labeling an image, adjusting the weights of different modalities, or localizing specific regions within the image, such as a nodule. The next state reflects the outcome of the agent's action and how the environment has evolved as a result. For instance, after labeling an image, the next state might include an updated training dataset or refined features for classification. Further details on the RL problem formulation for these tasks are provided in the next section.

Regarding the data used, in this work ~\citep{narmatha2023ovarian}, the authors constructed a dataset of US images of ovarian cysts obtained from the iStock portal ~\citep{istockphoto}. These images are classified into seven categories under the supervision of medical experts and are processed through filters to eliminate unnecessary noise and artifacts that could negatively affect the performance of the RL model.
Alternatively, for thyroid nodule classification, the authors in ~\citep{zhang2023detection} used the publicly available DDTI dataset to generate a dataset of benign and malignant nodules, which were then used to train and test their proposed RL model. To enhance the dataset size and introduce further variability, data augmentation techniques are also applied, as seen in ~\citep{zhang2023detection}, where the authors added Gaussian noise, rotated images, flipped them, and adjusted brightness levels.

In a different context, the authors ~\citep{wang2020auto} use a dataset of multimodal US images of breast nodules classified as benign or malignant. These images include B-mode, Color Doppler, Share Wave Elastography and Strain Elastography modalities, all validated by medical experts. 
Similarly, in the work ~\citep{huang2021aw3m} focused on classifying breast nodules, multimodal US data from volunteers were also used. The dataset was randomly divided into training, validation, and testing sets at the patient level (1022 for training, 100 for validation, and 438 for testing). In addition, a public dataset of B-mode US images was used for model validation. In both works, preprocessing steps included data augmentation, brightness adjustment, and resizing all images to uniform dimensions to ensure consistency in the analysis.

Image registration is the process of aligning images, whether taken using the same or different imaging methods (such as CT, MRI, or US), into a common coordinate system ~\citep{shanmuganathan2024two}. The goal of image registration is to find a transformation that links corresponding points between images using either feature-based or intensity-based methods ~\citep{shanmuganathan2024two}. Medical image registration is essential for several reasons. For instance, patients often undergo multiple scans over time using the same or different imaging modalities. Aligning these images helps assess disease progression or evaluate the effectiveness of treatments ~\citep{shanmuganathan2024two}. Additionally, some imaging methods, such as US, tend to produce noisy and blurred images, making them less suitable for certain analyses. In such cases, registration between the US images and clearer modalities, like CT scans, is necessary for accurate interpretation and comparison.
In the context of medical image registration, particularly for the application of DL techniques and specifically RL, a large volume of images is necessary for various reasons. This includes ensuring comprehensive coverage of various views and angles, particularly for dynamic organs like the heart, which require careful scanning from multiple perspectives ~\citep{shanmuganathan2024two}. Additionally, high-quality images are essential, as US images are often blurry, so it’s important to include enough high-resolution images to enable DL models to perform well in registration tasks. That being said, as in other applications, images used for registration can come in various formats and from different sources, such as simulated 3D scans, as done in this work ~\citep{zeng2020hierarchical}, or real 4D scans taken from multiple volunteers ~\citep{shanmuganathan2024two}. 

From an RL perspective, the state in an image registration task typically represents the image and its characteristics (e.g., current transformation or alignment between images). The action corresponds to any movement or transformation aimed at aligning the images, such as translation, rotation, or deformation. The next state is the updated transformation or the newly aligned image after the action is applied, representing a better alignment or closer match between the images.

In the context of medical diagnosis, we have identified additional studies that apply RL models for disease diagnosis, including the work by ~\citep{xie2023reinforced} for thyroid cancer diagnosis and ~\citep{li2020coronary} for diagnosing heart occlusion. In the following section, we will explain how the RL problems are formulated in each of these studies. Concerning the datasets, ~\citep{xie2023reinforced} used a dataset of US images collected from hospitals, which were labeled by medical professionals to include both benign and malignant cases for the training and validation of the RL model. This dataset is then transformed into RL components—state, action, and next state—to guide decision-making in diagnosing thyroid cancer. In contrast, ~\citep{li2020coronary} used data from hospitals, including heart color Doppler echocardiography reports, blood biochemical indicators, and basic patient information (such as gender, age, diabetes status, blood pressure, blood sugar levels, heart rate, chest pain, city, and family history of coronary heart disease). Based on this information, the authors constructed an RL state space that represents patient data, while the action space corresponds to the degree of occlusion in the heart arteries, and the next state reflects the resulting change in diagnosis decision.

From the studies reviewed, it is evident that data scarcity and variability remain significant challenges. Most of the research relied on private datasets, with public datasets being relatively rare, as shown in the table \ref{tab:data_diagnosis}. Furthermore, the datasets used were often small in size, which prompted some studies to apply data augmentation techniques to increase the diversity of the data. To improve the robustness of RL-based solutions, it is crucial for datasets to not only include variations in image quality but also encompass a variety of US image types. This inclusion of diverse variations will enhance the model's ability to generalize and perform effectively in different real-world US diagnosis scenarios.

\begin{table*}[htbp]
  \centering
   \caption{Overview of Data Sources for RL Model Development - Decision-Making \& Diagnosis}
   \resizebox{\textwidth}{!}{
  \begin{tabular}{|p{1.5cm}|p{2.0cm}|p{2.5cm}| p{2.0cm}| p{8.5cm}|}
\hline
    Reference & Scanned Organ &  Dataset Type & Data Source & Description \\ \hline

 ~\citep{liu2022video} & Fetal & Private dataset & Human & 50 labeled US video recordings of 13-65 minutes, from fetal screening examinations of 50 different patients, acquired between 24-30 weeks of gestation \\
\hline

 ~\citep{liu2020ultrasound} & Fetal & Private dataset & Human & 50 US videos of 13-65 minutes recordings fetal screening US examinations acquired between 24–30 weeks of gestation\\
\hline

~\citep{mathews2022unsupervised} & Lung &  Private dataset  & Human & 100 lung US videos taken from 40 subjects using different US machines and over various geographies (India and Spain) collected during the 2020–2021 period\\
\hline

 ~\citep{mathews2021vid} & Lung &  Private dataset & Human & US videos \\
\hline

 ~\citep{narmatha2023ovarian} & Ovary & Public dataset & Human & 87 US ovarian cyst images collected from the istock portal\\
\hline

 ~\citep{wang2020auto} & Breast & Private dataset & Human & 1616 multimodal US image sets of breast nodules from 835 patients, with each set containing B-mode, Color Doppler, SWE, and SE images from the same patient\\
\hline 

 ~\citep{huang2021aw3m} & Breast & Private \& Public datasets & Human & Multimodal US data: B-mode, Doppler, SWE, SE.
Private dataset: obtained from 835 patients. 1560 sets of multi-modal US images were acquired (419 benign cases and 416 malignant cases) with the size 448 × 320. 
Public dataset: contains only B-mode. 163 images (53 malignant cases and 110 benign cases) with an average size of 760 × 570
\\
\hline 

 ~\citep{zhang2023detection} & Thyroid & Public Dataset & Human & The open-source DDTI dataset includes 480 image data from 299 patients, consisting of 270 females and 29 males. The dataset contains malignant and benign nodules \\
\hline

 ~\citep{zeng2020hierarchical} & Musculoskeletal System & Private dataset & Simulation & 1600 US-CT image pairs. Volumetric US images are simulated using PLUS and 3D Slicer, with each volume containing 500 to 600 images. The simulated US data is based on 10 different CT scans obtained from 8 subjects \\
\hline

 ~\citep{shanmuganathan2024two} & Heart & Private dataset & Human & 7 volunteers (+18 years old) with no known cardiac disease, stable sinus rhythm. 124 4D echocardiography scans of dimensions n ×272 × 176 × 208 and n × 224 × 176 × 208 / n represents the number of volumes or frames within a sequence and ranges between 25 and 30 for each scan\\
\hline

 ~\citep{xie2023reinforced} & Thyroid & Private dataset & Human & 754 images from 245 cases. The dataset includes benign and malignant cases \\
\hline

 ~\citep{li2020coronary} & Heart & Private dataset & Human & the heart color Doppler echocardiography reports, 21 blood biochemical indicators and information about patients (gender, age, diabetes, blood pressure, blood sugar, heart rate, chest pain, city, family coronary heart disease history)\\
\hline

\end{tabular}
}
\label{tab:data_diagnosis}
\end{table*}

\subsection{RL Problem Formulation and Algorithms}

The formulation of the RL problem varies depending on the specific task at hand. In this section, we will explore how RL is applied across several diagnosis related-tasks, including video summarization, image classification, image registration, and other diagnostic processes. Each task leverages RL in unique ways, with distinct state spaces, action spaces, and reward structures designed to optimize performance for the given application.

In the context of video summarization, existing approaches adopt a similar formulation of the RL problem. Specifically, they leverage RL to decide whether to include or exclude a frame from the summary, creating a binary action space where the agent must either keep or discard a frame based on its relevance. The state space consists of a spatio-temporal representation of the video frames, where each state encodes spatial information about the frame's content and quality, along with temporal information that captures the frame’s position in the video sequence. This helps the RL agent understand how frames are interrelated over time, guiding the agent in deciding which frames to retain or reject. The reward signal in these methods is designed to optimize the selection of frames by focusing on their diversity and representativeness. The diversity reward assesses the dissimilarity between the frames chosen for the summary video, while the representativeness reward evaluates how well the summary reflects the original video ~\citep{liu2022video}. In ~\citep{liu2020ultrasound}, additional rewards are introduced, including a detection reward that calculates the likelihood of a frame being diagnostically valuable. Furthermore, the studies by ~\citep{mathews2022unsupervised, mathews2021vid} enhance the reward signal by incorporating additional metrics, such as the SSIM metric and classification scores related to lung health status.

The problem of image classification can be formulated in several ways. From the perspective of RL, the image classification problem is approached as a task where an agent makes decisions based on representations extracted from images. 
For instance, the authors in ~\citep{narmatha2023ovarian} describe the RL environment, using DQN, as having a state space that includes features extracted from ovarian cyst images using CNN, along with the corresponding prediction scores for these images. The action space consists of two primary actions: either labeling the current image and adding it to the training dataset or skipping it. The labeling process concludes when all available data has been processed or when the annotation budget is exhausted. To address the issue of delayed rewards, where rewards are assigned only after a task is completed and making it difficult to identify which actions contributed to success, the authors propose an intermediate reward approach. This reward function is designed to encourage the RL agent to avoid excessively selecting ovarian cyst images for annotation. By focusing on the most informative samples, the agent can develop effective strategies to maximize rewards while minimizing resource use. Furthermore, the reward function encourages the agent to annotate images where it has lower confidence, thus enhancing learning and improving overall model performance. As training progresses and more images are annotated, the accuracy and confidence of the RL model in its predictions increase.

In a different context, in ~\citep{zhang2023detection}, the authors used RL to automatically and accurately localize the region of thyroid nodules and subsequently classify whether the nodule is malignant or benign. In their approach, the RL state space consists of the thyroid US image and the coordinates of the area where the current nodule is located. The action space includes four possible actions (up, down, left, right), with a step size of 1 or -1. The reward is based on the calculation of the effective IoU ratio to assess the overlap between the target area where the nodule exists and the predicted area.

The diagnosis of breast cancer is a challenging task, which is why the authors in  ~\citep{wang2020auto} propose leveraging various US modalities, including B-mode, Color Doppler, Shear-wave Elastography (SWE), and Strain Elastography (SE), to improve the accurate identification of breast nodules. This work approaches classification as a multi-task learning problem, where each task focuses on a different US modality. RL is integrated into the framework to automatically determine the significance of each modality and assess their respective contributions. Here, the RL environment consists of US images from multiple modalities, and the state space is represented by a set of weights that indicate the importance of each modality. The action space includes the range of possible adjustments the agent can make to these weights, specifically between [-0.2, 0, +0.2]. The reward function evaluates the classification performance of the RL agent with higher accuracy leading to greater rewards. This mechanism directs the agent toward making optimal adjustments to the weights assigned to each modality.
In a similar approach within the breast cancer context, the authors in ~\citep{huang2021aw3m} present a RL-based framework to automatically learn the optimal weights for CNN models used in multimodal US classification, eliminating the need for manual hyperparameter tuning through multiple experimental iterations. The authors employ the REINFORCE algorithm, where the state space consists of normalized weights, with each element representing the weight of one of the modality streams (comprising 4 US modalities and their fusion). The action space involves actions that incrementally adjust these weights by values of -0.1, 0, or +0.1. The reward is defined as the cube of the classification accuracy on the validation set. The RL model employs 5 distinct agents, each with its own set of parameters, and operates under a bi-level optimization strategy. In this setup, several networks are trained with different weight configurations, and the best-performing network is selected for further training, while at the same time, the agent parameters are adjusted to maximize the reward, thus improving the overall model performance.

\begin{figure}[htbp]
    \begin{center}
    \centering
\includegraphics[width=1.0\linewidth]{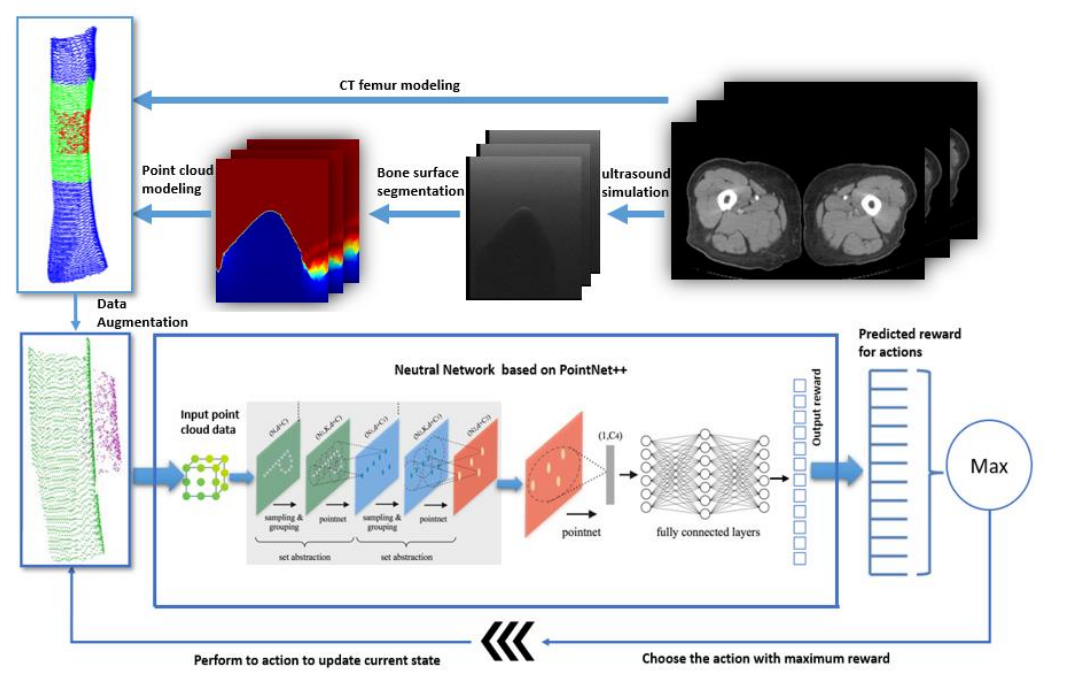}
    \caption{Detailed Overview of the Proposed Framework for Computed Tomography (CT)-Ultrasound Registration ~\citep{zeng2020hierarchical}.}
       \vspace{-35pt}
    \label{registration}
    \end{center}
\end{figure}

In the field of image registration, various techniques are employed for medical image registration ~\citep{chen2024survey}, which leads to diverse approaches for applying RL. For example, in the study by ~\citep{zeng2020hierarchical}, RL is used to develop an automatic intraoperative registration technique for US-based Computer-Assisted Orthopedic Surgery (CAOS), aimed at supporting real-time decision-making during surgery. This method aims to achieve cloud-based, real-time registration without the need for initial alignment. In this framework, the RL state space represents the current transformation in the registration process, while the action space comprises 12 possible actions, which involve translations or rotations (+/- 1). The reward function is designed to select actions that minimize the distance to the target position. A detailed illustration of the proposed framework is provided in Figure \ref{registration}.
In ~\citep{shanmuganathan2024two}, RL is applied for point-based registration between 3D US images from different windows to enhance image quality, recognize structures, and fuse images for a comprehensive view of the heart. The state is defined by a region of interest (ROI) consisting of 45×45×45 voxels centered around each agent, with the model input being the history of the previous four states. The action space includes six potential directions (up, down, left, right, forward, and backward), and the reward function is based on the Euclidean distance difference between the current and target landmarks, as well as the difference between the previous and target landmarks.

In the field of image registration, the application of RL varies depending on the type of registration and the specific objectives of the task. For example, ~\citep{xie2023reinforced} used RL Q-learning to guide and validate diagnostic decisions in the context of thyroid nodule classification. The RL problem involves a state space where each state represents the current diagnostic information the agent has about the patient, such as the classification result, demographic data, and test results. The action space allows the agent to take two possible actions after receiving a classification result: (1) make a diagnosis decision, where the agent determines whether the thyroid nodule is benign or malignant, or (2) ask for more information, where the agent can request additional diagnostic feedback (e.g., another US image) if it is unsure about the current diagnosis. The reward structure is designed as follows: if the agent makes the correct diagnosis (benign or malignant), it receives a positive reward (+1); if the agent makes an incorrect diagnosis, it receives a negative reward (-1); if the agent requests more information, it might receive no reward (0) since it has not yet made a final decision; and if the agent becomes doubtful after receiving enough feedback, it receives a negative reward (-1), indicating that no further useful information is available. In the context of heart occlusion, the authors in ~\citep{li2020coronary} propose an RL approach to diagnose cardiovascular occlusion using the A3C algorithm. This approach involves multiple independent agents working in parallel, with each agent having an auxiliary network that can update parameters in the primary structure. In this configuration, the state space consists of the patient's data, while the action space includes the predicted degree of occlusion for each of the eight coronary arteries. The reward is provided based on the accuracy of the model's predictions for the degree of occlusion, with higher rewards given for more accurate predictions.

The formulation of RL problems can take various forms, but several challenges persist, making it a complex task. One of the key limitations in RL problem design is the construction of an effective reward function, particularly in tasks such as video summarization. According to ~\citep{mathews2022unsupervised}, the rewards of representativeness and diversity are often considered generic; however, the authors enhance these by incorporating classification scores and a similarity index. Despite these improvements, a key question remains as to whether these reward functions can be generalized to other human organs, given that the solution was tested only on lung-related data.

\subsection{RL Implementation and Training}

The implementation of RL varies based on the nature of the input data and the specific task being addressed. In this section, we will explore the application of RL in tasks related to disease diagnosis and decision-making using US images or videos. We will then review the methods and technologies utilized in each study for different tasks within this domain.

\begin{figure}[htbp]
    \begin{center}
    \centering
\includegraphics[width=1.0\linewidth]{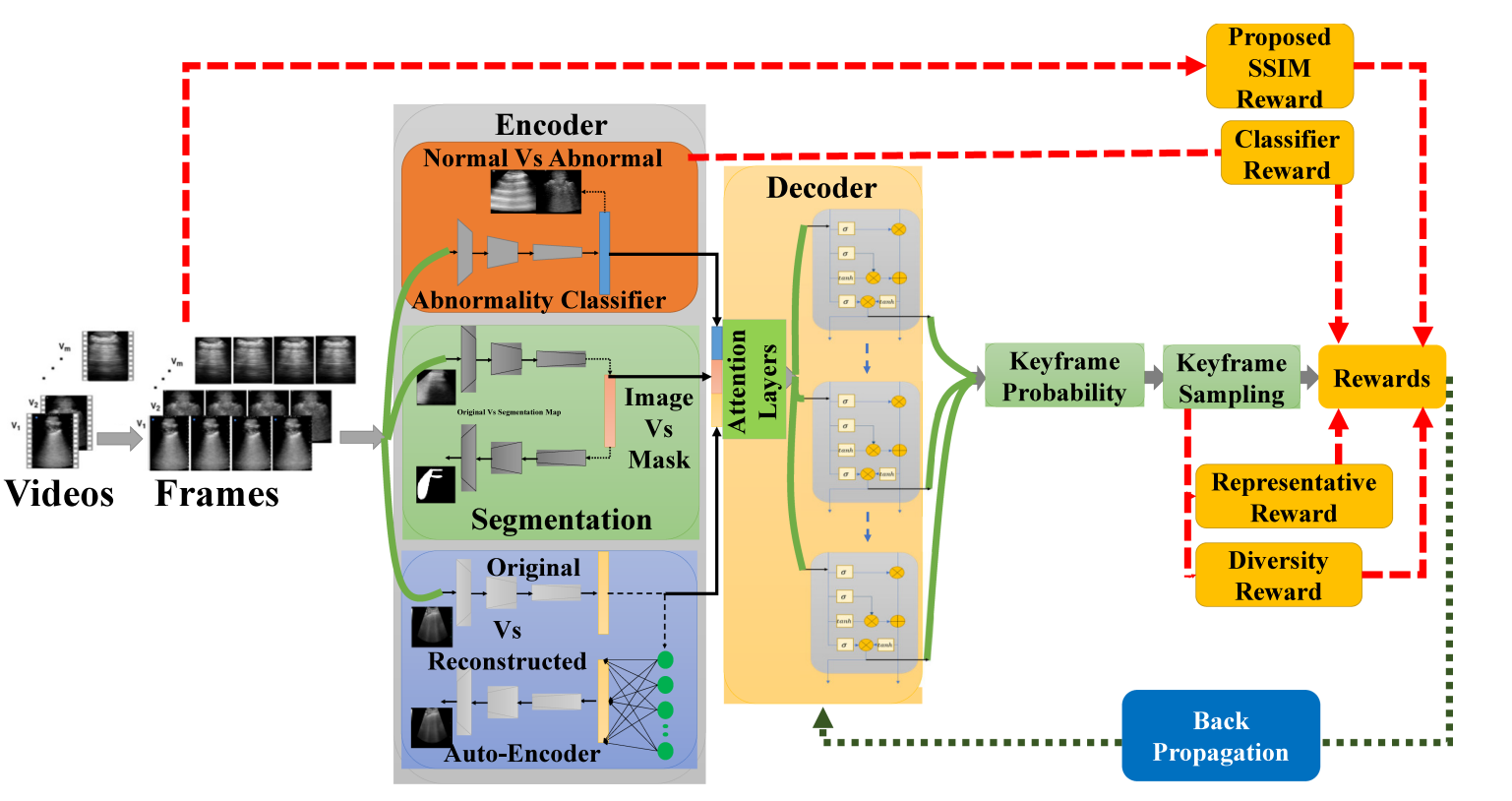}
   \caption{Multi-Latent Space Video Summarization Framework ~\citep{mathews2022unsupervised}.}
      \vspace{-30pt}
    \label{summarization}
    \end{center}
\end{figure}

In video summarization, several studies have explored different architectures to effectively capture and process spatio-temporal information. For instance, ~\citep{liu2020ultrasound} use a CNN for feature extraction, followed by a Bi-LSTM to analyze temporal dependencies from both past and future frames in US videos. This combination enables the model to capture complex temporal relationships and generate accurate frame importance scores for summarization. Similarly, ~\citep{liu2022video} propose a framework consisting of a spatio-temporal CNN with 3D convolutions for feature extraction, followed by a 3D Spatio-Temporal U-Net (3DST-UNet) to model both spatial and temporal dependencies within the video sequence. The 3DST-UNet outputs latent scores for each frame, which are then used by a RL agent to select frames for the summary. The study also employs two feature extraction methods: I3D, which inflates 2D convolutions pre-trained on large image datasets (ImageNet) into 3D convolutions, and ST3D, which uses 3D convolutions to directly capture spatio-temporal dependencies. Compared to the Bi-LSTM, the 3DST-UNet outperforms in capturing more complex spatio-temporal dependencies, making it more effective for video summarization tasks. Furthermore, the authors in ~\citep{mathews2022unsupervised,mathews2021vid} propose a combined encoder that integrates a classifier encoder to detect features linked to abnormal lung conditions, a segmentation encoder that highlights biological landmarks within the frames, and an autoencoder that generates a compressed representation of the frames. This architecture also includes a Multi-Head Self-Attention mechanism to highlight important information. In their decoder, they use both Bi-LSTM and transformer-based architectures, further enhancing the model’s ability to capture complex temporal and spatial dependencies. Figure \ref{summarization} provides a detailed overview of the proposed framework in ~\citep{mathews2022unsupervised}.

\begin{figure}
    \begin{center}
    \centering
\includegraphics[width=1.0\linewidth]{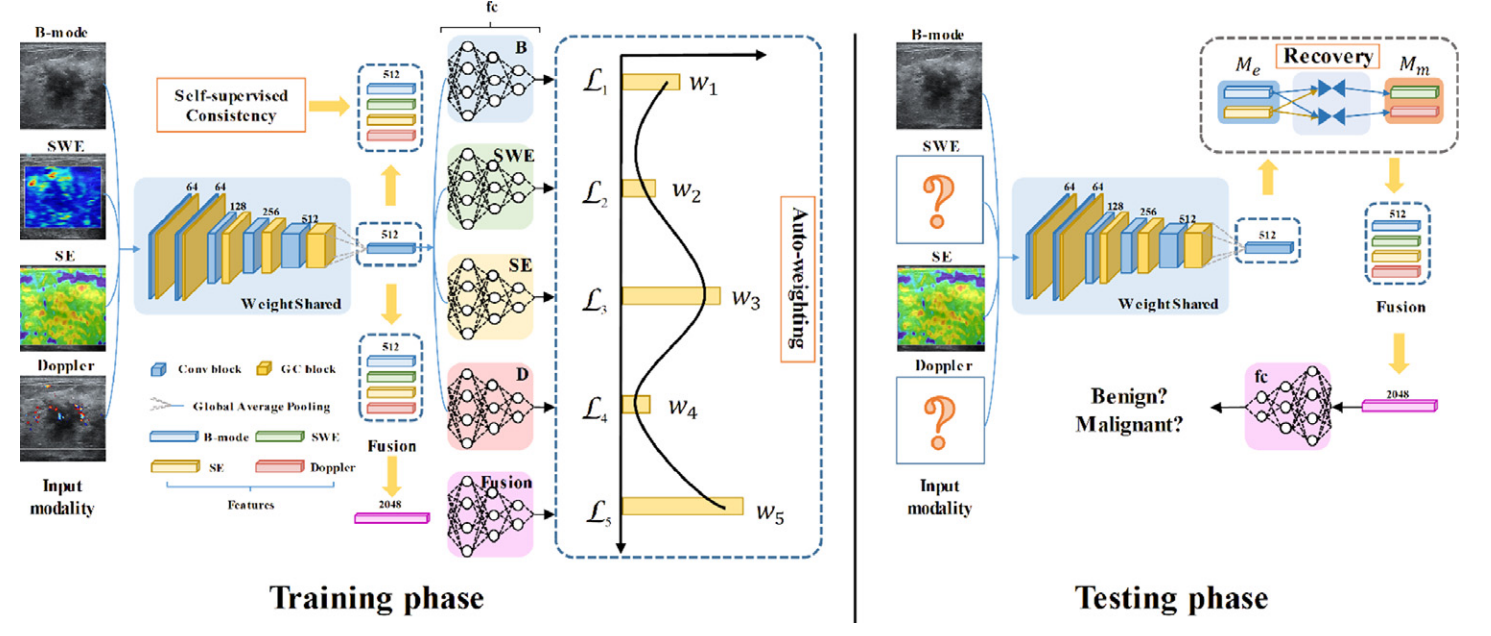}
    \caption{Auto-weighting Multi-modal with Missing Modalities Framework ~\citep{huang2021aw3m}.}
       \vspace{-30pt}
    \label{AW3M}
    \end{center}
\end{figure}

In the context of image classification, RL can be implemented, as proposed in this work ~\citep{narmatha2023ovarian}, where the authors developed a framework incorporating a CNN-based feature extractor and a DQN-based classifier. For the CNN model, the authors used five convolutional layers along with one fully connected layer, with input images sized at 224x224 pixels. To optimize the fine-tuning of hyperparameters, they applied Harris Hawks Optimization (HHO) to determine the optimal set of hyperparameters, including the discount factor, batch size, and dropout rate. Their framework demonstrated strong performance in terms of accuracy, recall, and precision when compared to traditional classification models such as ANN, CNN, and AlexNet, with a significant reduction in computation time attributed to the use of HHO.
Alternatively, RL can be integrated into the classification framework not as a classifier, as described in this work ~\citep{wang2020auto}. In this study, a classification framework is proposed to detect the presence of breast nodules using four US modalities. This framework employs a multi-task learning model that incorporates four ResNet18 networks. In addition to independent loss functions for each stream, an extra fusion loss is included to promote competition among the modalities. Within this framework, RL is utilized to automatically assess the significance of each modality and evaluate their respective contributions.
In the same context of breast nodule classification, ~\citep{huang2021aw3m} proposed a framework based on a multi-stream CNN and RL. This framework uses an ImageNet-pretrained ResNet18 as a feature extractor for US modalities, incorporating Global Context blocks to enable the model to capture global information from the input images. Additionally, the framework includes a fusion stream block to integrate information from multiple imaging modalities, and a recovery block to address potential missing modality data, restoring key features by leveraging the available modalities. Figure \ref{AW3M} provides an overview of the complete solution. For training this framework, a learning rate of 0.0003 and a mini-batch size of 16 were used. In the auto-weighting mechanism, each individual loss weight was initialized to 0.2 and updated throughout the training for 50 epochs. 

In another scenario, specifically for thyroid nodule classification, it is essential to accurately localize the region where the nodules are located. As a result, in this work ~\citep{zhang2023detection}, the authors employed a Region Recommendation Network to obtain the coordinates of the nodule area within the entire US image. They then used RL to adjust the predicted region coordinates to better match the actual location of the nodules. Subsequently, EfficientNet was used to process the image in these regions and classify it, determining whether the region contains a thyroid nodule.

Beyond the task of image registration, RL has shown significant potential in improving disease diagnosis by enhancing decision-making processes. For example, in the study by ~\citep{zeng2020hierarchical}, RL is used to develop an automatic intraoperative registration technique to assist decision-making during surgery. The authors employed hierarchical RL Q-learning in combination with the PointNet++ network and K-nearest neighbors (KNN) to perform 3D greedy registration between CT and US images. Prior to the registration, they enhanced the visibility of bone shadows and segmented the bone surfaces to facilitate the alignment process.

In another study by ~\citep{shanmuganathan2024two}, RL is applied for point-based registration of 3D US images from different acquisition windows, with the goal of recognizing anatomical structures, improving image quality, and fusing images to obtain a complete view of the heart. The authors utilized Multi-Agent Deep Reinforcement Learning (MADRL) for point-based registration, performing both rigid and non-rigid registration. For rigid registration, they used SimpleITK to calculate the optimal transformation that aligns the fixed and moving images based on landmarks predicted by the RL model. The RL model consists of three agents, each responsible for predicting three landmarks. Following rigid registration, non-rigid registration is applied using SimpleElastix to allow for more flexible, complex transformations that enable the images to deform and better align with the fixed image. This non-rigid registration is performed specifically on the end-diastolic frame in heart imaging, as it provides more stable results. Additionally, B-splines are used to map each voxel between the fixed and moving images, ensuring accurate alignment.

In a case of thyroid cancer diagnosis using RL-based approach,  the authors in ~\citep{xie2023reinforced} used Collaborative Training with ResNet50 and Inception\_v3 pretrained on ImageNet and to protect sensitive medical data, the proposed system uses differential privacy. This ensures that training parameters (such as model weights) are protected during transmission between hospitals, preventing indirect data leakage caused by attacks on the model parameters.
From cardiovascular perspective, ~\citep{li2020coronary} employed multi-task learning to account for the correlation between coronary arteries. Their research found that when one artery is severely blocked, it is likely that other arteries will also show some degree of occlusion. Multi-task learning allows the model to predict occlusions in multiple arteries simultaneously using two mechanisms: soft sharing, where each artery has its own prediction model with independent parameters, and hard sharing, which reduces overfitting by sharing some parameters between the models of different arteries. Additionally, they utilized transfer learning, transferring knowledge gained from arteries with more data to improve predictions for arteries with fewer training examples. The authors also incorporated continual learning to retain knowledge from previous tasks while adapting to new tasks, and progressive learning to prevent catastrophic forgetting, ensuring that the model does not lose previously learned information when learning new tasks. Each artery has its own neural network, and lateral connections enable knowledge sharing between the networks. In the RL block, the authors use an encoder-decoder structure with an attention mechanism in each A3C agent for feature extraction. The processed data is then passed through a BiLSTM (Bidirectional Long Short-Term Memory) network to better capture the context by considering both past and future information.

RL has shown promising results in medical image processing tasks such as video summarization, image classification, and image registration, but several limitations persist. RL models often require complex architectures to handle spatio-temporal dependencies or multi-modal data, which can result in high computational costs, difficulty in model convergence, and challenges in hyperparameter tuning. For example, in video summarization, architectures like CNNs combined with Bi-LSTM or 3DST-UNet are effective at capturing temporal and spatial relationships, yet they demand significant computational resources and large annotated datasets. Similarly, in image classification and registration, RL struggles with stability and efficiency, particularly when dealing with non-rigid deformations or real-time applications. Furthermore, while multi-task learning offers a potential solution for integrating multiple tasks into a single model, it introduces additional complexity and the risk of overfitting. The limitations identified in this review primarily relate to these issues with RL algorithms, such as DQN, which face convergence problems and require more robust solutions. Future research should focus on developing more efficient and stable RL algorithms, while also addressing the computational demands of DL models, to improve their feasibility and robustness for clinical applications.

\subsection{RL Validation and Fine-Tuning}

Validating RL models is essential to determine their ability to generalize to unseen data and and perform effectively across various scenarios. Typically in diagnosis tasks, a test dataset is employed for both model validation and fine-tuning. This section examines the metrics used to evaluate RL-based solutions for each US-related diagnostic task and also outlines the hyperparameter fine-tuning approaches applied in each case.

In video summarization, common evaluation metrics include F1 score, precision, and recall, with ~\citep{liu2022video} also incorporating summary length to assess the effectiveness of the generated video summaries. According to the authors, their approach preserves critical information for creating meaningful summaries and is applicable to other medical diagnostic videos, such as endoscopy or physiotherapy assessment videos. In ~\citep{liu2020ultrasound}, the authors assess their solution against state-of-the-art video summarization methods, demonstrating that their approach outperforms existing solutions. They also conduct experiments using different summary length constraints (15\%, 25\%, 35\%, and 45\%), with the results showing that increasing the number of selected key frames improves the F1 scores. Similarly, the approaches in ~\citep{mathews2022unsupervised, mathews2021vid} also demonstrate strong results, highlighting the novelty of their method, which integrates classification and segmentation during the summarization process. Their solution is particularly suitable for telemedicine and Point-of-Care Ultrasound (POCUS), as it reduces video size by approximately 75\%, making it ideal for environments with limited bandwidth and compatible with standard CPUs, without requiring high-end GPUs. Regarding hyperparameter tuning, most studies do not provide specific details, except for ~\citep{liu2020ultrasound}, where stochastic gradient descent with momentum of 0.9 and a weight decay of 0.00001 is used. The initial learning rate is set to 0.0001 and reduced by half every 50 epochs, with a maximum of 300 training epochs.

For image classification tasks, standard evaluation metrics such as accuracy, sensitivity, precision, specificity, and F1-score are commonly used. In the work by ~\citep{narmatha2023ovarian}, the authors validate their model using additional metrics, including Intersection over Union (IoU), F-measure, and confusion matrix. Their approach outperforms traditional deep neural networks for classification, as the HHQ framework saves computational time by efficiently deriving factors and simplifying the fine-tuning process of the DQN. In ~\citep{wang2020auto}, the authors emphasize the importance of multi-modal US data, noting that single modality is insufficient for accurate breast cancer diagnosis, leading to lower classification accuracy. The RL-based approach is more effective because it enables the model to learn and adjust the weights of each modality based on its relevance to the classification task, rather than treating all modalities equally. Similarly, ~\citep{huang2021aw3m} demonstrates the effectiveness of multi-modal classification, highlighting how RL facilitates dynamic weight learning for different modalities. Regarding hyperparameter fine-tuning, not all studies explicitly describe their tuning process. For instance, ~\citep{huang2021aw3m} used a learning rate of 0.0003 and a mini-batch size of 16. In their auto-weighting framework, each individual loss weight was initialized at 0.2 and updated over 50 epochs. The agent parameters were optimized using the Adam optimizer, with a learning rate of 0.0003 and a total of 50 epochs. In ~\citep{zhang2023detection}, additional hyperparameters were specified, including an epsilon value of 0.9. The model was trained with a batch size of 32 over 100 epochs. The Adam optimizer was used with a learning rate of 0.0002. The loss function employed for training was categorical cross-entropy.

In the context of image registration, the authors in ~\citep{shanmuganathan2024two} assess their RL-based solution for aligning 4D echocardiogram images obtained from various sonographic windows using two validation approaches. First, they perform visual validation using the Slicer3D tool to check the accuracy of the alignment between the registered 4D image sequences and the corresponding left ventricle annotations. Then, they validate the solution using quantitative metrics, including the Dice score (to measure the overlap between registered images), Hausdorff distance (to assess the alignment of image boundaries), mean squared error, and Mattes mutual information, all of which evaluate the quality of the image registration process. Their method demonstrates favorable results compared to the ANTs library, a widely used toolkit for medical image registration and segmentation, especially in terms of Dice overlap and Hausdorff distance metrics.
In contrast, the study by ~\citep{zeng2020hierarchical} evaluates the performance of their model for registering CT and US images using metrics like Target Registration Error (TRE), success rate, and average registration time. They compare their RL-based solution with three well-known point cloud registration methods: Iterative Closest Point (ICP), Coherent Point Drift (CPD), and Normal Distribution Transform (NDT). While these methods performed well in general, they struggled with challenging cases, particularly those involving significant perturbations in the vertical axis, resulting in relatively low success rates. In contrast, the RL-based solution showed promising results, although it had a higher run time compared to these state-of-the-art methods.

Regarding hyperparameter configurations in image registration tasks, different settings were used for training the RL models. For instance, in the study ~\citep{zeng2020hierarchical}, a learning rate of 0.000006 was used, and the total training time was 1 week and 2 days for the proposed method to converge. In contrast, in the study ~\citep{shanmuganathan2024two}, the batch size was set to 128 for the apical data and 64 for the parasternal data, with the number of episodes set to 75,000. 

In another diagnostic decision case, ~\citep{xie2023reinforced} validated their RL-based solution using accuracy, which measures the overall correctness of the diagnosis, sensitivity, which assesses the model's ability to identify malignant nodules, and specificity, which indicates its ability to predict benign nodules. Their RL-based approach for thyroid cancer diagnosis demonstrated promising results compared to traditional methods like majority voting. By integrating multiple diagnostic information sources while preserving privacy, their approach holds potential for future improvements, particularly in performing multi-level fusion to better replicate real-world diagnostic processes. Similarly, in ~\citep{li2020coronary}, an RL-based solution for predicting coronary artery occlusions, when combined with multi-task learning, outperformed several machine learning algorithms, including Naive Bayes, decision trees, single RNN, single GRU, single LSTM, bidirectional LSTM, and bidirectional GRU, even with variations in the number of arteries being diagnosed.

Overall, the proposed solution demonstrates promising results across various tasks, but it still has some limitations. One key limitation is the limited diversity of the datasets used for testing, as well as the lack of variation in the US machines and probes employed for data collection. Additionally, there is insufficient clarity regarding the hyperparameters used in training the RL models, which remains another area of uncertainty. Another limitation concerns the computation time, particularly in the case of multi-modal data, which requires optimization. Additionally, the robustness of future RL solutions in handling missing data remains an area that requires further attention.

\section{Challenges and Future Directions}
\label{Sec: Challenges}
\begin{figure*}
    \begin{center}
    \centering
\includegraphics[width=0.8\linewidth]{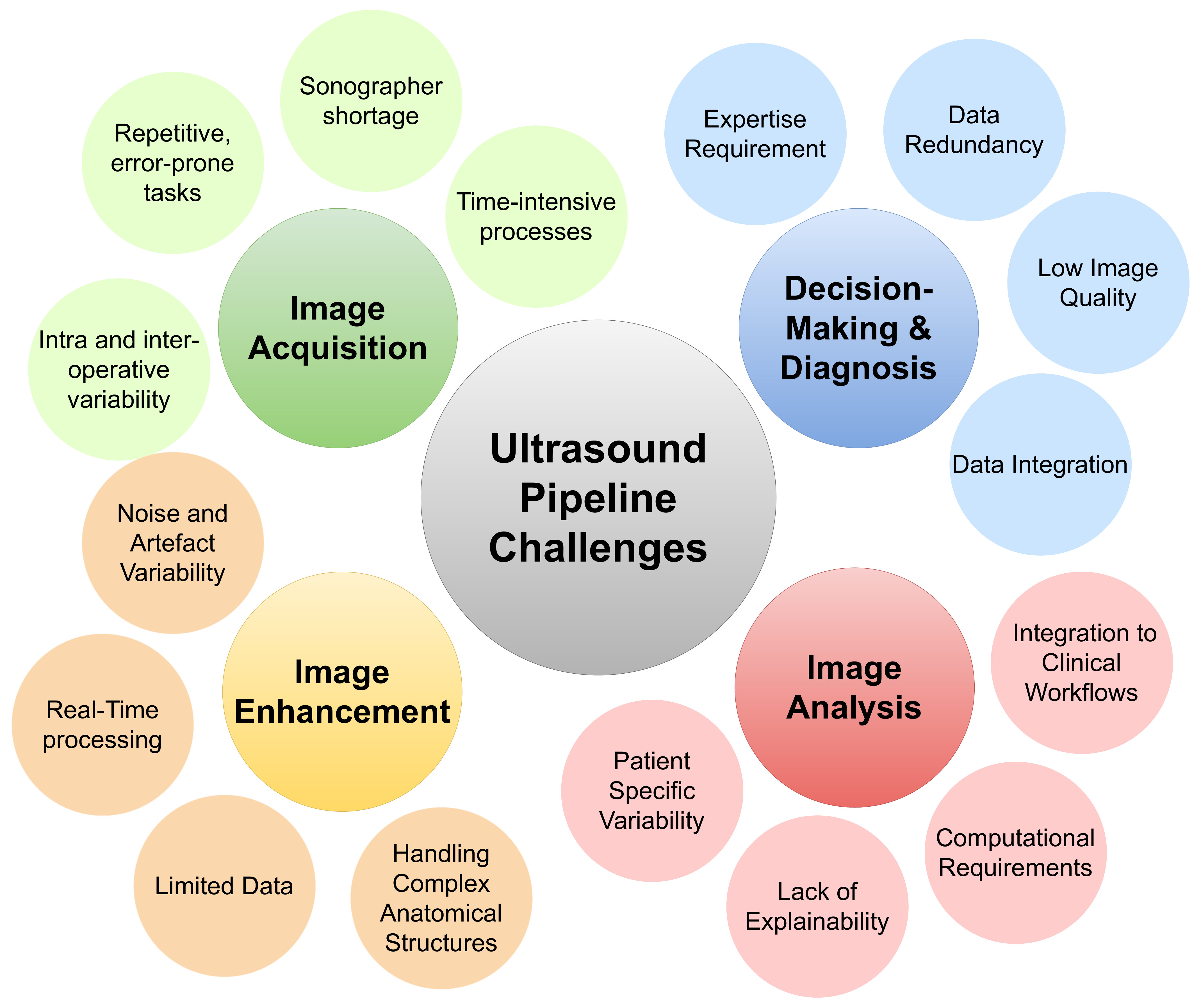}
    \caption{Overview of Challenges in the Ultrasound Pipeline Addressed by Reinforcement Learning Applications.}
       \vspace{-30pt}
    \label{us_challenges}
    \end{center}
\end{figure*}

The integration of RL into US imaging has shown great promise in enhancing various stages of the imaging pipeline, from image acquisition to diagnosis. Figure \ref{us_challenges} provides a summary of the current challenges in the US pipeline that are being addressed by RL applications discussed in this survey. However, several challenges remain unaddressed, which limit its full potential and broader adoption. In this section, we delve into the challenges related to RL application in US imaging and provide an overview of potential future directions for improving this field.

\subsection{Data Availability}

The availability of data remains one of the most significant challenges in applying RL to US imaging. A review of the literature on RL applications in US imaging reveals that public datasets containing US images are extremely limited. For instance, in the area of US image acquisition, only ~\citep{macuradynamic}, ~\citep{hu2024probe}, and ~\citep{shen2023towards} utilized publicly available datasets, while other studies relied on private datasets collected from medical institutions, data from phantoms, or simulated data. However, phantom and simulated data can sometimes have limitations and may not always fully replicate the complexity of real anatomical structures in human organs. Similar to other stages of the imaging pipeline, the use of public datasets is constrained by their scarcity. The primary barrier to the availability of US imaging data is the ethical and legal constraints surrounding the sharing of human data.

The issue of data availability is not the only challenge; variations in the data and its scalability also pose significant obstacles in RL-based studies. Several papers highlight this issue. For example, ~\citep{shida2024robotic} emphasizes the limitations of their study, noting that their dataset consists solely of data from male subjects in their teens and twenties, meaning that their solution has not been validated across a broader range of patient ages and genders. Similarly, ~\citep{su2024fully} acknowledges that their system faces limitations due to the lack of variation in their training dataset, which does not include diverse nodule sizes or artifacts. In another example, ~\citep{luo2024multi} reports failures in their solution due to insufficient real-world data, with manual adjustment of arm positioning during robotic US scanning being used to mitigate this limitation. In the context of autonomous US scanning, ~\citep{deng2024portable} highlights the absence of open-source US datasets that include probe pose data under consistent force conditions, which prevents the RL solution from being generalized across all scenarios. Furthermore, ~\citep{macuradynamic} points to the lack of 3D imaging, which occasionally forces researchers to rely on 2D images instead.

\subsection{Human Anatomical Variability}

A significant challenge that hinders the application of RL in the US image scanning process is the variability of human anatomy. Human anatomy is considered highly complex from an RL perspective due to factors such as deformability, the soft nature of the environment, and the lack of standardized surfaces. For example, the surface of the abdomen is flat, while the arm has a sloped surface, and there are notable differences between male and female anatomies. An RL model must learn to account for all these factors in order to generalize across the diverse range of human anatomies. This task is particularly difficult due to the limited availability of data, as previously discussed. Data collected from volunteers typically involves only 20 to 30 participants per study, and when using simulated data or data from phantoms, the representation is constrained to specific cases that cannot be generalized to all human anatomies. For instance, in the work by ~\citep{luo2024multi}, the system encounters difficulties with visual disturbances such as artifacts and occlusions, which negatively impact its performance.
To address this challenge in autonomous US image acquisition, some studies propose the use of a RGB camera to capture the entire environment, allowing the robot to learn from the camera’s input and better understand the anatomy being scanned. This approach helps validate the system across multiple anatomical regions, such as the abdomen and lumbar spine ~\citep{ning2021autonomic}. Additionally, ~\citep{su2024fully} used depth cameras to assist the robot in directing its focus toward the target, the thyroid, although the scanning process remained primarily based on the US images themselves.

\subsection{Sim-to-Real Transferability}
There are significant differences between the simulated environments used for training RL models and real-world scenarios. These differences can cause RL solutions to fail during validation in real-world settings. For example, in the work by ~\citep{luo2024multi}, the robot became trapped in local optima due to a maximum step limit, which led to the need for reconfiguring the experiment by increasing the step limit. To mitigate the impact of these challenges related to transferability, some studies explore alternative approaches for training RL models. In ~\citep{raina2024coaching}, for instance, a ``human coach" intervened during the training process by adjusting the probe's position, orientation, or the force applied to the phantom. After these interventions, the reward from the resulting policy showed improvements at each step of the RL model's training process. IRL is also proposed as a solution, leveraging expert knowledge to ensure stable transferability from the training to the test environment ~\citep{ning2023inverse}.
Another issue related to the transferability from simulation to real-world scenarios is the subjectivity of labeling, which can hinder the validation of models in real-life scenarios. As highlighted in the work by ~\citep{liu2022video}, the subjective annotation of videos makes video summarization tasks particularly challenging.

\subsection{Lack of interpretability (Explainability)}
The lack of interpretability in RL models poses a significant challenge in US image analysis and enhancement, as these systems often function as ``black boxes", making their decision-making processes opaque. This limits trust and adoption by clinicians, who need to understand how and why certain features are emphasized or altered, especially in high-stakes diagnostic scenarios. Furthermore, this opacity complicates error detection, regulatory approval, and clinical training, as the rationale behind the model's outputs remains unclear. Addressing this issue requires integrating explainable AI (XAI) techniques such as attention mechanisms, saliency maps, or feature visualization to provide insights into model behavior. Simplifying reward structures to align with measurable clinical improvements and embedding domain knowledge into the system can further enhance transparency. By offering intuitive, interpretable outputs and involving clinicians in the development process, RL-based US systems can improve trust, usability, and clinical relevance.


\subsection{Future Directions}

RL has shown promising results as a solution for decision-making in autonomous US scanning across several organs. However, existing studies still face certain limitations that can be addressed in future work.
In real-world US scanning, medical professionals often need additional image processing tools once the target area is reached, such as image segmentation and classification. For example, during cardiac US scans, experts typically classify images based on the acquired view. Therefore, integrating real-time analysis to identify specific areas, such as segmenting regions or classifying images according to specific features, would be beneficial. However, patient safety concerns currently limit robotic scanning to fixed, slow speeds. Incorporating real-time image analysis could further slow the process, requiring patients to hold still for longer periods, which may cause discomfort. Consequently, a balance must be struck between scanning speed and the feasibility of real-time image processing.
Another area for improvement is the design of the action space, as existing works generally use discrete actions, while human scanning typically involves continuous movements. Future work should focus on developing continuous action spaces to better handle more complex scanning tasks.
Additionally, the studies reviewed use a variety of methods for training models, such as using grids to represent scanned areas, or relying on physical simulators and other image modalities. Therefore, developing a standardized simulation environment that can be adapted to any human organ is crucial for advancing the training and validation of RL models in this field. This approach would also help address ethical challenges related to obtaining real human data.

Despite the promising potential of RL in US image enhancement and analysis, several critical tasks remain underexplored. For example, image registration, which is essential for multi-view or multi-modal image fusion, has not been extensively investigated in the context of RL. This is particularly important for deformable tissues in US imaging, where accurate registration can enhance the precision of diagnostic procedures. Another area that remains relatively unexplored is the adaptive image enhancement, where RL could dynamically adjust image parameters in real-time to optimize image quality based on the specific features being analyzed. On the other hand, image analysis tasks like multi-target segmentation, which involves the simultaneous identification of multiple anatomical structures, a challenging tasks that could greatly benefit from RL methods, particularly in complex scenarios such as fetal imaging or cardiac US. Additionally, inter-frame consistency in real-time US video processing is another domain where RL could offer significant advantages by ensuring consistent tracking of landmarks or lesions across frames. While RL has been applied in landmark detection and segmentation, there is a considerable opportunity to explore its potential for dynamic decision-making in US image analysis, where RL systems could adapt to diverse imaging conditions or patient-specific characteristics. These underexplored areas present significant opportunities for advancing US imaging enhancements and analysis tasks through the application of RL.

The application of RL in the context of diagnosis and decision-making using US images remains limited. Based on the related work reviewed in this survey, existing studies have primarily focused on specific cases, such as video summarization and image classification, and have addressed only a few human organs. This highlights significant potential for further exploration in US-based diagnosis. For instance, in the context of valvular diagnosis, cardiologists currently perform manual tracing of aortic Doppler US, a process that is both time-consuming and prone to errors. Developing RL-based solutions to automate this task would represent a valuable contribution to the field. Another example is risk stratification and phenogrouping, where DL solutions have been applied to cardiovascular risk assessment and phenogrouping, such as for atherosclerotic cardiovascular disease or identifying asymptomatic patients at risk of adverse cardiovascular events ~\citep{esau2022approach, sabovvcik2021echocardiographic}. However, RL has yet to fully exploit the significant potential of US imaging, which holds valuable diagnostic information. Moreover, there is a clear need for multi-task learning to handle the complex, multi-faceted tasks required for accurate and efficient diagnosis, as it can integrate multiple diagnostic aspects and improve performance across different tasks simultaneously.

The application of RL in US imaging pipeline is still in its early stages, and there are many potential areas yet to be explored, such as its use in surgery. In surgical cases, US images are sometimes used by surgeons to guide their interventions. However, US images are often blurry and noisy, which reduces their effectiveness. As a result, surgeons typically rely on other pre-operative imaging modalities with higher resolution and clearer images for their procedures. Consequently, a promising area of research would be to explore how US images could be effectively utilized during surgery, despite their inherent limitations.
One study by ~\citep{ao2025saferplan} proposed an RL-based solution for using US during robotic spine surgery. In this delicate procedure, the goal is to implant pedicle screws to stabilize the spine while avoiding nerve damage. The proposed RL solution continuously adjusts the surgical plan based on real-time intraoperative US images. The RL agent is responsible for determining actions that move the surgical drill through the patient's tissues until it reaches the optimal position for screw placement, all while avoiding bone breaches and soft tissue injuries. This work presents a specific example of how RL can be applied in the context of US imaging for surgery and encourages the exploration of other potential applications. While real-time validation may be challenging, the creation of simulated environments for testing RL models provides a promising starting point for further research.

\section{Conclusion}
\label{Sec: Conc}
This survey aims to highlight the advancements and challenges in the application of RL within the context of US image modalities. The US pipeline, which begins with image acquisition, followed by image enhancement, image analysis, and concluding with decision-making and diagnosis, is explored. We review existing works on RL, discussing the key stages necessary for the development of an RL solution. These stages include data preparation and processing, as no DL or RL model can function effectively without data; RL problem formulation and algorithm definition; RL implementation and training, where we describe DL solutions used for training RL models as well as hyperparameter definition; and finally, RL validation and fine-tuning, illustrating how RL models are validated within the US image context.
Despite the advantages of applying RL in the US pipeline, the adoption of RL is not without its challenges. Therefore, we identify the critical challenges and open issues associated with RL in the US context. Finally, we provide insights into future research directions aimed at overcoming these challenges and advancing RL integration in US applications.

\bibliographystyle{bst/sn-basic}
\bibliography{sn-bibliography}

\end{document}